\documentclass[useAMS,usenatbib]{mn2e} 
\usepackage{epsfig,lscape,amsmath,amssymb} 
\usepackage{graphicx,longtable,times,threeparttable,multirow} 
\usepackage{rotating,color}
\usepackage[T1]{fontenc} 
\usepackage{aecompl}
\usepackage{bm}

\DeclareGraphicsExtensions{.eps,.eps.gz,.epsi}

\newcommand\aj{AJ} \newcommand\apj{ApJ} %
\newcommand\aap{A\&A} \newcommand\pasp{PASP} \newcommand\mnras{MNRAS}%
\newcommand\apjs{ApJS}  %

\begin{document}

\title[]{The LAMOST Stellar Parameter Pipeline at Peking University --- LSP3}
\author[Xiang et al.]{M. S. Xiang$^{1}$\thanks{E-mail: xms@pku.edu.cn},
        X. W. Liu$^{1,2}$\thanks{E-mail: x.liu@pku.edu.cn}, H. B. Yuan$^{2}$,
        Y. Huang$^{1}$, Z. Y. Huo$^{3}$, H. W. Zhang$^{1}$, 
        \newauthor B. Q. Chen$^{1}$, H. H. Zhang$^{1}$, N. C. Sun$^{1}$, C. Wang$^{1}$, 
        Y. H. Zhao$^{3}$, J. R. Shi$^{3}$, A. L. Luo$^{3}$, 
        \newauthor G. P. Li$^{4}$, Y. Wu$^{3}$, Z. R. Bai$^{3}$, Y. Zhang$^{4}$, 
         Y. H. Hou$^{4}$, H. L. Yuan$^{3}$, G. W. Li$^{3}$
\\ \\ $1$ Department of Astronomy, Peking University, Beijing 100871, P. R. China \\
$2$ Kavli Institute for Astronomy and Astrophysics, Peking University, Beijing 100871, P. R. China \\
$3$ Key Laboratory of Optical Astronomy, National Astronomical Observatories, 
    Chinese Academy of Sciences, Beijing 100012, P. R. China \\
$4$ Nanjing Institute of Astronomical Optics \& Technology, National Astronomical Observatories, 
    Chinese Academy of Sciences, Nanjing 210042, P. R. China \\}

\date{Received:}

\pagerange{\pageref{firstpage}--\pageref{lastpage}} \pubyear{2013}

\maketitle

\label{firstpage}

\begin{abstract}
{We introduce the LAMOST Stellar Parameter Pipeline at Peking University --- LSP3, 
developed and implemented for the determinations of radial velocity $V_{\rm r}$ 
and stellar atmospheric parameters (effective temperature $T_{\rm eff}$, 
surface gravity log\,$g$, metallicity [Fe/H]) for the LAMOST Spectroscopic 
Survey of the Galactic Anti-center (LSS-GAC). We describe the algorithms 
of LSP3 and examine the accuracy of parameters yielded by it. The precision 
and accuracy of parameters yielded are investigated by comparing results of 
multi-epoch observations and of candidate members of open and globular clusters, 
with photometric calibration, as well as with independent determinations 
available from a number of external databases, including the PASTEL archive, 
the APOGEE, SDSS and RAVE surveys, as well as those released in the LAMOST DR1. 
The uncertainties of LSP3 parameters are characterized 
and quantified as a function of the spectral signal-to-noise ratio (SNR) and 
stellar atmospheric parameters. We conclude that the current implementation 
of LSP3 has achieved an accuracy of 5.0\,km\,s$^{-1}$, 150\,K, 0.25\,dex, 
0.15\,dex for the radial velocity, effective temperature, surface gravity 
and metallicity, respectively, for LSS-GAC spectra of FGK stars of SNRs per 
pixel higher than 10. The LSP3 has been applied to over a million LSS-GAC 
spectra collected hitherto. Stellar parameters yielded by the LSP3 will be 
released to the general public following the data policy of LAMOST, together 
with estimates of the interstellar extinction $E(B-V)$ and stellar distances, 
deduced by combining spectroscopic and multi-band photometric measurements 
using a variety of techniques.}
\end{abstract}

\section{Introduction}
\label{sect:intro}

The structure and origin of the Galactic disk(s) are among the hottest 
debating issues of the Galactic astronomy. An archeological approach to 
the problems relies on the collection of information in multi-dimensional 
phase space for large samples of stars. The on-going LAMOST Spectroscopic 
Survey of the Galactic Anti-center (LSS-GAC), will collect medium-to-low 
resolution ($R$ $\sim$\,1800) spectra of millions of stars and deliver 
fundamental stellar parameters, including radial velocity ($V_{\rm r}$) 
and stellar atmospheric parameters (effective temperature $T_{\rm eff}$, 
surface gravity log\,$g$, metallicity [Fe/H]), as well as elemental abundance 
ratios (the $\alpha-$element to iron abundance ratio [$\alpha$/Fe], and 
the carbon to iron abundance ratio [C/Fe]), deducible from the spectra 
(Liu et al. 2014). Combined with accurate optical and infrared (IR) photometry, 
the LSS-GAC will also deliver estimates of the interstellar extinction 
and distances to individual stars (Yuan et al. 2014, submitted; Paper\,III hereafter). 
Together with determinations of proper motions, either from existing catalogs, 
such as the PPMXL (Roeser et al. 2010) and UCAC4 (Zacharias et al. 2013),
or from the forth coming measurements of Gaia of unprecedented accuracy 
(Perryman et al. 2001), as well as accurate parallaxes (distances) also 
provided by Gaia, the LSS-GAC will provide an unprecedented large stellar 
database in multi-dimensional phase space to study the stellar populations, 
kinematics and chemistry of the Galactic disk and its assemblage and evolution 
history. Deriving accurate stellar parameters from millions of medium-to-low 
resolution fiber spectra in an efficient way is thus of fundamental importance 
to fulfill the scientific goals of LSS-GAC.  

Various methods have been developed in the past to derive stellar atmospheric 
parameters from large number of medium-to-low resolution spectra 
(Recio-Blanco et al. 2006; Lee et al. 2008a, Wu et al. 2011). 
The approaches generally fall into two main categories of method (Wu et al. 2011): 
the minimum distance method (MDM) and non-linear regression method. 
Both categories of method have been applied to large stellar spectroscopic 
surveys, including the SEGUE (Yanny et al. 2009), RAVE (Steinmetz et al. 2006), 
APOGEE (Majewski et al. 2007) and LAMOST (Zhao et al. 2012). 
The MDM is usually based on spectral template matching, and searches for 
the template spectrum that has the shortest distance in parameter space 
from the target spectrum. The $\chi^2$ minimization, cross-correlation, 
weighted mean algorithm, and the $k$-nearest neighbor (KNN), are thought 
to be specific cases of MDM (Wu et al. 2011). Softwares and pipelines developed 
based on those algorithms include the TGMET (Katz et al. 1998), MATISSE 
(Recio-Blanco et al. 2006), SSPP (Lee et al. 2008a), ULySS (Koleva et al. 2009), 
that of Allende Prieto et al. (2006) and of Zwitter et al. (2008). 
The non-linear regression method is sometimes also referred to as the 
artificial neural network (ANN). The method constructs a functional mapping 
between the spectra and stellar atmospheric parameters by training a library 
of template spectra with non-linear algorithms such as the principal component 
analysis (PCA),  and then apply the mapping to target spectra. Related work 
can be found in Re Fiorentin et al. (2007) and Lee et al. (2008a).
In addition to the above two categories of method, other approaches 
have been developed, for example, the line-index method based on the 
relations between the stellar atmospheric parameters and the equivalent 
widths of spectral features and/or photometric colours (Wilhelm et al. 1999; 
Beers et al. 1999; Cenarro et al. 2002). More recently, a Bayesian approach 
to determine stellar atmospheric parameters combing spectral and photometric 
measurements has been developed by Sch\"onrich \& Bergemann (2013). 

Different methods usually have different valid parameter ranges, outside 
which the methods perform poorly. For example, the line-index method usually 
loses sensitivity when the adopted metallic lines are either saturated or 
too weak. Because stars are widely distributed in the parameter space and 
different spectral features have different sensitivity to the atmospheric 
parameters, one can hardly rely on one single method to derive stellar 
atmospheric parameters with a uniform accuracy for all types of star. 
A ``multi-method'' approach, which takes averaged stellar atmospheric 
parameters deduced from a variety of methods that utilizes different 
spectral wavelength ranges, is adopted by the SSPP (Lee et al. 2008a). 
Since systematic errors from different methods cannot be easily compared 
and combined, the systematic errors of the final parameters are 
difficult to estimate.

Almost all of the methods determine the stellar atmospheric parameters
via either direct or indirect comparisons between the target and template 
spectra, a set of comprehensive templates of known parameters covering a 
broad parameter space are thus of fundamental importance. 
Both libraries consisting of empirical and synthetic spectral templates 
have been used in the literature. Lists of the currently available empirical 
and synthetic libraries can be found in Wu et al. (2011) and at a  
website on stellar spectral 
libraries\footnote{\it http://pendientedemigracion.ucm.es/info/Astrof/invest/actividad/spectra.html}. 
An advantage of the empirical libraries is that they consist of spectra 
of ${\it real}$ stars. The disadvantage is that it is often laborious and 
time-consuming to build an empirical spectral library of stars of accurately 
known parameters that cover wide parameter ranges with sufficient resolution 
and homogeneity. It is clear that the parameter space encompassed by the existent 
empirical libraries are limited by our current knowledge of stars in the solar 
neighborhood and the available observations. On the other hand, while it may be 
straightforward to construct a set of synthetic spectra covering homogeneously 
a wide parameter space, it is difficult to assess the robustness of the spectra, 
especially those of very low ($T_{\rm eff}<4500$\,K) or high temperatures 
(for instance, stars of the O, B or A spectral types). Properties of stellar 
spectra are determined not only by basic stellar parameters such as $T_{\rm eff}$, 
log\,$g$, [Fe/H] and [$\alpha$/Fe], but also depend on other parameters and 
processes such as the micro-turbulence and rotation velocities as well as convection. 
Observational uncertainties combined with inadequacies in our understanding 
of stellar atmospheres may lead to unrealistic parameters {$T_{\rm eff}$, log\,$g$, 
[Fe/H] and [$\alpha$/Fe]} by matching an observed medium-to-low resolution 
spectrum with a library of synthetic spectra. 

Thanks to the efforts involving many observers, several empirical spectral 
libraries, including the ELODIE (Prugniel \& Soubiran 2001; Prugniel et al. 2007) 
and MILES (S\'anchez-Bl\'azquez et al. 2006; Falc\'{o}n-Barroso et al. 2011), 
are now available. They cover a wide range of stellar parameters, 
accurately determined with high resolution spectroscopy. 
A pipeline, the LAMOST stellar parameter pipeline (LASP), which is mainly 
based on the Universit\'{e} de Lyon Spectroscopic Analysis Software 
(ULySS; Koleva et al. 2009; Wu et al. 2011) and makes use of the
ELODIE library, has been developed and applied to the LAMOST spectra
at the LAMOST Operation and Development Center of the National Astronomical
Observatories of Chinese Academy of Sciences (NAOC; Wu et al. 2014).
The library is used to determine radial velocities as well as atmospheric parameters. 
Parameters thus determined have been made available via the LAMOST official 
data release (Luo et al. 2012, Bai et al. 2014).

As parts of the LSS-GAC survey, a pipeline, the LAMOST Stellar Parameter 
Pipeline at Peking University -- LSP3, has been developed in parallel. 
Similar to the LASP, the LSP3 determines stellar atmospheric 
parameters by template matching, but using the MILES rather than the 
ELODIE empirical library instead. Compared to the ELODIE spectra which are secured 
using an echelle spectrograph with a very high spectral resolution 
($R\sim42,000$; Prugniel \& Soubiran 2001; Prugniel et al. 2007), 
the MILES spectra are obtained using a long-slit spectrograph at a spectral 
resolution (FWHM$\sim2.4$\,{\AA}) comparable to that of the LAMOST spectra, 
and are accurately flux-calibrated to an accuracy of a few per cent over the 
$\lambda\lambda$3525--7410 wavelength coverage. The stellar atmospheric 
parameters of MILES spectra, determined in most cases using high resolution 
spectroscopy, have been calibrated to a uniform reference (Cenarro et al. 2007). 
On the other hand, the radial velocities of MILES stars are not as accurately 
determined as those in the ELODIE library, given the fairly low spectral 
resolution of MILES spectra. Thus for radial velocity determinations, 
the LSP3 continues to make use of the ELODIE library. 

The LSP3 has been successfully applied to hundreds of thousands spectra 
collected for the LSS-GAC survey. Radial velocities and atmospheric parameters, 
together with other additional parameters such as estimates of 
interstellar extinction and distance to individual stars are released 
as value-added catalogs supplementary to the LAMOST official data release (Paper\,III).  
In this work, we introduce the algorithm and implementation of LSP3 
in detail, and examine the accuracy of stellar parameters yielded by the LSP3, 
by applying the LSP3 to the spectral templates themselves, to LAMOST multi-epoch 
spectra of duplicate stars and to LAMOST and SDSS spectra of member candidates of 
open and globular clusters. Parameters yielded by the LSP3 are compared extensively  
with independent determinations from a number of external databases, 
including the PASTEL archive and the APOGEE, SDSS and RAVE surveys, 
as well as with values published in the LAMOST first data releases (DR1; Bai et al. 2014). 

The paper is organized as follows. In Section\,2, we introduce template 
libraries adopted by the LSP3. Section\,3 describes the methodology of 
LSP3 in detail. In Section\,4, we examine the LSP3 algorithm by applying 
it to the template spectra themselves. In Section\,5, we discuss the 
precision and accuracy of LSP3 by comparing the results yielded by different 
algorithms and by multi-epoch observations of duplicate stars. 
In Section\,6, LSP3 stellar parameters are compared extensively 
with independent determinations from external databases. Calibration 
and error estimates of LSP3 parameters are presented in Section\,7. 
In Section\,8, we discuss the error sources of LSP3 stellar parameters. 
We close with a summary in Section\,9. 

\section{The spectral templates}

\subsection{The MILES and ELODIE libraries}
The MILES library consists of 985 stars spanning wide range 
of stellar atmospheric parameters.
The spectra are obtained with the 2.5\,m Issac Newton Telescope, covering 
the wavelength range 3525 -- 7500\,{\AA} 
at an almost constant resolution of full width at half maximum (FWHM) of about 
2.5\,\AA\, (S\'anchez-Bl\'azquez et al. 2006; Falc\'on-Barroso et al. 2011), 
which is slightly smaller than the typical FWHM of LAMOST spectra ($\sim$\,2.8\,\AA). 
The high accuracy of (relative) flux calibration (S\'anchez-Bl\'azquez et al. 2006) 
and wide coverage of stellar parameters that are homogeneously calibrated  
(Cenarro et al. 2007) make the MILES an ideal 
empirical spectral library for stellar parameter determinations. 
The spectra are converted to match the LAMOST resolution 
by convolving with Gaussians and resampled to 1.0\,{\AA} per pixel. 
The width of the Gaussian, which is allowed to vary with wavelength, 
is taken to be the mean of all the 4000 fibers of LAMOST. 

The MILES spectra are wavelength-calibrated to an 
accuracy of only approximately 10\,km\,s$^{-1}$, not good enough for 
the purpose of radial velocity determinations for the LAMOST spectra. 
We have thus decided to use the ELODIE library as radial velocity 
templates. The library contains 1959 high-resolution spectra of 1388 stars, 
obtained with the ELODIE echelle spectrograph mounted on the Observatoire 
de Haute-Provence 1.93\,m telescope, covering wavelength range 
3900 -- 6800\,{\AA} at a resolving power of 42,000 
(Prugniel \& Soubiran, 2001; Prugniel et al. 2007). 
In addition to spectra of the original resolving power, 
the library also provides another set of spectra, degraded to a 
resolving power of 10,000. We use the latter set of spectra. 
The spectra are further degraded in resolution to match that 
of the LAMOST and resampled to 1.0\,{\AA} per pixel. 
For stars with multiple spectra, only the one flagged as the best is used. 
Note that both the MILES and ELODIE libraries provide spectra in rest 
laboratory wavelengths, calculated using radial velocities determined from 
the spectra. As a test of the accuracy of radial velocities 
adopted by the ELODIE, we cross-correlate the spectra with synthetic 
ones (Munari et al. 2005) of identical stellar atmospheric parameters, 
and find an average velocity residual and standard deviation of 
$-2.8\pm$0.7\,km\,s$^{-1}$. 
The small value of standard deviation reflects the high resolution 
of ELODIE spectra and that the spectra are wavelength-calibrated 
to high a precision. The offset, $-2.8$\,km\,s$^{-1}$, is however significant. 
Its origin is unclear. As shall be shown in Section\,6.1, we correct 
for any systematics in radial velocities determined with the ELODIE 
by calibrating the results against external databases. For comparison, 
a similar exercise for the MILES library yields a residual 
of $-2.6\pm$6.5\,km\,s$^{-1}$. The above exercise also finds a few 
spectra in the ELODIE library that have very large velocity residuals. 
For spectra with residuals in excess of 3$\sigma$ of the mean, 
we have applied corrections to the wavelengths 
using the above determined residuals. 
Finally, given the scarce of stars of  temperatures higher than 7000\,K, 
we have added 360 synthetic spectra (Munari et al. 2005) with temperatures 
between 7000 and 12,000\,K to the ELODIE library as radial velocity templates. 

\subsection{Interpolation of spectra}
Although the MILES library has a decent coverage of the stellar 
parameter space, the coverage is not homogeneous and there are 
clusters and holes in the distribution of stars in the parameter space.
Fig.\,1 shows the distributions of MILES stars in the  
$T_{\rm eff}$ -- log\,$g$ and $T_{\rm eff}$ -- [Fe/H] planes. 
At $T_{\rm eff}\sim$\,5700\,K, for example, a number of stars 
cluster around [Fe/H] $\sim$\,0.1 and $\sim$\,$-0.5$\,dex, 
but few at [Fe/H] $\sim$\,$-0.2$\,dex. 
The presence of clusters and holes in the distributions introduces 
patterns and biases in the resultant stellar atmospheric parameters 
derived by template matching. 

An observational campaign to fill the holes and to further expand 
the parameter space coverage, as well as to expand the template spectral 
wavelength coverage to 9200\,{\AA} to utilize the full potential of 
the LAMOST spectra, especially those from the red-arm, is well under way, 
using the NAOC 2.16\,m telescope and the 2.4\,m telescope of the Yunnan 
Astronomical Observatory.
As a remedy for the time being, we interpolate the MILES spectra 
to fill up the apparent holes in parameter space. To do this, 
we first exclude 85 out of the 985 MILES template stars that do not have 
a complete set of high quality parameters  ($T_{\rm eff}$, log\,$g$ and [Fe/H]). 
Of the 900 stars left, 14 fall close the low log\,$g$ boundaries 
of the distribution in the $T_{\rm eff}$ -- log\,$g$ plane and 
are also not used for the interpolation. 
To interpolate the spectra, the remaining 886 stars are divided 
into four groups in the $T_{\rm eff}$ -- log\,$g$ plane 
(Table\,1). For each group of stars, a third-order polynomial of 
20 coefficients, is used to fit the spectral flux density normalized 
to unity at 5400\,{\AA} at each wavelength as a function of stellar 
atmospheric parameters, $T_{\rm eff}$, log\,$g$ and [Fe/H]. 
Here a third-order polynomial is selected as a compromise 
considering the fact that stellar spectra are a complicated function of 
atmospheric parameters and the limited number as well 
as parameter coverage of the MILES templates. 

To examine the uncertainties of interpolated spectra, 
we have applied the interpolation scheme to the templates themselves. 
Specifically, we drop a template from the library, and fit the 
rest with the polynomial. The fit is then used to calculate the 
dropped spectrum at the given parameters. To characterize the goodness 
of fit, the dispersion of the relative differences between the interpolated 
and observed spectra, is calculated between 4320 -- 5500\,{\AA}, 
the window the LSP3 adopts for estimating stellar parameters. 
When calculating the relative differences, the interpolated spectrum 
is allowed to scale with a third-order polynomial to match the SED 
of the observed one. The exercise is repeated for all template in the library. 
For FGK dwarfs, giants, cool dwarfs and hot stars, the mean and scatter 
of dispersion thus calculated for all templates in the library are 0.007$\pm$0.003, 
0.013$\pm$0.007, 0.023$\pm$0.012 and 0.020$\pm$0.010, 
respectively. Fig.\,2 compares the interpolated and observed template 
spectra for four example stars, one from each of the four groups. 
Compared to typical uncertainties of LAMOST spectra, the errors of 
interpolated templates due to fitting uncertainties are marginal, 
except for LAMOST spectra of very high SNRs. We have also tried to 
interpolate the templates using the formula of Prugniel et al. (2011). 
The results are generally comparable with each other. 

To fill up some of the apparent holes in parameter space 
covered by the MILES library, some fiducial spectra are created by 
interpolating the existing templates in parameter space. 
In total, 416 fiducial spectra are interpolated using the fits generated above 
and added to the MILES library.
The distributions of parameters of the interpolated spectra are 
over-plotted in Fig.\,1 along with those of the original spectra. 
The resultant parameter coverage in $T_{\rm eff}$ -- [Fe/H] plane, 
though still not fully homogeneous, is much improved.
Finally, for the purpose of spectral classification only, 
we have added 18 spectra of white dwarfs (WDs), carbon stars and 
late-M/L type stars retrieved from the SDSS database 
to the final library of spectral templates used by the LSP3.

\begin{table}
\begin{center}
\caption{Partition of the MILES stars for spectral interpolation}
\begin{tabular}{cccccccc}
\hline
Group & $T_{\rm eff}$ (K) & log\,$g$ (cm\,s$^{-2}$) & Number of stars \\
\hline
Dwarfs & 4500 -- 7500 & $>$ 3.2  & 360 \\
Giants & 3000 -- 5500 & $<$ 3.4 &  354 \\
Hot stars & $>$ 7000 & -- &  125 \\
Cool dwarfs & $<$ 5000 & $>$ 3.2  & 51 \\
\hline
\end{tabular}
\label{Table1}
\end{center}
\end{table}

\begin{figure*}
\centering
\includegraphics[width=160mm]{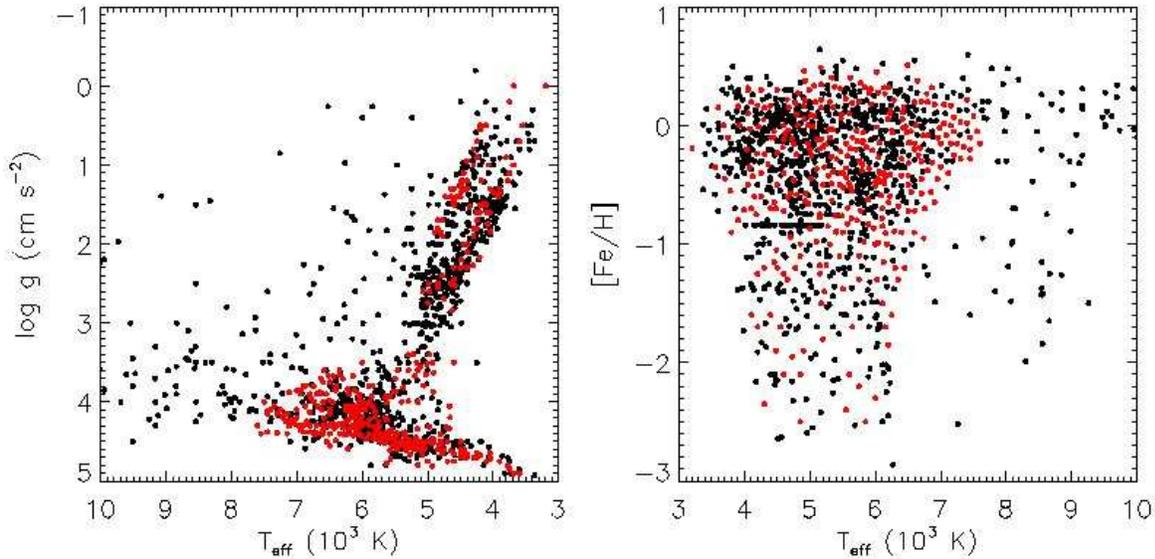}
\caption{Distributions of stellar atmospheric parameter of the MILES spectral 
         templates adopted by the LSP3 in the planes of 
         $T_{\rm eff}$ -- log\,$g$ (left) and $T_{\rm eff}$ -- [Fe/H] (right). 
         Black dots represent the original 900 MILES 
         template spectra, while those in red represent the 416 interpolated spectra.}
\label{Fig1}
\end{figure*}

\begin{figure}
\centering
\includegraphics[width=90mm]{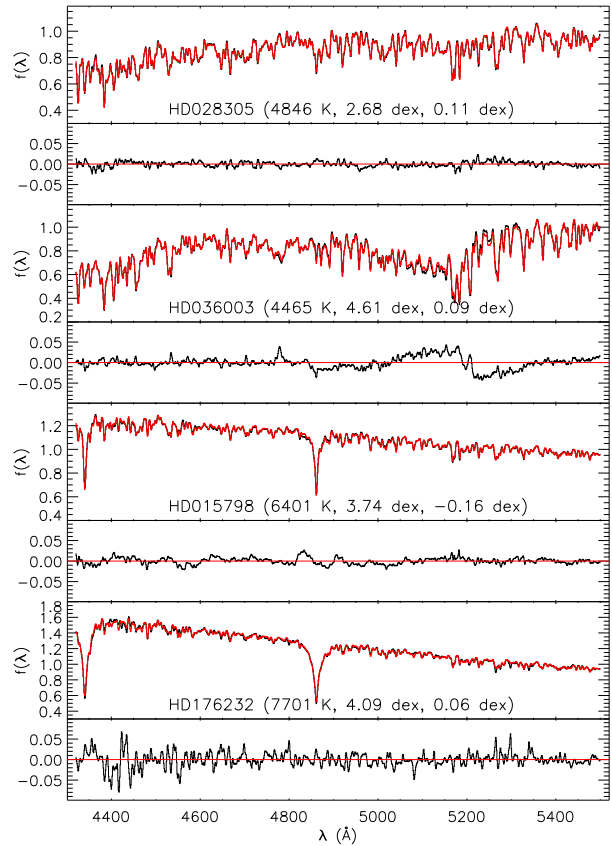}
\caption{Comparison of interpolated (red) and observed (black) MILES template spectra  
         for four example stars, one from each of the four groups of stars listed  
         in Table\,1. The flux densities are in arbitrary units. The name and 
         atmospheric parameters (in the sequence of $T_{\rm eff}$, log\,$g$, 
         [Fe/H]) of the star are marked in each panel. 
         For each star, the residuals between the interpolated and 
         observed spectra are also plotted at the bottom of each panel.}
\label{Fig2}
\end{figure}

\section{METHODOLOGY}
The LSP3 adopts a cross-correlation algorithm to determine stellar radial 
velocities. For the determinations of stellar atmospheric parameters, 
LSP3 uses two approaches: the weighted means of parameters of the 
best-matching templates and values yielded by $\chi^2$ minimization. 
Both methods are based on $\chi^2$ values calculated from the target 
and matching template spectra.  
$\chi^2$ is defined as 
\begin{equation}
   \chi^2 = \sum_{i=1}^{N}\frac{(O_i - T_i)^2}{\sigma_i^2} 
\end{equation} 
where $O_i$ and $T_i$ are respectively flux densities of the 
target and template spectra of the $i$th pixel. $N$ is the 
total pixel number used to calculate $\chi^2$, and $\sigma_i$ 
is the error of flux density of the target spectrum of the $i$th pixel. 
Note that here we have neglected the errors of flux density of the template spectrum.  
The LSP3 is designed to match the LAMOST blue- and red-arm spectra 
with templates separately. However, limited by the wavelength 
coverage of the template spectra, in the current version of LSP3, 
only results derived from the blue-arm spectra are used. 

\subsection{Flowchart}
\begin{figure}
\centering
\includegraphics[width=90mm]{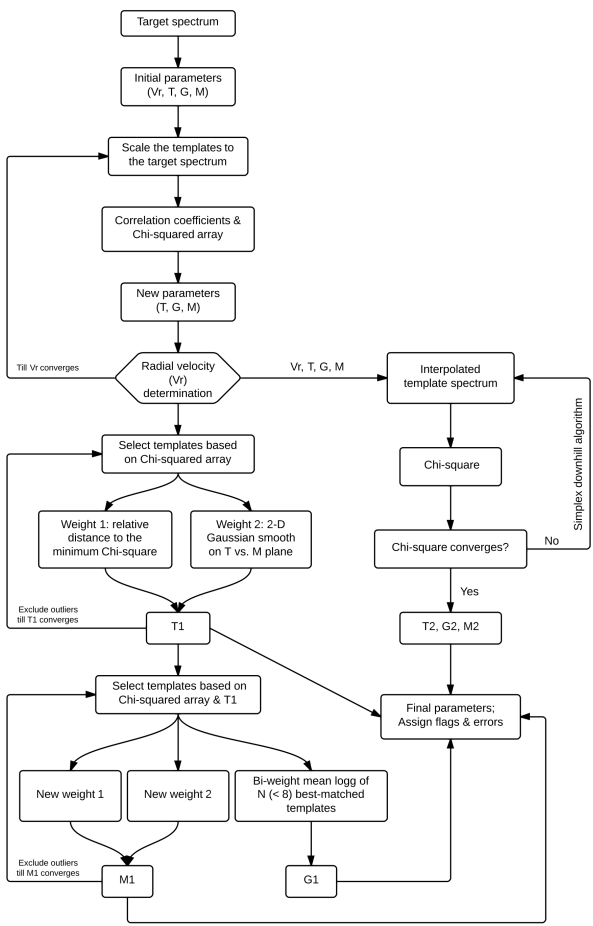}
\caption{LSP3 Flowchart.}
\label{Fig3}
\end{figure}

Fig.\,3 illustrates a flowchart of the LSP3. 
For a given target spectrum, a set of initial parameters ($V_{\rm r}$, 
$T_{\rm eff}$, log\,$g$, [Fe/H]) are first estimated by matching the 
normalized target spectrum with normalized MILES templates (cf. Section\,3.3 for detail). 
A subset of templates that fall in a given parameter box centered on 
the initial estimates are then selected and scaled to match the 
spectral energy distribution (SED) with the $V_{\rm r}$-corrected 
target spectrum to calculate values of $\chi^2$.
From those $\chi^2$ values, a new set of atmospheric parameters are 
estimated by taking the biweight mean values of parameters of 
the four best-matching templates. 
Several ELODIE spectra of stellar atmospheric parameters closest 
to the newly estimated ones, are then selected to determine a revised 
value of radial velocity using a cross-correlation algorithm. This newly 
derived radial velocity is then used to convert the target spectrum 
to laboratory wavelengths and re-calculate values of $\chi^2$. 
This process is iterated until values of $V_{\rm r}$ from two consecutive 
iterations differ by less than a predefined margin (default 3.0\,km\,s$^{-1}$ 
in the current version of LSP3). 

Once $V_{\rm r}$ has been determined, two algorithms: a) weighted means 
of parameters of best-matching templates and, b) $\chi^2$ minimization, 
are applied to further improve estimates of the target atmospheric 
parameters. For the weighted mean method, two weights are assigned to 
each MILES template, one accounts for the degree of similarity between the 
target and template spectra (i.e. values of $\chi^2$), another accounts 
for the local density of templates in the $T_{\rm eff}$ -- [Fe/H] plane of the 
parameter space. $T_{\rm eff}$ is first derived since it is the parameter 
that $\chi^2$ is most sensitive to. Values of log\,$g$ 
and [Fe/H] are then determined using only templates that fall in a 
narrow (predefined) range of the afore determined value of $T_{\rm eff}$. 
The processes are iterated to exclude obvious outliers of templates 
of parameters that deviate significantly from the weighted mean values. 
For the $\chi^2$ minimization method, a simplex downhill algorithm 
(Nelder \& Mead 1965) is used to search for the minimum $\chi^2$ between 
the $V_{\rm r}$-corrected target spectrum and templates in the parameter space. 
Note that the template spectrum of a given set of atmospheric parameters 
is calculated using the fits deduced in Section\,2.2. 
In the current version of LSP3, results derived from the weighted mean 
method are adopted as the final parameters of the pipeline. Specific flags (cf. Section\,3.8) 
are assigned to each target spectrum analyzed to indicate the quality and/or 
any warning of the parameters deduced. 
Errors of the final parameters are estimated by combining the 
random and systematic errors, and are functions of the SNR, 
$T_{\rm eff}$, log\,$g$ and [Fe/H]. Here the random errors are 
estimated by comparing results derived from multi-epoch observations of duplicate targets, 
while the systematic errors are derived by applying the LSP3 to 
the MILES templates.

\subsection{Selection of matching wavelength range}
LAMOST spectra are split with a dichroic into blue and red parts 
and are collected with two arms, the blue-arm spectra covering 3700 -- 5900\,{\AA}, 
and the red spectra covering 5700 -- 9000\,{\AA} (Cui et al. 2012). 
The blue- and red-arm raw spectra are processed separately in the LAMOST 
2-D pipeline and are pieced together after flux calibration (Xiang et al. 2014, submitted, hereafter Paper\,I).

The LSP3 is designed to determine stellar parameters with the blue- and 
red-arm spectra separately, although the current version of LSP3 makes 
use of the results from the blue-arm spectra only. This approach is based on the following 
considerations. First, the LSS-GAC targets stars of a wide range of colours, 
thus depending on the colour, the blue- and red-arm spectra may have very 
different signal-to-noise ratios (SNRs). Second, the accuracy of 
wavelength calibration of the blue- and red-arm spectra are 
different, since they are calibrated separately and corrected for systematics 
using different sets of sky emission lines (Luo A.-L., private communication). 
Finally, the blue- and red-arm spectra sometimes do not
piece together smoothly due to uncertainties in flat-fielding, 
sky subtraction, and flux calibration. 
The problem is most acute for spectra of low SNRs. 
There are some rare cases where the current pipeline of flux-calibration fails to 
yield a reliable set of spectral response curves (SRCs). Spectra of those 
plates are processed with a nominal set of SRCs (Paper\,I), leading to 
large uncertainties in the SEDs of those 
spectra, in particular around the cross-over wavelength of the dichroic.

As a default, the current version of LSP3 uses the 4320 -- 5500\,{\AA} 
wavelength region of the blue-arm spectra to derive stellar parameters. 
The region is selected in order to exclude the wavelength range beyond 5500\,{\AA} 
where the instrumental sensitivity drops rapidly due to the cutoff 
of the dichroic, and to avoid prominent atomic lines such as the 
Ca\,{\sc ii} H, K lines at 3967 and 3933\,{\AA}, often strongly 
saturated in stars of solar metallicity, 
and strong molecular absorption bands such as the CH G-band at 4314\,{\AA}.  
Excluding the wavelength region shorter than 4320\,{\AA} however does 
pose some problems, in particular for metal-poor stars, for which the 
Ca\,{\sc ii}\,K line at 3933\,{\AA} serves as an important metallicity indicator. 
Also the Ca\,{\sc i} $\lambda$4226 line is an important indicator of the 
stellar surface gravity (Gray \& Corbally, 2009), while the G-band provides 
information of the [C/Fe] abundance ratios (Lee et al. 2013). 
As such, an option of matching target spectra with templates over a wider 
range of wavelengths, from 3900 -- 5500\,{\AA}, is also implemented in the 
LSP3. For most targets, stellar parameters derived from the spectral range 
4320 -- 5500\,{\AA} differ little from those from the range of 3900 -- 5500\,{\AA}. 
Stellar parameters derived from the 3900 -- 5500\,{\AA} wavelength range will be 
presented in the next release of LSP3, 
along with estimates of [$\alpha$/Fe] and [C/Fe] abundance ratios. 

For the red-arm spectra, the spectral range available for template matching 
is currently limited by the wavelength coverage of MILES spectra that extends 
only to 7410\,{\AA}. 
An option to determine stellar parameters by matching the 6100 -- 6800\,{\AA} 
red-arm spectra is also implemented in the LSP3. 
However, few spectral features are available in this wavelength regime 
to constrain the stellar parameters robustly. As such parameters yielded 
by this option are not provided for the moment. 
An observational campaign to extend the MILES spectra to 9200\,{\AA} is 
currently in progress (Wang et al. 2014).
We expect that the next version of LSP3 will include the wavelength region 
of the Ca\,{\sc ii} triplet lines in the red for template matching.

\subsection{Initial parameters}

Good estimates of initial parameters are important for two reasons. 
Firstly, an initial value of $V_{\rm r}$ is needed to convert the 
observed wavelengths of a target spectrum to laboratory values when 
calculating the $\chi^2$ value of the target and matching template spectra. 
Secondly, the initial values of $T_{\rm eff}$, log\,$g$ and [Fe/H] 
are used to limit the parameter range of template spectra in order 
to reduce the number of templates for $\chi^2$ calculations and 
thus to speed up the optimization. 

The initial parameters are estimated by matching the continuum-normalized 
target spectrum with similarly normalized MILES template spectra.   
To obtain the continuum, the blue- (3900 -- 6000\,\AA) and red-arm 
(5900 -- 9000\,\AA) spectra are fitted, separately, with a fifth-order 
polynomial. The approach is similar to that used in the SSPP (Lee et al. 2008a).  
Note that continuum-normalized spectra are used to estimate 
the initial parameters only. When deriving the final parameters, spectra 
without continuum normalization are used (cf. Section\,3.4). 
The normalized target spectrum is shifted in velocity with discrete 
values between $-1000$ and 1000\,km\,s$^{-1}$, at a step of 10\,km\,s$^{-1}$ 
within $\pm$300\,km\,s$^{-1}$ and a step of 50\,km\,s$^{-1}$ beyond. 
A Bessel interpolation is adopted to interpolate $V_{\rm r}$-shifted spectra. 
$\chi^2$ values and correlation coefficients of the 
$V_{\rm r}$-shifted target spectrum with all the MILES templates are calculated. 
To estimate the initial value of $V_{\rm r}$, the template that yields 
the maximum correlation coefficient is selected out. Then for this template, 
its correlation coefficient with the target spectrum as a function of 
velocity is fitted with a Gaussian plus a second-order polynomial to 
find the exact value of $V_{\rm r}$ where the correlation coefficient peaks. 
In doing so only a few discrete values of velocity shift around the 
maximum correlation coefficient are used for the fitting. 
As for the initial values of $T_{\rm eff}$, log\,$g$ and [Fe/H], 
we select the 20 templates that give the smallest $\chi^2$ 
(at certain discrete value of velocity shift). 
The biweight means of atmospheric parameters of those 20 best-matching templates 
are then taken to be the initial parameters of the target spectrum, and the 
standard deviations of parameters of those 
templates are adopted as the errors of the initial parameters. 
Note that 20 is simply an empirical value based on an examination 
of the distribution of $\chi^2$. The final parameters deduced are found to be 
insensitive to this value given that a large box (0.2$\times$$T_{\rm eff}$ 
in $T_{\rm eff}$, 3.0\,dex in log\,$g$ and 1.0\,dex in [Fe/H]; Section 3.4) 
is set to re-do the template matching iteratively when estimating the final 
parameters with the weighted mean algorithm (Section 3.5).  

\subsection{Radial velocity determination and the final $\chi^2$ array}

As described in \S{3.1}, an iterative process is implemented to determine 
radial velocity and atmospheric parameters. 
It is designed to minimize the effects of uncertainties in $V_{\rm r}$ 
on the calculation of $\chi^2$ array of the target spectrum with the MILES 
templates that are used to derive atmospheric parameters on the one hand, 
and, on the other hand to ensure radial velocity is determined by cross-correlating 
with an ELODIE template that has atmospheric parameters 
closest to the target. 

Unlike most of the previous work where template matching 
is carried out using continuum-normalized spectra (e.g. Lee et al. 2008a), 
the LSP3 uses non-normalized spectra. 
One reason is that there is important information (in particular that of 
the effective temperature) encoded in the observed continuum shape (i.e. SED) 
of a target spectrum. Another reason is that accurate estimate of the continuum 
level over a wide wavelength range for medium-to-low resolution spectra 
 is often quite difficult, especially for spectra of low SNRs or for 
stars of late-types whose spectra are dominated by prominent and broad 
molecular absorption bands. 
As designed, the LSS-GAC targets include many late-type stars and 
a significant fraction of the spectra accumulated so far have SNRs lower 
than 20 per pixel in the blue (3700 -- 5900\,\AA) (cf. Paper\,III). 
To account for effects such as reddening by the interstellar dust grains 
and uncertainties in spectral flux calibration, a low-order polynomial is 
however allowed to scale the SEDs of template spectra to match that of the 
target spectrum of concern when calculating values of $\chi^2$. 
Based on extensive tests and tries, we find that a third-order polynomial 
is high enough to account for possible effects due to reddening and flux 
calibration for the wavelength ranges of concern (4320 -- 5500\,{\AA} 
in the blue and 6100 -- 6800\,{\AA} in the red), and at the same time 
low enough to avoid inducing undesired artifacts. 

To save computation time, for a given set of initial atmospheric parameters 
($T_{\rm eff}$, log\,$g$ and [Fe/H]), a 3-D box in the parameter space 
centered on the initial values and of dimensions 3 times the corresponding 
errors is defined. To ensure the box contains a sufficiently large 
number of templates, the box is required to have a minimum side of   
0.2$\times$$T_{\rm eff}$, 3.0\,dex and 1.0\,dex in the dimension of 
$T_{\rm eff}$, log\,$g$ and [Fe/H], respectively.
Typically, a box contains 100 -- 500 templates, depending on
the initial values of parameters. Values of $\chi^2$ between
the $V_{\rm r}$-corrected target spectrum and 
the MILES templates of parameters falling inside the box are then 
calculated, after scaling the SEDs of templates to match 
that of the target using a third-order polynomial. 

From the $\chi^2$ values of target spectrum with MILES templates, 
the biweight mean values of parameters of the 4 best matching templates 
are adopted as the new set of parameters. 
Five ELODIE spectra that have parameters ``closest'' to 
the newly derived set of parameters are then selected and used 
to determine $V_{\rm r}$ by cross-correlation. Here, ``closest''
is defined by distances in the atmospheric parameter space 
assuming a distance of 75\,K in temperature is equivalent to a 
distance of 0.1\,dex in log\,$g$ or in [Fe/H]. 
To determine $V_{\rm r}$, we first scale the 5 ``nearest'' ELODIE 
spectra to match the SED of the target spectrum using a third-order 
polynomial, shift the wavelengths in velocities between
$-1000$\,km\,s$^{-1}$ and 1000\,km\,s$^{-1}$ with a step of 5\,km\,s$^{-1}$, 
and then calculate the correlation coefficients between the 
target and the continuum-rectified, velocity-shifted ELODIE 
spectra. For the ELODIE template showing the highest correlation, 
the correlation coefficient as a function of velocity shift is fitted 
with a Gaussian plus a second-order polynomial to determine the best 
matching radial velocity. 

The above process is iterated until values of $V_{\rm r}$ from two 
consecutive iterations differ by less than 3.0\,km\,s$^{-1}$. 
Typically, 2 -- 3 iterations are sufficient. 
The final array of $\chi^2$ values is recorded for a further 
iteration of atmospheric parameter determinations with a weighted 
mean method.

\subsection{Parameters estimated by weighted mean}

In this approach, the LSP3 adopts the weighted mean of the 
atmospheric parameters of a subsample of templates selected based 
on the $\chi^2$ array as the parameters of the target spectrum. 
Considering that $\chi^2$ has different sensitivity to different parameters, 
and in general $\chi^2$ is 
more sensitive to $T_{\rm eff}$ than to [Fe/H] and log\,$g$, 
the LSP3 estimates $T_{\rm eff}$ first, and then determine 
[Fe/H] and log\,$g$ within a constrained range of $T_{\rm eff}$.  

To define a subsample of templates used to calculate the target parameters 
using the weighted mean algorithm, we first define a threshold value of 
$\chi^2$ such that templates with $\chi^2 < a\chi^2_{\rm min}$ are 
included in the subsample. Here $a$ is a free parameter, and we require 
that $a$ should be large enough to enclose sufficient templates 
(in terms of both the number and the parameter coverage), yet small enough 
to exclude those obviously ``inappropriate'' templates. Here, ``inappropriate'' means 
that given the parameters of a template, the probability that 
the target spectrum has the same parameters is lower than a predefined 
critical value. Given a degree of freedom ($\sim$\,1160) of the blue-arm 
spectra employed ($\lambda\lambda$4320 -- 5500\,{\AA}, with 
the Hg\,{\sc i} $\lambda$4358 city light emission line masked out), 
we set $a=1.05$. For the $\chi^2$ range of concern here, the probability 
is sensitive to the value of $\chi^2$, and the probability decreases 
almost linearly with increasing $\chi^2$. In the cases where only 
a couple of templates ($< 10$) fall within the aforementioned $\chi^2$ 
cut, $a$ is increased in order to encompass more ($\gtrsim 10$) 
templates. Fig.\,4 shows the distributions of the number of templates, 
as well as the spreads of parameters covered by the templates 
satisfying the $\chi^2$ cut for the LSS-GAC targets processed with 
the current version of LSP3. 
It shows that for most stars about 10 -- 20 templates are used 
to calculated the target parameters using the weighted mean method 
and the parameters of those templates spread over a range of 
about 400\,K, 0.6\,dex and 0.5\,dex in $T_{\rm eff}$, log\,$g$ and [Fe/H], respectively. 
Those numbers seem reasonable given the accuracy of parameters 
achievable with the current design of LSP3 (cf. Sections\,5 and 6). 
\begin{figure}
\centering
\includegraphics[width=80mm]{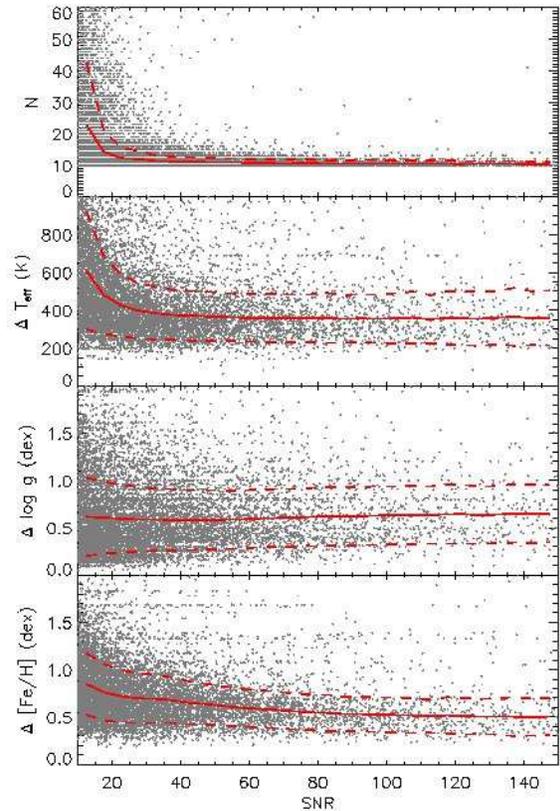}
\caption{Distributions of the number of templates satisfying the 
         $\chi^2$ cut (top panel), as well as of the spreads 
         (i.e. from the maximum to the minimum) of parameters 
          of those templates in $T_{\rm eff}$, $\log\,$ and [Fe/H] 
          (lower three panels), as function of the spectral SNR. 
         In each panel, 10,000 stars randomly selected from the current 
         sample are shown. The solid and dashed lines in each panel delineate 
         the mean and the mean plus and minus the standard deviation 
         as a function of the SNR. In the top panel, 
         only the mean and the mean plus the standard deviation 
         are shown.}
\label{Fig4}
\end{figure}

The LSP3 assigns two weights to each template used to calculate 
the weighted mean parameters. One weight accounts for the 
degree of similarity between the target and template spectra, defined as 
\begin{equation}
 w_1\equiv1.0-\frac{\chi^2-\chi^2_{\rm min}}{\chi^2_{\rm max}-\chi^2_{\rm min}}\times f. 
\end{equation}
Here $\chi^2_{\rm max}$ is the maximum value of $\chi^2$ of templates 
in the subsample selected by the aforementioned criteria, and $f$ is 
a fudge factor that defines the weight of template with the maximum 
value of $\chi^2$. Note that here we have effectively used 
a linear relation to replace the real likelihood distribution, 
which is proportional to exp($-\chi^2)$. This is a temporary measure, 
based on trials and tests,  and reflects the fact that, as discussed 
in Section\,8.3, the current error estimates of LAMOST spectra are not 
reliable, and, consequently, the absolute values of $\chi^2$. Nevertheless, 
a linear relation should be a good approximation to the likelihood distribution 
for $\chi^2$ value around 1.0, its most expected value.
By this definition, the template having the minimum 
value of $\chi^2$ has the highest weight of unity. If one assumes that 
the minimum value of $\chi^2$ equals 1.0, the most possible statistical value, 
then the probability that a given correct model (template) has a value 
of $\chi^2$ larger than $a \times \chi^2_{\rm min} = 1.05$ is $\sim$\,0.1 
for a degree of freedom of 1160. The probability that a correct model has a 
$\chi^2$ value larger than $\chi^2_{\rm min}$ is 
0.5, half the weight assigned to the template with the minimum $\chi^2$. 
Thus if we take $f$ to be 0.8, then the template with the maximum $\chi^2$ 
will have a weight $w_1 = 0.2$. 
Considering that the parameters of templates themselves have uncertainties, 
we have decided to give a higher weight to template with the maximum $\chi^2$ 
and assign $f$ as $0.5$ in the current version of LSP3. 

The second weight, $w_2$, accounts for the inhomogeneity of template 
distribution in the $T_{\rm eff}$ -- [Fe/H] parameter plane. 
Even though we have interpolated the templates to fill the 
obvious holes in parameter coverage, local inhomogeneities in parameter 
space remain. The interpolated spectra are also subjected to interpolation errors. 
Based on those considerations, we have introduced a Gaussian kernel to smooth 
the parameters of the selected templates that are used to calculate the 
stellar parameters in the $T_{\rm eff}$ -- [Fe/H] plane.
Firstly, we define a box selection function $f$($T_{\rm eff}$, [Fe/H]) 
such that $f=1$ for templates belonging to the subsample used to calculate 
the weighted mean parameters, and $f=0$ for all other templates. 
Then weight $w_2$ for template $i$ in the subsample is given by, 
\begin{equation}
w_2 = \frac{1} {F(T_{\rm eff}^i, {\rm [Fe/H]}^i)},  
\end{equation}

where

\begin{equation} 
\begin{split}
   F(T_{\rm eff}^i,{\rm [Fe/H]}^i) = 
   \sum_{j=1}^{N}f(T_{\rm eff}^j,{\rm [Fe/H]}^j) 
   \times g(T_{\rm eff}^{ji},{\rm [Fe/H]}^{ji}),  
\end{split}
\end{equation}

\begin{align}
 &g(T_{\rm eff}^{ji},{\rm [Fe/H]}^{ji}) = & \nonumber\\
&{\rm exp}(-\frac{(T_{\rm eff}^j-T_{\rm eff}^i)^2}{2.0\sigma_{T_{\rm eff}}^2})
  \times {\rm exp}(-\frac{({\rm [Fe/H]}^j-{\rm [Fe/H]}^i)^2}{2.0\sigma_{\rm [Fe/H]}^2}).& 
\end{align}
Here $N$ is the number of templates in the subsample. 
In principle, values of $\sigma_{T_{\rm eff}}$ and $\sigma_{\rm [Fe/H]}$ 
should be assigned based on local density of templates in the parameter plane. 
In the current version of LSP3, $\sigma_{T_{\rm eff}}$ and $\sigma_{\rm [Fe/H]}$ 
are simply assumed to be constants and equal 50\,K and 0.05\,dex, respectively.   

The final weight of a template in the subsample is given by, 
\begin{equation} 
W = w_1 \times w_2.
\end{equation} 
The weighted mean of temperatures adopted for the target spectrum is thus,
\begin{equation} 
 T_{\rm eff} = \frac{\sum\limits^{N}_{i=1}W^i \times T_{\rm eff}^i}{\sum\limits^{N}_{i=1}W^i}.
\end{equation}

The above process is iterated. In each iteration, templates with 
values of $T_{\rm eff}$ that differs from the weighted mean 
in excess of two times the standard deviation of the subsample 
are excluded. Note that the LSP3 always keeps the template with 
the minimum $\chi^2$ in the weighting box. If the distance between 
the weighted mean $T_{\rm eff}$ and that of template with the 
minimum $\chi^2$, $d_{\rm min}$, becomes larger than twice the 
standard deviation of templates in the box, then templates with 
$T_{\rm eff}$ that differ from the weighted mean by more than 
$d_{\rm min}$ are excluded in the next iteration. 
Fig.\,5 shows an example of this process. 

Once $T_{\rm eff}$ is determined, the LSP3 selects templates 
that have $T_{\rm eff}$ between (1.0$\pm$0.05) $\times$ $T_{\rm eff}$ 
and $\chi^2<a\chi^2_{\rm min}$ to calculate a weighted mean 
value of [Fe/H]. The weights of the individual templates 
are assigned in the same way as for $T_{\rm eff}$ 
estimation discussed above. The estimate adopted for the target 
spectrum is thus,  
\begin{equation}
{\rm [Fe/H]} = \frac{\sum\limits^{N}_{i=1}(W_{\rm [Fe/H]}^i \times {\rm [Fe/H]}^i)} 
{\sum\limits^{N}_{i=1}{W_{\rm [Fe/H]}^i}}.
\end{equation}
For the estimate of log\,$g$, the LSP3 simply adopts a biweight mean 
of $\log\,g$ of the $N$ templates that have the highest $W_{\rm [Fe/H]}$. 
Here we require $N \lesssim 8$. 
This is because log\,$g$ is usually the parameter that $\chi^2$ is  
least sensitive to, and the log\,$g$ values of templates selected with 
the above $\chi^2$ cut may spread over a wide range, sometimes 
even encompassing those of giants and dwarfs. 
By taking the biweight mean of values of only the few 
best-matching templates, we avoid the risk 
of averaging log\,$g$ values of giants and dwarfs. 
An iteration process similar to that for $T_{\rm eff}$ is also 
applied for the estimates of [Fe/H] and log\,$g$. 
Parameters determined with this method are denoted by
$T_{\rm eff1}$, log\,$g_1$, [Fe/H]$_1$ for effective temperature, surface gravity
and metallicity, respectively.

\begin{figure*}
\centering
\includegraphics[width=180mm]{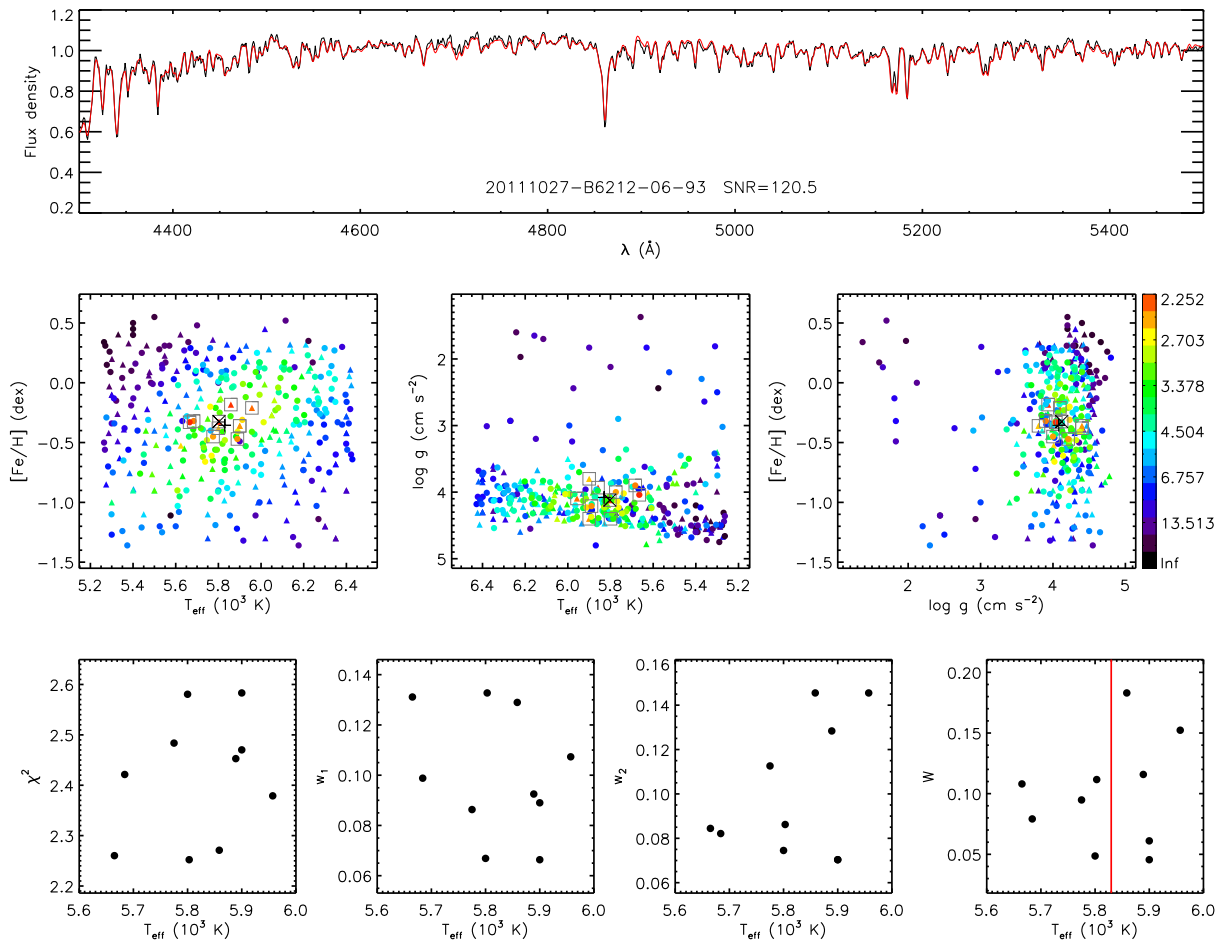}
\caption{An example of estimating $T_{\rm eff}$ with the weighted mean algorithm. 
         The top panel plots the target (black) and the best-matching 
         template (red) spectra. The middle panel shows the colour coded distribution 
         of $\chi^2$ values of the individual templates. Dots 
         represent the original MILES templates and triangles those interpolated. 
         The cross indicates the best-matching template (i.e. the one with 
         the minimum $\chi^2$). Open squares in grey denote templates 
         selected to calculate the weighted mean of parameters, marked by 
         the bold plus. 
         From left to right the three plots show the distributions in the 
         $T_{\rm eff}$ -- [Fe/H], $T_{\rm eff}$ -- log\,$g$ and log\,$g$ -- [Fe/H] 
         plane, respectively. 
         The bottom panel illustrates the process of calculating the 
         weighted mean for $T_{\rm eff}$ estimation. From left to right, 
         the four plots are, respectively, values of $\chi^2$, normalized $w_1$, $w_2$ and $W$  
         as a function $T_{\rm eff}$ for templates in the subsample 
         (weighting box). The red line in the fourth panel indicates 
         the final value of $T_{\rm eff}$ adopted. For this particular target 
         spectrum, $a=1.18$ (cf. \S{3.5}).}
\label{Fig5}
\end{figure*}

\subsection{Parameters estimated by $\chi^2$ minimization}

This approach searches for the minimum $\chi^2$ 
in the stellar atmospheric parameter space, and adopts the parameters 
of the template that has the minimum $\chi^2$ as those of the target.  
As shown in Fig.\,3, the target spectrum is first converted to 
laboratory wavelengths using $V_{\rm r}$ deduced by cross-correlating 
with templates in the ELODIE library. Taking the biweight means 
of parameters of the 4 best-matching MILES templates as the initial values, 
the LSP3 searches the parameter space for the minimum $\chi^2$ between the 
$V_{\rm r}$-corrected target spectrum and the MILES templates using 
a downhill simplex minimization algorithm (Nelder \& Mead 1965). 
Here, the template spectrum for given set of $T_{\rm eff}$, log\,$g$ and 
[Fe/H] is created using the parameter -- spectra relation deduced 
by fitting the MILES spectral flux density at each wavelength  
as a function of stellar atmospheric parameters (cf. Section\,2.2).
Parameters derived from this method are denoted as $T_{\rm eff2}$,
log\,$g_2$, [Fe/H]$_2$ for effective temperature, surface gravity and metallicity,
respectively.

\subsection{The final parameters}
Compared with parameters determined by the weighted mean, those deduced by 
$\chi^2$ minimization are less affected by the inhomogeneous distribution 
of MILES templates in the parameter space. 
However, the parameters derived from the $\chi^2$ minimization method 
are affected by the uncertainties of fiducial 
spectra calculated using the parameter -- spectral flux density relations. 
The uncertainties vary with wavelength and depend on the location 
of the template in the parameter space. 
As shall be shown in Sections\,5 and 6, the $\chi^2$ minimization 
method yields $T_{\rm eff}$ and [Fe/H] with an precision comparable 
to that of the weighted mean approach. The results for log\,$g$ are inferior, 
yielding larger scatters compared to those derived by the weighted mean method. 
Given the nature of $\chi^2$ minimization method and the complex behaviors 
of spectral flux density for varying parameters, it is difficult to ensure that the 
optimization converges to the global minimum rather than a local one, 
yielding wrong parameters as a consequence. 
With the above considerations, the current version of LSP3 simply adopts 
the parameters derived from the weighted mean method, $T_{\rm eff1}$ , log\,$g_1$, 
[Fe/H]$_1$, as the final parameters of the target spectrum. 
Parameters given by the $\chi^2$ minimization method are provided for 
comparison only, and various flags are assigned based on the degree of discrepancy between 
the parameters deduced from the two approaches (cf. Section\,3.8).
Errors of the final parameters are estimated by combining the random errors, 
estimated by comparing the results yielded by multi-epoch observations of duplicate stars,  
and the systematic errors, estimated by applying the LSP3 to the MILES 
spectra themselves. Clearly, the errors are functions of the spectral SNR, $T_{\rm eff}$,
log\,$g$ and [Fe/H] (cf. Section\,7).

\subsection{Flags}

The LSP3 assigns 9 integer flags to the final parameters adopted for 
each star to mark potential anomalies of the derived values. The flags 
are listed in Table\,2. Except the first one, all other flags are cautionary, 
and the smaller the values, the higher the quality 
the parameters derived. 

\begin{table*}
\centering
\caption{Flags assigned to the final parameters}
\begin{tabular}{p{1cm}cp{12cm}cl} 
\hline 
Flag & Value & Description   \\
\hline 
1 & $1,2,3$ & The best-matching template is one of the original MILES spectra (1), 
   a fiducial spectrum calculated from the fitting parameters (2), or a spectrum 
         of special type (3). \\
2 & $1,2,3 ...$ & $\chi^2_{\rm min}$ larger than the median value of stars in the 
   corresponding SNR, $T_{\rm eff}$, log\,$g$ and [Fe/H] bins by less than $n \times {\rm MAD}^1$ . \\
3 & $1,2,3 ...$ & The peak correlation coefficient smaller than the median  
             value of stars in the corresponding SNR bin by less than $n \times {\rm MAD}$. \\
4 & $1,2,3 ...$ & The final $T_{\rm eff}$ differs from the value  
                    of the best-matching template by less than $n\sigma$, 
     where $\sigma$ is the estimated error$^2$ of $T_{\rm eff}$. \\
5 & $1,2,3 ...$ & Same as Flag 4 but for log\,$g$. \\ 
& &  \\
6 & $1,2,3 ...$ & Same as Flag 4 but for [Fe/H]. \\ 
& & \\
7 & $1,2,3 ...$ & The difference between values of $T_{\rm eff}$ derived 
                    from the weighted mean  
                     and $\chi^2$ minimization methods is smaller than 
                    $n\sqrt{(\sigma_1^2 + \sigma_2^2)}$, where $\sigma_1$ and 
                    $\sigma_2$ are the estimated 
                    errors given by the two methods, respectively. \\
8 & $1,2,3 ...$ & Same as Flag 7 but for log\,$g$. \\ 
& & \\
9 & $1,2,3 ...$ & Same as Flag 7 but for [Fe/H]. \\ 
\hline
\end{tabular}
\begin{tablenotes}
\item[]$^1$ Mean absolute deviation; $^2$ See Section\,7.
\end{tablenotes}
\label{Table1}
\end{table*}

The first flag describes which category of templates the best-matching 
one (the one with the minimum $\chi^2$) belongs to. The templates 
are divided into 3 categories: the ``normal'' templates from the original 
MILES library, the fiducial templates calculated from the parameter -- spectral 
flux density relations, and the templates of `special' types. 
The special templates include 18 SDSS templates and 14 MILES templates 
that fall in specific regions in the $T_{\rm eff}$ -- log\,$g$ plane as 
discussed in Section\,2.2. 
The remaining 886 MILES templates are referred to as normal. 
Depending on the category that the best-matching template belongs to, 
the LSP3 assigns an integer 1, 2 or 3 to Flag\,1. 
Parameters of stars with Flag\,1 = 3 should be treated with caution as the 
stars may well have a peculiar spectral type (e.g. white dwarfs, carbon stars, 
late-M/L type stars, BHB stars). 

The second flag describes the anomalies of the minimum $\chi^2$ 
of the best-matching template. A target spectrum can have an abnormally 
large $\chi^2_{\rm min}$ for a number of reasons, including contamination 
of cosmic rays, poor background (sky and scattered light) subtraction,
incorrect error estimates of spectral flux density, or the star is of 
some special spectral type that no template in the library can matches with. 
The second flag aims to 
signal out such possibilities. We divide the LSS-GAC targets observed 
hitherto into bins of the SNR, $T_{\rm eff}$, log\,$g$ and [Fe/H], 
and calculate the median and mean absolute deviation (MAD) of  
$\chi^2_{\rm min}$ values of stars in each bin. Then we construct relations 
between the median/MAD of $\chi^2_{\rm min}$ and the above parameters 
(SNR, $T_{\rm eff}$, log\,$g$ and [Fe/H]) by linear interpolation. 
For a star of a given set of SNR, $T_{\rm eff}$, log\,$g$ and [Fe/H], 
if $\chi^2_{\rm min}$ is larger than the median value by less than 
$n \times {\rm MAD}$ predicted by the relations, then the LSP3 assigns 
an integer $n$ to the second flag of that star. Note that a lower limit of 
1 is set to $n$ for all the 9 flags. 

The third flag describes the correlation coefficient for $V_{\rm r}$ estimation. 
We construct a relation between the median/MAD values of peak correlation 
coefficients for $V_{\rm r}$ estimation and the spectral SNR. For a star of 
given SNR, if the peak correlation coefficient is smaller than the median 
value by less than $n \times {\rm MAD}$ predicted by the relation, 
the LSP3 assigns an integer $n$ to the third flag of that star. 

Flags 4 to 6 describe the differences between the final parameters and 
the parameters of the best-matching template, for $T_{\rm eff}$, 
log\,$g$ and [Fe/H], respectively. 
If the difference is $n$ times smaller than the estimated uncertainty 
of the parameter concerned (cf. Section\,7), the LSP3 assigns 
an integer $n$ to the corresponding flag. 

Flags 7 to 9 describe the difference between the parameters derived 
from the weighted mean and from the $\chi^2$ minimization methods, for $T_{\rm eff}$, 
log\,$g$ and [Fe/H], respectively. 
Let $\Delta X$ denote the difference of a given parameter, where $X$ represents 
$T_{\rm eff}$, log\,$g$ or [Fe/H]. If $\Delta X < n\sqrt{{\sigma_1}^2+{\sigma_2}^2}$, 
then the LSP3 assigns an integer $n$ to the corresponding flag. 
Here $\sigma_1$ and $\sigma_2$ are the estimated uncertainties of parameter 
yielded by the two methods, respectively.

\section{Test with the MILES LIBRARY}
\begin{figure*}
\centering
\includegraphics[width=160mm]{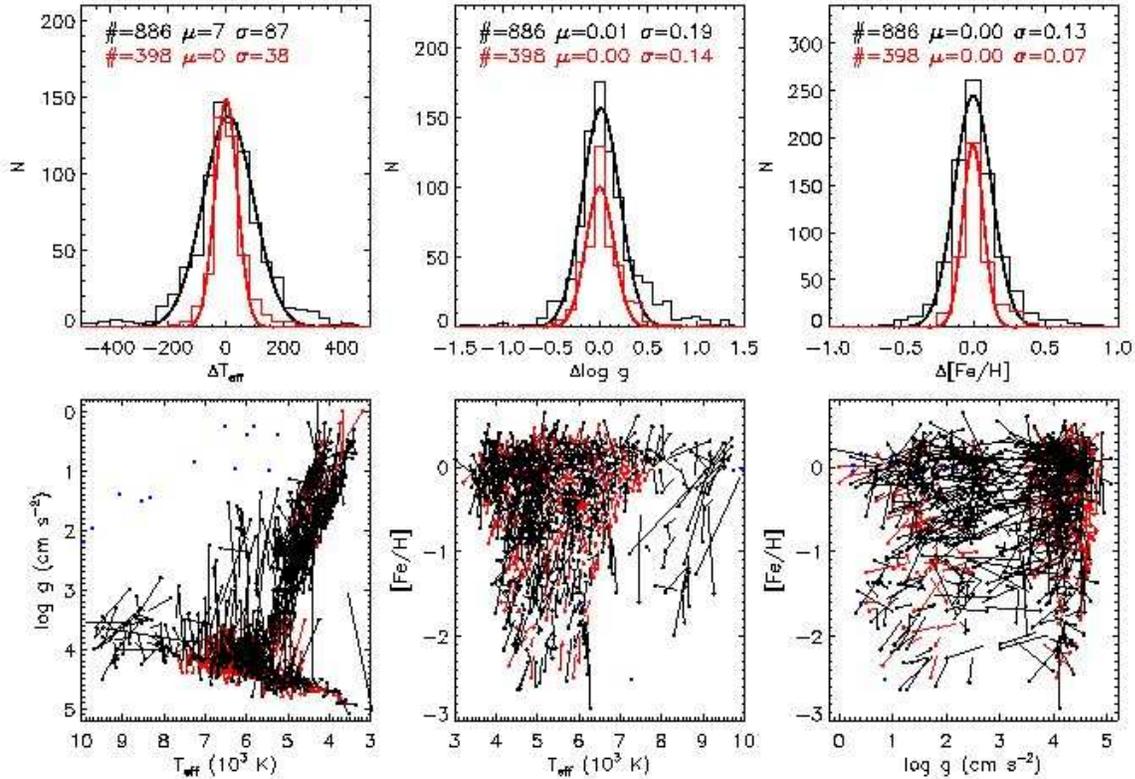}
\caption{The upper panels plot distributions of differences between 
         the parameters yielded by the LSP3 with the weighted mean algorithm 
         and those from the MILES library. Also over-plotted are Gaussian 
         fits to the distributions. The number of templates, 
        the mean and dispersion of the Gaussian fit are marked. 
        The lower panels connect the parameters yielded by the LSP3 and those from 
         the library by lines in 2-D parameter 
        planes, with those from the LSP3 marked by arrows and those 
        from the library by dots. Again, black and red symbols/lines refer 
        respectively to the original MILES and the `interpolated' templates, whereas 
        blue dots denote templates of special spectral types (cf. Section\,2.2).}
\label{Fig6}
\end{figure*}

We apply the LSP3 to the MILES template spectra to quantify the intrinsic 
errors of the algorithms. Fig.\,6 plots the comparison of the LSP3 results 
deduced from the weighted mean algorithms with parameters from the MILES library. 
The upper panels shows the distributions of the parameter differences, 
and the low panels are direct comparsons in the 2-D parameter planes. 
An overall dispersion of 87\,K, 0.19\,dex and 0.13\,dex for $T_{\rm eff}$, log\,$g$ 
and [Fe/H] respectively, is found for the original MILES spectra, while 
as for those interpolated spectra the corresponding values are much smaller, 
about 38\,K, 0.14\,dex, 0.07\,dex. 
Both errors in the MILES parameters themselves and those introduced by 
the LSP3 algorithms contribute to those values, and they are the 
lower limits of errors of the LSP3 parameters. 

In the upper middle panel of Fig.\,6, there is a non-Gaussian tail in the 
distribution of differences of log\,$g$. The values of log\,$g$ for stars in the tail 
could be systematically overestimated by as much as 0.5\,dex or more. 
Those stars correspond to data points connected by long arrows in the 
bottom left panel, and are mostly F/G-type subgiants/giants/supergiants 
of log\,$g < 3.0$\,dex or subgiants/turn-off stars of log\,$g$ slightly 
larger than 3.0\,dex. Their values of log\,$g$ have been overestimated 
by the LSP3 due to the boundary effects of the weighted mean algorithm: 
near the boundary of parameter coverage of the library, the weighted mean 
algorithm tend to yield parameters that are biased toward the `inner' 
region of the parameter coverage where most of the templates fall.  
Such systematic effects are primary defects of the current LSP3. 
The `gaps' seen in the deduced values of log\,$g$ presented in Figs.\,17, 
18 and 21 are probably partly due to such boundary effect. Some similar 
but weaker ($\sim$\,0.1\,dex) boundary effects may also be present in 
the case of [Fe/H] values deduced for super-metal-rich stars. 
Currently, we are carrying out a large campaign to expand the extent and 
homogeneity of parameter coverage of the MILES template library (Wang et al., in preparation).

A similar examination shows that the $\chi^2$ minimization approach 
yields results comparable to those from the weighted mean algorithm. 
Note the $\chi^2$ minimization method is sensitive to the initial values assigned. 
As described in Section\,3.5, the LSP3 adopts the biweight means of 
parameters of the 4 best-matching templates as the initial values. 
Tests show that if we simply assign the initial values to those of 
{\em the} best-matching template, we get significantly different 
results for some stars, presumably for those stars $\chi^2$ 
converges to some local rather than the 
global minimum, a potential risk inherent to the method.

\section{Precisions and Uncertainties of the LSP3 Algorithms}

\begin{figure*}
\centering
\includegraphics[width=180mm]{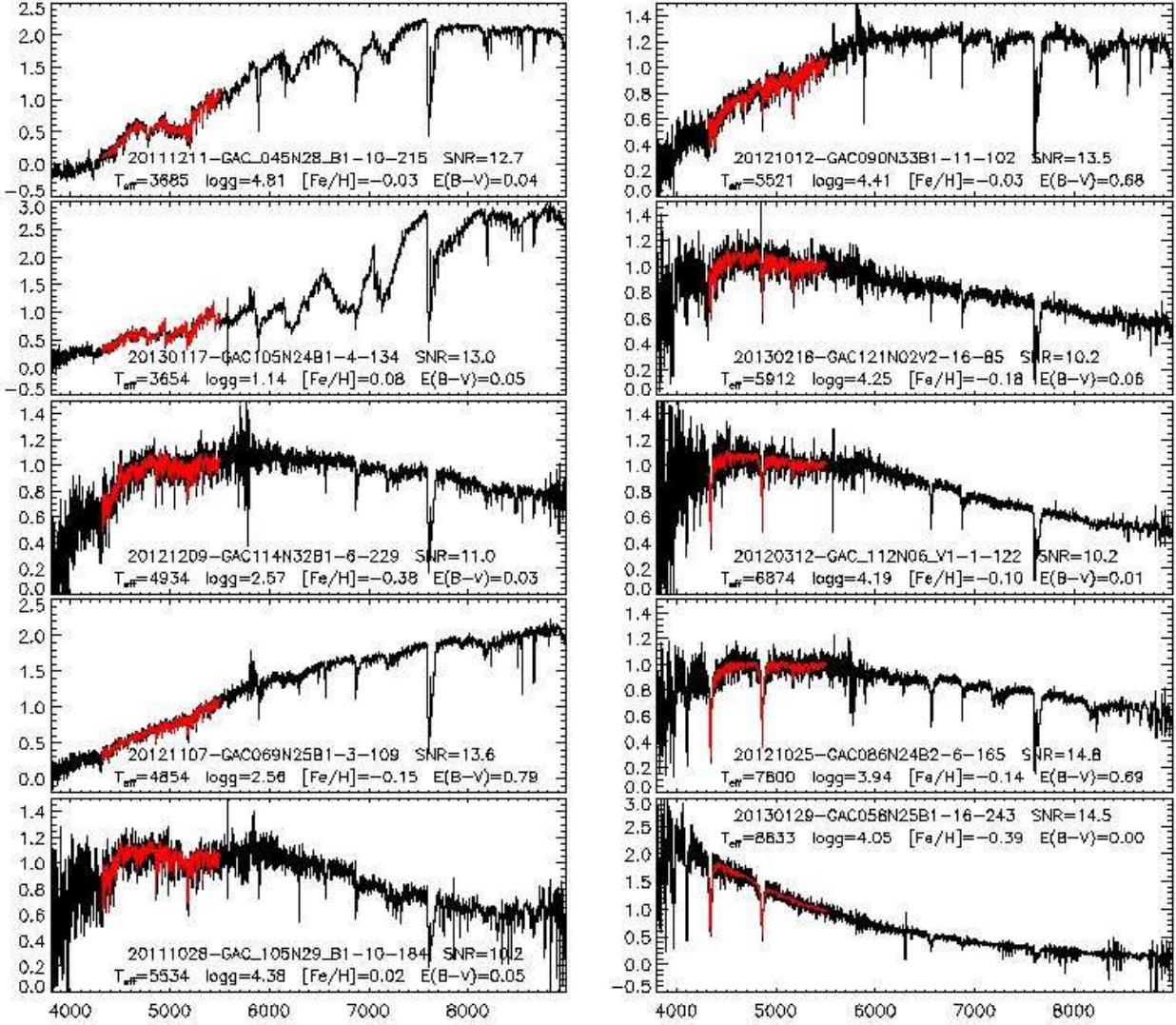}
\caption{Examples of LSS-GAC spectra (black) of SNR between 10 and 15.     
         Also over-plotted spectra in red are the best-matching template spectra 
         for the wavelength range of 4320--5500\,{\AA}. 
         The spectral flux densities are plotted in arbitrary scale. 
         The LSS-GAC spectral ID (`date-plate-spec-fiber'), SNR, 
         LSP3 stellar atmospheric parameters ($T_{\rm eff}$, log\,$g$, [Fe/H]), 
         as well as reddening $E(B-V)$ from Yuan et al. (2014b) are marked.} 
\label{Fig7}
\end{figure*}

Preceded by one-year-long Pilot Surveys, the LAMOST Regular Surveys 
were initiated in October 2012. By June 2013, 1.8 million spectra of 
about 1.3 million LSS-GAC targets have been collected in total, with 
750,867 (1,042,586) spectra having SNR in the blue (red) higher than 10 
(Liu et al. 2014; Paper\,III). For spectra with an median SNR per pixel 
better than 3, parameters $V_{\rm r}$, $T_{\rm eff}$, log\,$g$ and [Fe/H] 
are determined with the LSP3. Unless specified otherwise, the SNRs are 
calculated per pixel in a wavelength range of 100\,{\AA} centered at 
4650\,{\AA}, where one pixel corresponds to $\sim$\,1.07\,{\AA}. 
Fig.\,7 shows example LSS-GAC spectra of SNRs between 10 
and 15. Also over-plotted in the Figure are the best-matching template 
spectra for the wavelength range 4320--5500\,\AA. The spectral flux densities 
are plotted in arbitrary scale, and a third-order polynomial is allowed 
to correct for the SED differences between the LSS-GAC and template spectra.   
In this Section, we investigate the precision of LSP3 parameters by 
comparing the results yielded by different algorithms (Section\,5.1) 
as well as by comparing parameters deduced from multi-epoch observations 
of duplicate targets (Section\,5.2). 

\begin{figure*}
\centering
\includegraphics[width=160mm]{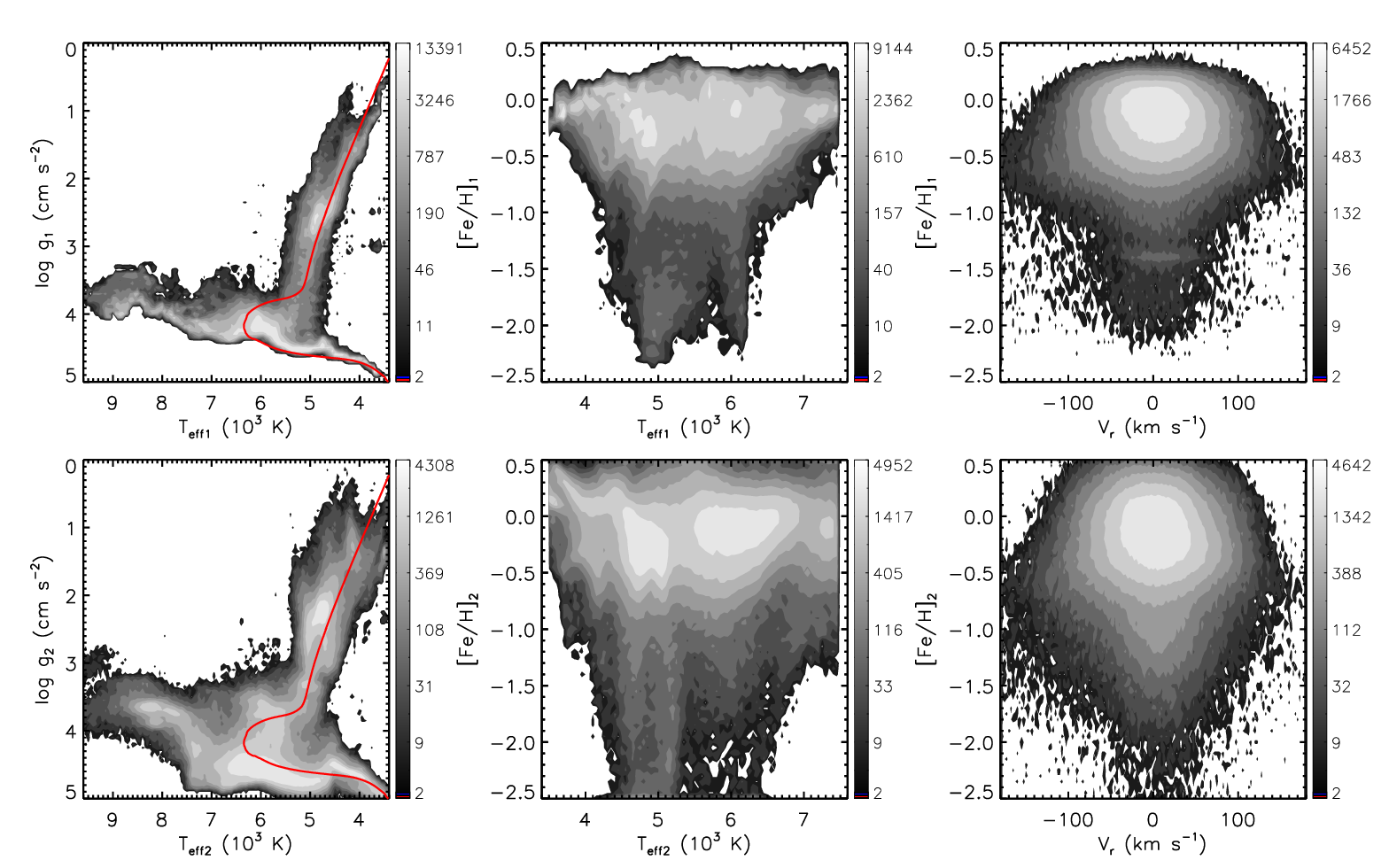}
\caption{Contour distributions of LSP3 parameters derived from
         the weighted mean (upper panels) and from the $\chi^2$ minimization
         (lower panels) methods. The plots include parameters deduced from
         1,091,301 spectra of SNRs $> 5$ for 869,741 stars. The isochrones
         plotted in the $T_{\rm eff}$ -- log\,$g$ planes are from Dotter et al. (2008),
         and have an age of 5\,Gyr, [Fe/H] of $-0.2$\,dex and [$\alpha$/Fe] of 0.0\,dex.}
\label{Fig8}
\end{figure*}

\subsection{The weighted mean versus the $\chi^2$ minimization methods}

As described in Section\,3, the LSP3 determines stellar atmospheric 
parameters with two methods: the weighted mean ($T_{\rm eff1}$, log\,$g_1$,
[Fe/H]$_1$) and the $\chi^2$ minimization ($T_{\rm eff2}$,log\,$g_2$, [Fe/H]$_2$). 
Fig.\,8 compares the distributions of parameters derived 
with the two methods in the $T_{\rm eff}$ -- log\,$g$, 
$T_{\rm eff}$ -- [Fe/H] and $V_{\rm r}$ -- [Fe/H] planes. 
Note that only parameters derived from spectra of SNRs $> 5$ are 
shown in the Figure. The SNR cut leads 1,091,301 spectra of 869,741 stars. 

In Fig.\,8, in the two panels of plot in the $T_{\rm eff}$ -- log\,$g$ 
plane (the HR diagram), a Dartmouth isochrone (Dotter et al. 2008) of 
age 5\,Gyr, metallicity [Fe/H] = $-0.2$\,dex and $\alpha$-element to iron 
ratio [$\alpha$/Fe] = 0.0\,dex is over-plotted. 
Fig.\,8 shows that on the whole the weighted mean algorithm yields  
parameters in good agreement with the isochrone. 
At a given $T_{\rm eff}$, the $\chi^2$ minimization method 
yields a log\,$g$ distribution that looks ``fatter'' than the weighted 
mean algorithm, largely a consequence of the employment of extrapolated 
templates in the former approach. 
For dwarfs of effective temperatures between 4800 and 7500\,K, 
values of log\,$g_2$ are probably 
over-estimated by $\sim$\,0.2\,dex. An artificial feature (``branch'') 
of decreasing log\,$g$ with decreasing $T_{\rm eff}$ is also seen 
for dwarfs between 4400 and 5000\,K, as well as around 6000\,K. 
The systematic overestimation of log\,$g_2$, as well as the artifact 
branches, are likely caused by uncertainties in the fiducial 
templates calculated using the fitted parameter -- spectral 
flux density relations (cf. Sections\,2.2 and 3.6). 
In fact, a similar but more significant artificial branch of dwarf stars 
of $T_{\rm eff} < 4800$\,K is also seen in the HR diagram constructed 
using the adopted parameters of SDSS DR9, presumably 
a consequence of the usage of synthetic spectra by the SSPP in the analysis 
of those late type stars. 

In the $T_{\rm eff}$ -- [Fe/H] plots of Fig.\,8, only stars 
of effective temperatures between 3400 and 7600\,K are shown. 
Overall the distributions from the two methods resemble each other. 
The $\chi^2$ minimization gives slightly smoother distribution 
than the weighted mean, in particular near the edge of the distributions. 
Again, this is a natural consequence of using the extrapolated  
templates in the $\chi^2$ minimization method.  
For stars of $T_{\rm eff1}$ hotter than 7600\,K, 
values of [Fe/H] yielded by the current pipeline are probably unreliable 
and better calibration is needed for those hot stars (cf. Section\,6.7). 

\begin{figure*}
\centering
\includegraphics[width=160mm]{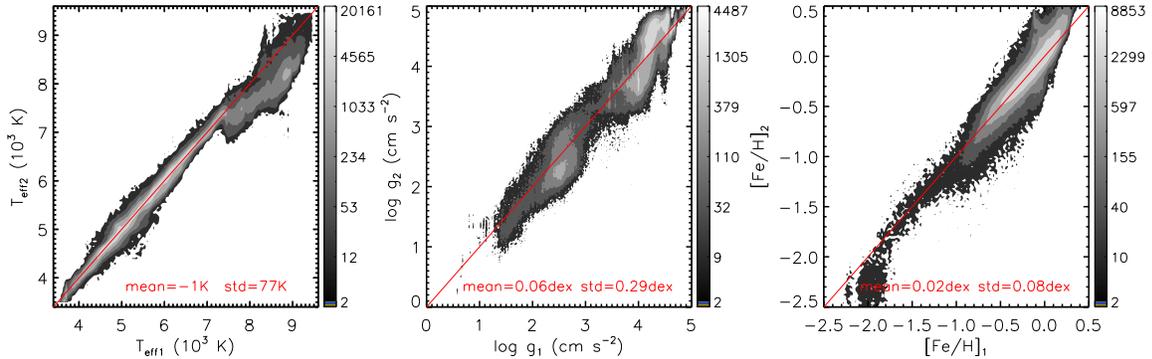}
\caption{Comparison of stellar atmospheric parameters derived with 
         the weighted mean and the $\chi^2$ minimization algorithms. 
         The mean and standard deviation of the differences are 
         marked in each panel.}
\label{Fig9}
\end{figure*}

A direct comparison of parameters derived from the two algorithms 
is shown in Fig.\,9. Here only results derived from spectra of 
SNRs better than 10 are shown. Fig.\,9 shows that on the whole values 
of $T_{\rm eff}$ yielded by the two methods are consistent 
with each other, with an overall dispersion of 77\,K and 
negligible systematic differences. However, for stars of $T_{\rm eff1} > 7500$\,K, 
values of $T_{\rm eff2}$ are a few hundred Kelvin systematically 
lower than $T_{\rm eff1}$. For log\,$g$, while there are no significant 
overall differences, there are clear systematic patterns of differences. 
We believe that those patterns are mostly likely caused by the problematic 
values of log\,$g$ yielded by the $\chi^2$ minimization method, as 
discussed above. For [Fe/H], only results of stars of $T_{\rm eff1}$ between 
3500 and 7500\,K are compared. Values of [Fe/H]$_1$ and [Fe/H]$_2$ 
agree with each other very well, with a mean and standard deviation of 
differences of 0.02$\pm$0.08\,dex. There are some evidence that for  
some stars of super-solar metallicity, [Fe/H]$_1$ is $\sim$\,0.1\,dex lower than
[Fe/H]$_2$. This is probably caused by some weak boundary effects 
of the weighted mean algorithm.  

Given that the $\chi^2$ minimization method yields erroneous values of 
log\,$g$, probably due to the inadequacies of the parameter -- spectral 
flux density relations used to interpolate (and extrapolate) templates,
the current version of LSP3 has adopted the stellar atmospheric parameters 
derived from the weighted mean method, $T_{\rm eff1}$, log\,$g_1$ 
and [Fe/H]$_1$ as the final estimates of parameters $T_{\rm eff}$, log\,$g$ 
and [Fe/H]. Values of $T_{\rm eff2}$, log\,$g_2$ and [Fe/H]$_2$ are provided 
for comparison only. Flags are however assigned to reflect the magnitudes of 
differences between the two sets of parameters derived respectively with the two 
algorithms (cf. Section\,3.8).

\subsection{Comparison of results from multi-epoch duplicate observations}

Owing to the overlapping of FoVs of adjacent plates, 
about 23 percent stars are targeted more than once in the LSS-GAC 
survey (Liu et al. 2014). The number of stars with duplicate 
observations is further enlarged by repeated observations, either 
because the original observations failed to pass the quality control 
(60 per cent of the targeted sources meet the SNR requirements), 
or for some other reasons (cf. Paper\,III). 
Those multi-epoch observations of duplicate targets provide an opportunity to 
test the precision of parameters delivered by the LSP3 at different SNRs and 
for stars located at different positions in the parameter space. 

To investigate the parameter precision, we first select spectra of duplicate stars 
obtained at different nights that have similar SNRs, and compare 
the stellar parameters yielded by the LSP3 as a function of the SNR for stars of 
different $T_{\rm eff}$, log\,$g$ and [Fe/H]. The results are presented 
in Figs.\,10 -- 13, which plot the precisions, i.e., the dispersions of 
parameters deduced from the two epoch observations divided by square 
root of two, as a function of the SNR. 
In the plots stars are grouped into bins of different temperatures 
and metallicities, as well as into dwarfs and giants.
Note that unless specified otherwise,  
all stellar atmospheric parameters presented hereafter refer to 
the final adopted values, i.e. those derived from 
the weighted mean method (cf. Section\,5.1). 

Fig.\,10 shows that the precision of $V_{\rm r}$ is a steep function 
of the SNR and $T_{\rm eff}$, but depends only moderately on [Fe/H]. 
Cooler or more metal-rich stars have better precision. For stars of 
$T_{\rm eff} < 6000$\,K and [Fe/H] $> -0.6$\,dex, $V_{\rm r}$ can be 
determined to a precision of 5\,km\,s$^{-1}$ at a SNR of $\sim$\,15 
and $\sim$\,7.0\,km\,s$^{-1}$ at a SNR of $\sim$\,10. 
For stars of $T_{\rm eff} < 6000$\,K but [Fe/H] $< -0.6$\,dex, the 
corresponding value at a SNR of $\sim$10 is about 10\,km\,s$^{-1}$. 
For hot dwarfs of $T_{\rm eff}$ between 7000 and 9000\,K, the precision 
decreases to $\sim$\,20 and 15\,km\,s$^{-1}$ at a SNR of $\sim$\,15 and 20, 
respectively. 

Fig.\,11 shows that the precision of $T_{\rm eff}$ are mainly sensitive 
to the SNR and $T_{\rm eff}$. In general, except for metal-poor giants, 
the precision of $T_{\rm eff}$ is better than 120\,K at a SNR higher than 
10 for $T_{\rm eff} < 7000$\,K. 
For metal-poor ([Fe/H] $< -0.6$\,dex) giants of $T_{\rm eff} < 7000$\,K, 
the precision is about 200\,K. For the hot, metal-poor stars, 
the precision is visibly worse. 

Fig.\,12 shows that the precision of log\,$g$ is most sensitive to 
the SNR and log\,$g$, but also has some dependence on $T_{\rm eff}$. 
The precision is higher for dwarfs than for giants. For dwarfs of 
SNRs better than 10, the precision ranges from 0.05 to 0.1\,dex, 
depending on $T_{\rm eff}$. For giants, the precision ranges from 
0.2 to 0.4\,dex at a SNR of $\sim$\,10 and becomes better than 0.2\,dex 
at a SNR of $\sim$\,15 for metal-rich stars of $T_{\rm eff} < 5000$\,K. 

Fig.\,13 shows that the [Fe/H] precision is sensitive to the SNR 
and [Fe/H], and to a less degree to $T_{\rm eff}$. 
For stars of [Fe/H] $> -0.6$\,dex, the precision is better than 0.1\,dex 
when the SNR is about 10. For metal-poor ([Fe/H] $< -0.6$\,dex) stars, 
the precision decreases to $\sim$\,0.15\,dex for dwarfs and 0.2\,dex for 
giants at a SNR of better than 10. 
Again, hot, metal-poor stars have poor precision. 

To examine the possible systematic errors introduced 
by low SNRs, we compare parameters deduced from spectra of duplicate targets 
obtained at different epochs but having different SNRs. 
We require that the spectrum of higher quality has a SNR better than 40. 
Fig.\,14 plots the differences of parameters deduced from the two 
epoch observations as a function of the SNR of the spectrum of 
lower quality. Fig.\,14 shows that as long as the SNR of the lower 
quality spectrum is better than 15, 
there are no systematic errors induced by the limited SNR of the 
lower quality spectrum. This is true for the four parameters. 
At lower SNRs, systematic errors occur for  
$T_{\rm eff}$ and [Fe/H], in the sense that the spectra of lower SNRs 
yields higher values of $T_{\rm eff}$ and [Fe/H]. 
At a SNR of $\sim$\,7.5, the systematic errors are about 
50\,K and 0.05\,dex for $T_{\rm eff}$ and [Fe/H], respectively. 

\begin{figure*}
\centering
\includegraphics[width=160mm]{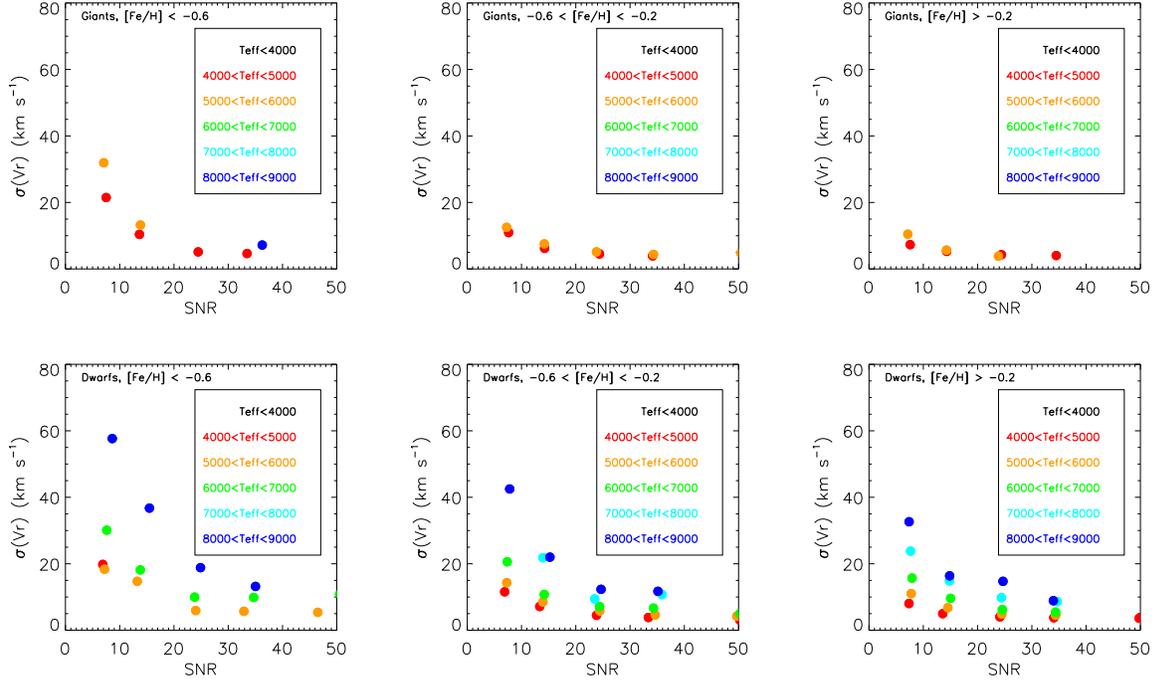}
\caption{Precision of radial velocity determinations, estimated by comparing 
         results yielded by duplicate observations 
         of similar SNRs, as a function of 
         the SNR. Stars are grouped into different bins of $T_{\rm eff}$ (colour-coded), 
         [Fe/H] (different columns) and into giants [log\,$g < 3.5$ (cm\,s$^{-2}$); upper row] 
         and dwarfs [log\,$g$ $\geq$ 3.5 (cm\,s$^{-2}$); lower rows].} 
\label{Fig10}
\end{figure*}

\begin{figure*}
\centering
\includegraphics[width=160mm]{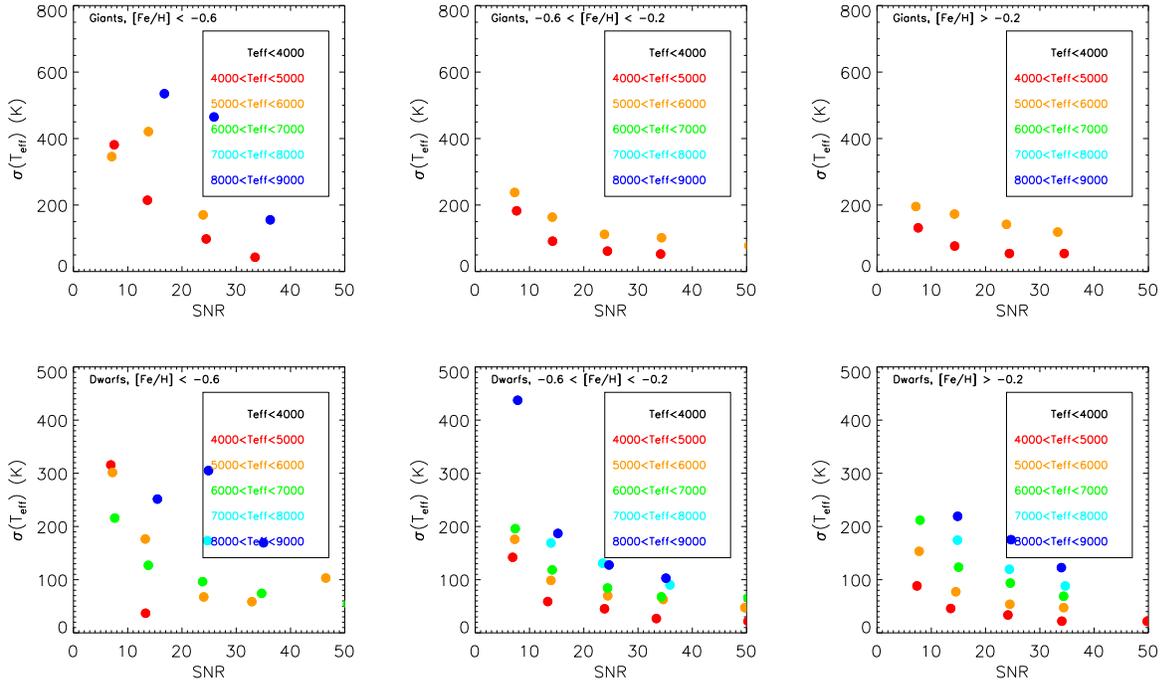}
\caption{Same as Fig.\,10 but for $T_{\rm eff}$. }
\label{Fig11}
\end{figure*}

\begin{figure*}
\centering
\includegraphics[width=160mm]{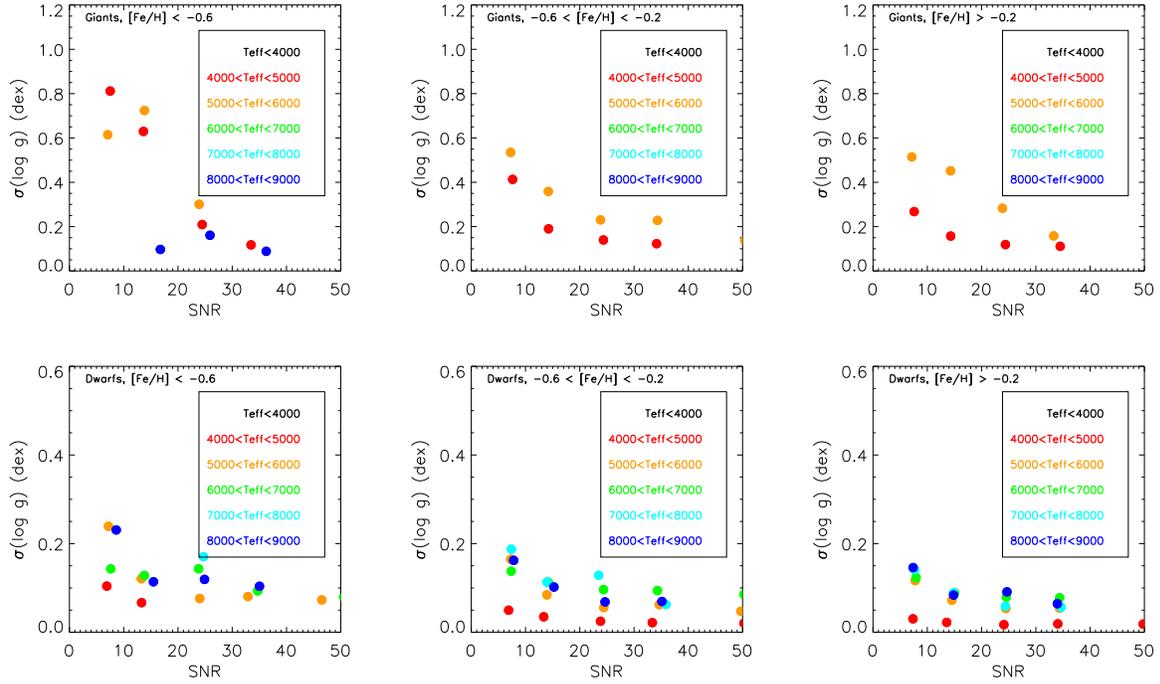}
\caption{Same as Fig.\,10 but for log\,$g$.}
\label{Fig12}
\end{figure*}

\begin{figure*}
\centering
\includegraphics[width=160mm]{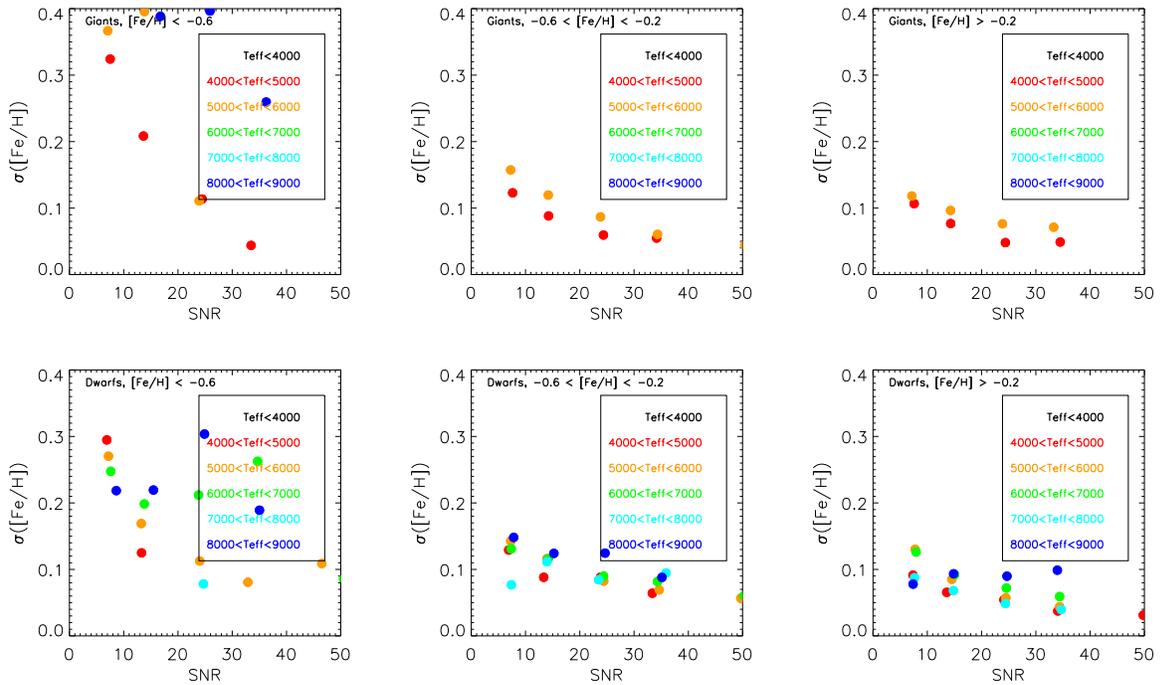}
\caption{Same as Fig.\,10 but for [Fe/H].}
\label{Fig13}
\end{figure*}

\begin{figure}
\centering
\includegraphics[width=80mm]{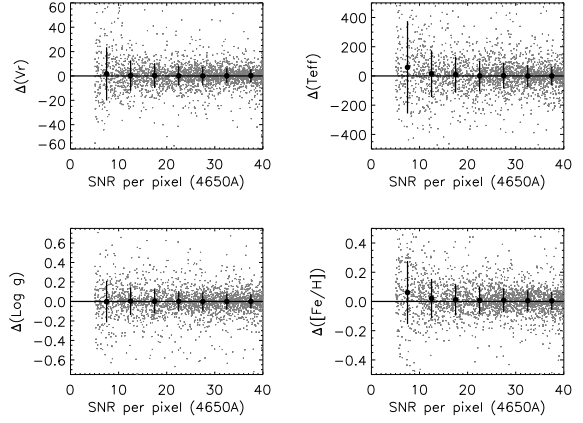}
\caption{Differences of parameters deduced from two duplicate observations 
         of very different SNRs as a function of the SNR of the spectrum 
         of lower quality. The spectrum of higher SNR has SNR better than 40 
         per pixel. The differences refer to parameters derived from the spectra 
         of lower SNRs minus those from the spectra of higher SNRs. 
         The means (black dots) and standard deviations (vertical error bars) 
         for the individual bins of SNR are also shown.} 
\label{Fig14}
\end{figure}

\section{Comparison of the LSP3 Parameters with EXTERNAL DATABASES}
\subsection{Radial velocities}
\begin{figure}
\centering
\includegraphics[width=80mm]{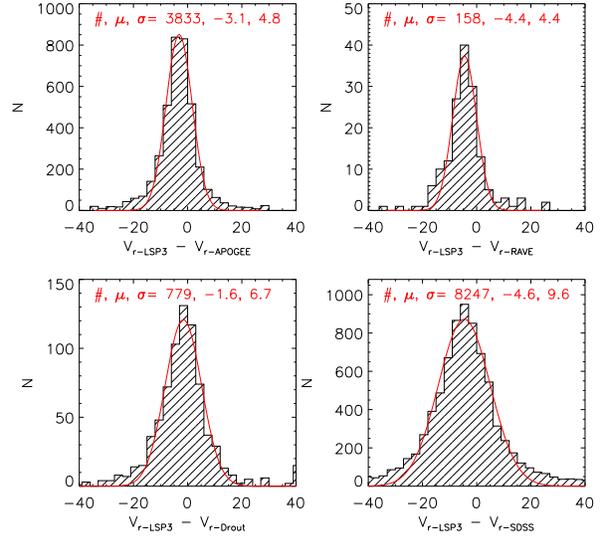}
\caption{Histograms of differences of radial velocities of common stars 
         derived from the LAMOST spectra with the LSP3 and those given by 
         the APOGEE (upper left) and by the RAVE surveys (upper right), 
         those from Drout et al. (bottom left) and from the SDSS survey (bottom right). 
         Over-plotted in red are Gaussian fits to the distributions, 
         with the mean and dispersion of the Gaussian marked in each panel.}
\label{Fig15}
\end{figure}

\begin{figure}
\centering
\includegraphics[width=90mm]{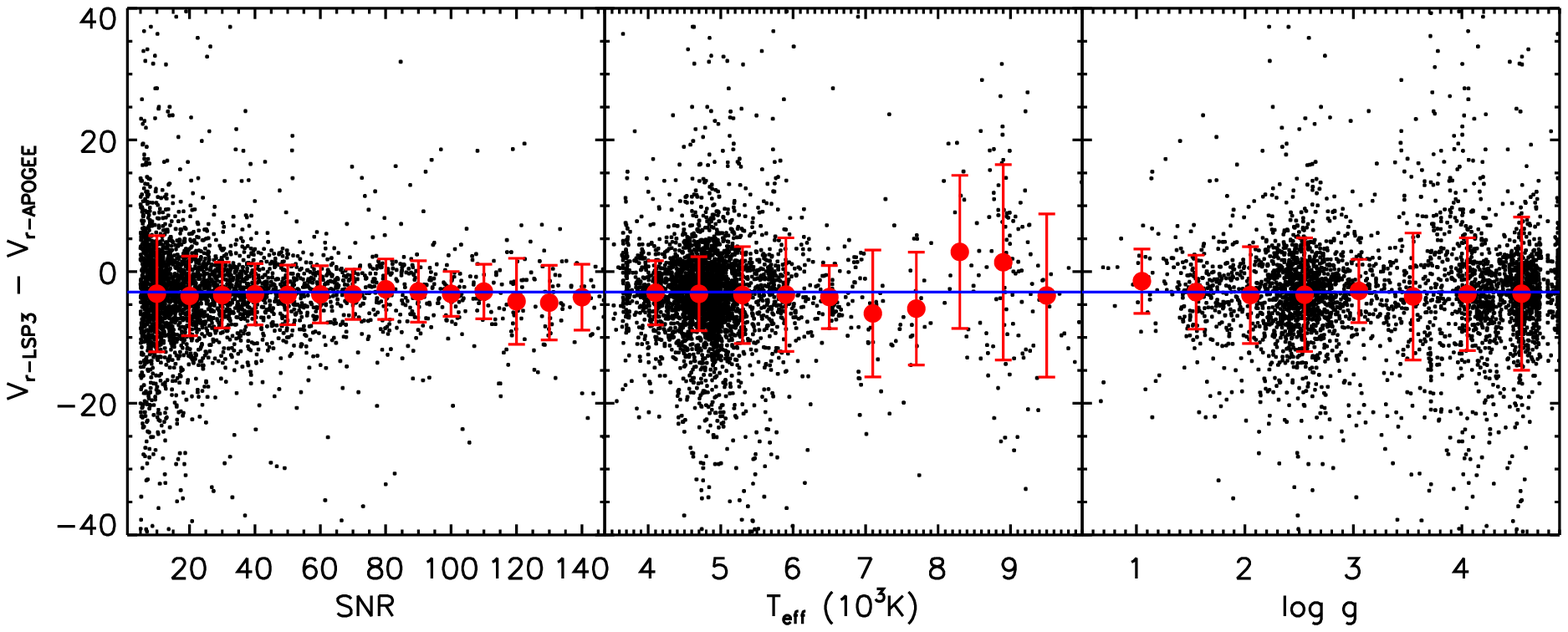}
\caption{The radial velocity difference between the LSP3 and the APOGEE 
         as function of SNR, $T_{\rm eff}$ and log\,$g$. 
         The red dots and the error bars are the mean differences and 
         dispersions in different SNR bins. 
         The blue line represents the $-3.1$\,km\,s$^{-1}$
         offset presented in Fig.\,15.} 
\label{Fig16}
\end{figure}

To examine the accuracy of LSP3 radial velocities, we compare them 
with measurements from a number of independent surveys, including the 
APOGEE (Ahn et al. 2013), RAVE (Steinmetz et al. 2006), and SEGUE (Yanny et al. 2009).

The APOGEE survey collects near-infrared (1.514 -- 1.696\,$\mu$m) spectra 
with a resolving power $R$\,$\sim$\,22,500 (Majewski et al. 2010; 
Ahn et al. 2013). Radial velocities of 57,454 stars have been 
released in the SDSS DR10, with a typical accuracy of $\sim$\,100\,m\,s$^{-1}$ 
(Ahn et al. 2013). We have cross-identified our sample with that of the APOGEE, 
and found 4009 LAMOST spectra of 3035 stars in common after discarding 
those of SNRs $< 5$ in our sample. 
Note that for the comparison we have excluded 176 stars with 
radial velocities differing from the corresponding APOGEE values by 
more than 40\,km\,s$^{-1}$. 
The origin of those large differences is unclear. However, 
it is found that for most of those stars, the templates for radial velocity 
determinations used by the APOGEE have effective temperatures that differ 
significantly ($>500$\,K) from the templates adopted by the LSP3.
A Gaussian fit to the differences of radial velocities derived from the 
remaining 3833 LAMOST spectra and those of the APOGEE yields an offset of 
$-3.1$\,km\,s$^{-1}$ and a dispersion of 4.8\,km\,s$^{-1}$ (Fig.\,15). 
The origin of the small offset is unclear, but the offset is found 
to be independent of the SNR and stellar atmospheric parameters (Fig.\,16). 
As Fig.\,16 shows, the magnitude of the dispersion is primarily controlled 
by the limited SNRs of LAMOST spectra. At high SNRs, the dispersion 
becomes about 4.0\,km\,s$^{-1}$. 
The hot stars have larger scatter, with typical values of about 15\,km\,s$^{-1}$ 
for stars of $T_{\rm eff} > 8000$\,K.
 
The RAVE survey collects medium-resolution ($R$\,$\sim$\,7500) spectra 
of stars of $9 < I < 12$\,mag. around the Ca\,{\sc ii} triplet region 
(8410 -- 8795\,\AA), and delivers radial velocities accurate to 
2.0 -- 3.0\,km\,s$^{-1}$ (Steinmetz et al. 2006). 
There are 83,072 radial velocity measurements for 77,461 stars  
in the RAVE third data release (Siebert et al. 2011). Most RAVE targets 
are in the southern celestial hemisphere and are significantly brighter than 
those targeted by the LSS-GAC. Only 158 RAVE stars are found in common 
with our sample. 
A Gaussian fit to the distribution of differences of velocities 
measured by the two surveys for those common stars yields an offset 
of $-4.4$\,km\,s$^{-1}$ and a dispersion of 4.4\,km\,s$^{-1}$ (Fig.\,15). 
Again, the dispersion arises mainly from the LAMOST measurement uncertainties.

Amongst the LSS-GAC targets of a LAMOST spectral SNR better than 5 and 
an effective temperature between 4000 and 9000\,K, we find 8247 stars 
in common with the SDSS DR9 (Ahn et al. 2012), 
most of which are from the SEGUE survey (Yanny et al. 2009). 
The SDSS spectra have a wavelength coverage and resolving power 
almost identical to those of LAMOST. 
As shown in Fig.\,15, a Gaussian fit to the velocity differences 
yields an offset of $-4.6$\,km\,s$^{-1}$ and a dispersion of 
9.6\,km\,s$^{-1}$. Both the LAMOST and SDSS measurements contribute, 
probably equally, to the dispersion. 

We have also compared the LSP3 radial velocities with datasets available 
from the literature, including velocities measured for stars in the M\,31 
and M\,33 direction (Drout et al. 2009, 2012). A total of 779 stars in 
our sample are found to be in common with those of Drout et al. 
A comparison of those stars (almost all the stars have an LSP3 $T_{\rm eff} < 7000$\,K) 
yields an average difference of $-1.6\pm6.7$\,km\,s$^{-1}$. 

The above comparisons show that the LSP3 radial velocities appear to 
have been underestimated by a small amount, between $\sim$\,$-5$ and $-2$\,km\,s$^{-1}$. 
Considering that the APOGEE yields radial velocities of the highest accuracy 
amongst all measurements discussed above, and that it also has a large
number of stars in common with our sample, we have adopted an offset of 
$-3.1$\,km\,s$^{-1}$ for the LAMOST velocity measurements as 
yielded by the above comparison with the APOGEE measurements. 
A constant of $+3.1$\,km\,$^{-1}$ is then added to all radial velocities yielded by the LSP3. 
The uncertainties of LSP3 radial velocities depend mainly on the 
spectral SNR and type ($T_{\rm eff}$) of the stars. 
As discussed in Section\,5.2, log\,$g$ and [Fe/H] have only 
moderate effects on the $V_{\rm r}$ determinations. 
For FGK stars, the LSP3 radial velocities are probably accurate 
to 5 -- 10\,km\,s$^{-1}$ for SNRs better than 10. The uncertainties 
increase to 10 -- 15\,km\,s$^{-1}$ at a SNR of $\sim$\,5. 
For early type stars, a 15\,km\,s$^{-1}$ accuracy is expected 
for SNRs better than 10.  
The assignment of errors to individual radial velocity 
measurements is described in Section\,7.

\subsection{Testing the stellar atmospheric parameters with the ELODIE 
spectral library}
\begin{figure*}
\centering
\includegraphics[width=180mm]{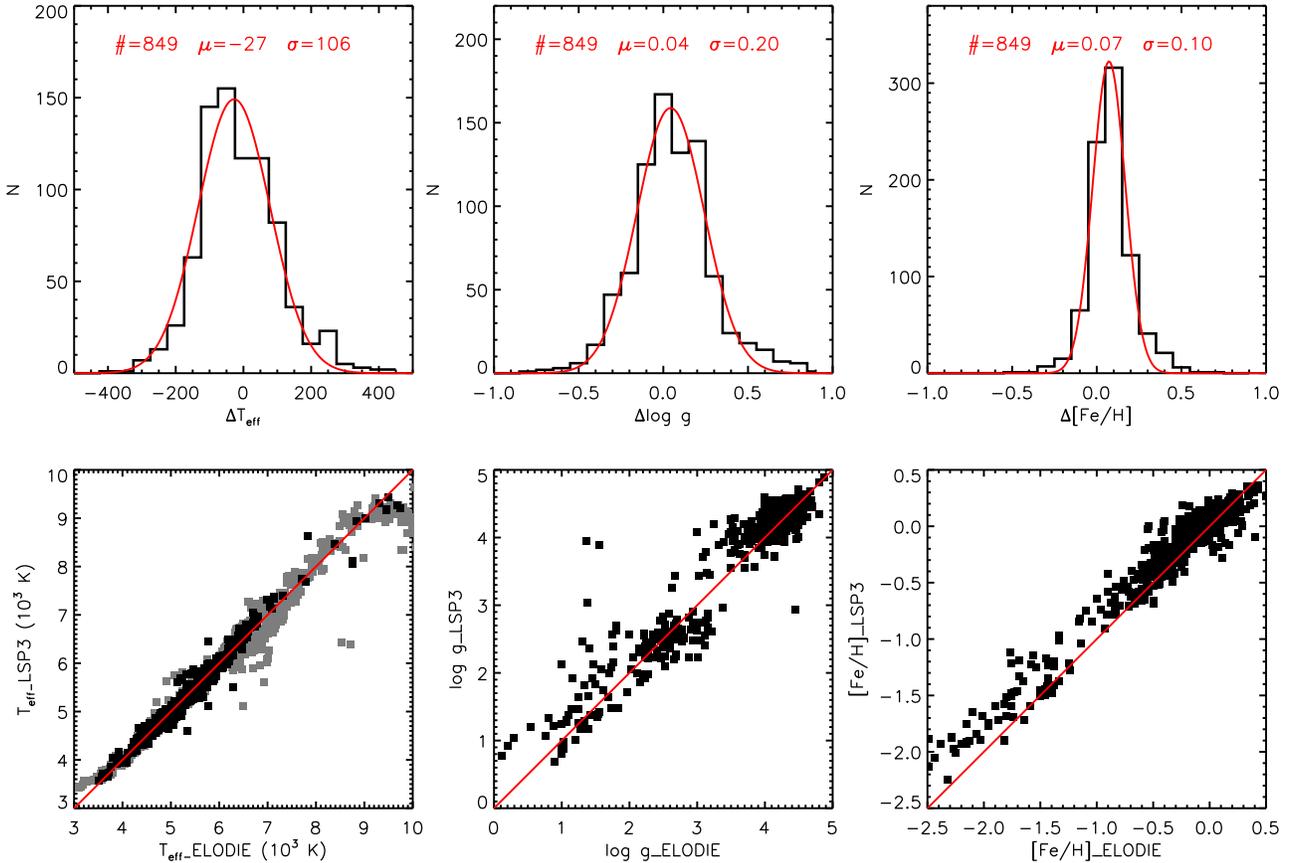}
\caption{Comparison of stellar parameters derived from the ELODIE 
         spectra with the LSP3 with values from the ELODIE library. 
         In the upper panels, for each parameter, 
         a Gaussian fit to the distribution of differences is over-plotted, 
         with the number of stars, the mean and dispersion of the fitted 
         Gaussian marked. Also in the upper panels, only stars with high quality 
         parameters available from the literature, mostly determined with 
         high resolution spectroscopy, are included, and those stars are marked 
         by black dots in the lower panel. For the comparison of $T_{\rm eff}$ 
         in the lower panel, the ELODIE stars without high quality parameters 
         from the literature are also shown, as marked by grey squares. For 
         those stars, the effective temperatures derived with the 
         TGMET software are used.}
\label{Fig17}
\end{figure*}
As described in Section\,2.1, the ELODIE library contains 
1959 spectra of 1388 stars obtained with an echelle spectrograph 
mounted on the Observatoire de Haute-Provence 193\,cm, covering 
the wavelength range 3900 -- 6800\,{\AA} at a resolving power of 
42,000 (Prugniel et al. 2007). More than half of the stars have stellar 
atmospheric parameters collected from the literatures and assigned 
flags ranging from 1 to 4 designating the quality of the parameters, 
with 4 being the best. The catalog also contains 
stellar atmospheric parameters derived using the TGMET software for 
all stars (Prugniel \&  Soubiran 2001; Prugniel et al. 2007). 

The ELODIE spectra are degraded to the LAMOST resolution and 
processed with the LSP3. Fig.\,17 compares the 
resultant LSP3 parameters with the ELODIE values. 
Only ELODIE stars in the temperature range 
$3500 < T_{\rm eff} < 10,000$\,K that have all three stellar 
atmospheric parameters ($T_{\rm eff}$, log\,$g$, [Fe/H]) 
available from the literature are included in the comparison. 
The Figure shows that in general the agreement is very good. 
A Gaussian fit to the distribution of their differences yields an 
average of $-27$$\pm$106\,K, 0.04$\pm$0.20\,dex, 0.07$\pm$0.10\,dex 
for $T_{\rm eff}$, log\,$g$ and [Fe/H], respectively. 
Nevertheless, some systematic discrepancies are seen in [Fe/H]: 
For metal-poor stars, the LSP3 values are $\sim$\,0.1 -- 0.2\,dex 
higher than those of ELODIE. A linear fit yields 
\begin{equation}
{\rm [Fe/H]}_{\rm ELODIE} = -0.07 + 1.08 \times {\rm [Fe/H]}. 
\end{equation}

For ELODIE stars that do not have high quality parameters from the 
literature, many of them are either very hot or cool stars, 
we compare the LSP3 effective temperatures with the TGMET values. 
Those are shown by grey dots in the lower left panel of Fig.\,17. 
The agreement is very good  
for stars of $7000 < T_{\rm eff} < 8500$\,K, with a scatter of  
less than 200\,K. For stars cooler than 3500\,K, effective temperatures 
given by the LSP3 are about 100 -- 200\,K higher. At both low ($\sim$\,3500\,K) 
and high ($\sim$\,9000\,K) temperatures, the LSP3 effective temperatures 
may have been affected by the boundary effects. 

\subsection{Comparison of LSP3 stellar atmospheric parameters with the PASTEL database}

\begin{figure*}
\centering
\includegraphics[width=160mm]{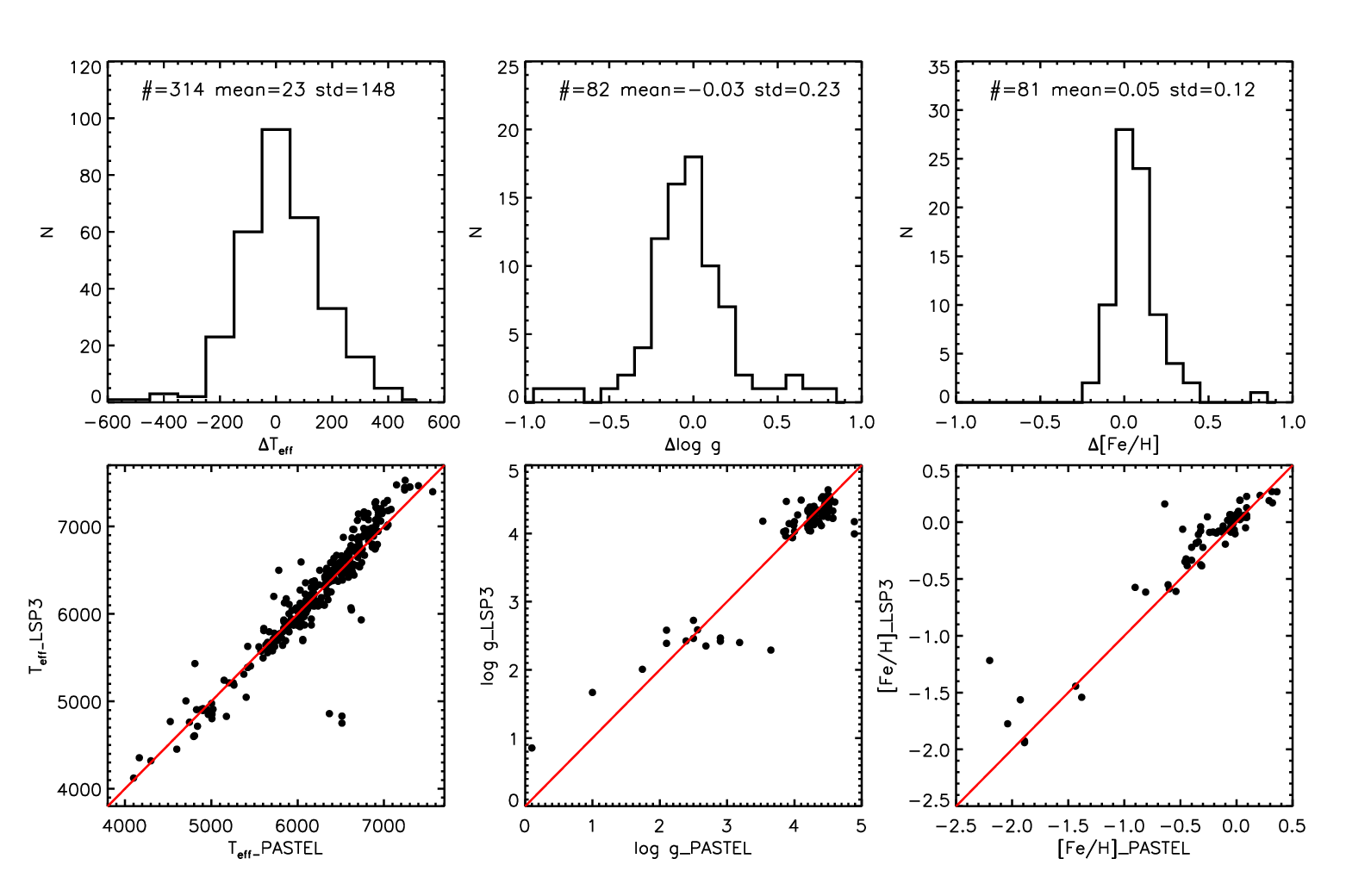}
\caption{Comparison of LSP3 stellar atmospheric parameters with
         those from the PASTEL database. The upper panel 
         shows the distributions of differences for the three parameters. 
         The number of stars, the mean 
         and standard deviation of the distribution are marked in the three top panels.}
\label{Fig18}
\end{figure*}

The PASTEL database (Soubiran et al. 2010) archives 
stellar atmospheric parameters ($T_{\rm eff}$, log\,$g$, and [Fe/H]) 
published in the literatures that are determined with high-resolution 
and high-SNR spectra. The current archive contains more than 30,000 measurements 
of $T_{\rm eff}$ for 16,649 stars. About 6000 of them have 
all the three parameters available. 

Imposing a SNR cut of 10, we find respectively 314, 82 and 81 stars 
in our sample that have values of $T_{\rm eff}$, log\,$g$ and [Fe/H] 
recorded in the PASTEL database. For the comparison, 
a few stars with LSP3 values of $T_{\rm eff}$ 
cooler than 4000\,K or hotter than 7500\,K have been discarded. 
A few stars have more than one records in the PASTEL database. 
For those, we have excluded measurements published 
before 1990, and average the remaining ones with equal weights. 
The comparisons are shown in Fig.\,18.  The distributions of differences 
have mean and standard deviations of $23\pm148$\,K, $-0.03\pm0.23$\,dex 
and $0.05\pm0.12$\,dex for $T_{\rm eff}$, log\,$g$ and [Fe/H], respectively. 
Except for a few obvious outliers, there is 
no systematic trend of difference in $T_{\rm eff}$ for stars between 4000 -- 6700\,K. 
Beyond 6700\,K, the LSP3 yields temperatures of about 100 -- 200\,K higher. 
For dwarfs as well as giants of log\,$g$ between 2 -- 3\,dex, the 
LSP3 log\,$g$ values match those of PASTEL well. There are only a couple of 
stars in the current sample that have log\,$g$ values below 2\,dex or between 
3 and 4\,dex. 
For those of log\,$g < 2$\,dex, the LSP3 seems to have overestimated the values. 
The log\,$g$ values of the few stars with a PASTEL log\,$g$ value between 3 and 4\,dex 
seem to have been either over- or under-estimated by the LSP3. 
However, the numbers of stars are too small to allow for a detail investigation.
For [Fe/H], the LSP3 values are on average 0.05\,dex higher than those of PASTEL. 
More data are needed for a more robust comparison. 

\subsection{Applying the LSP3 to candidates of cluster members}

Stars of a given open cluster (OC) are believed to form almost 
simultaneously from a single gas cloud with a small velocity 
dispersion and have almost the same metallicity. OCs thus serve as a good testbed 
to check the accuracy of radial velocity and metallicity determinations. 

Several OCs have been targeted with the LAMOST. We select member candidates 
of those clusters based on the celestial coordinates, positions on the 
colour-magnitude diagram (CMD) and radial velocities derived with the LSP3 
of the targets. Fig.\,19 illustrates the process of selecting candidates 
of the OC M\,67 as an example. We obtain the basic information of the clusters 
(e.g. coordinates of cluster centers, cluster angular radii) from the DIAS 
database\footnote[2]{http://www.astro.iag.usp.br/ocdb/} (Dias et al. 2002), 
and select stars within twice the angular radius of the cluster. 
Then we draw a line manually delineating the cluster isochrone on 
the CMD, and set a colour cut at each magnitude bin to select possible 
candidates of cluster members. The XSTPS-GAC photometric catalog (Liu et al. 2014) 
is used, and if unavailable, the 2MASS catalog (Skrutskie et al. 2006) is used instead. 
Stars selected from the CMD are cross-identified with targets observed 
with the LAMOST. The distribution of radial velocities derived from the 
LAMOST spectra with the LSP3 of the selected stars is fitted with  
a Gaussian. Stars of velocities within 2$\sigma$ of the mean are adopted as 
candidates of the cluster members. Finally, after imposing the CMD 
and radial velocity cuts, we double the circular radius in celestial coordinates to include 
more member candidates. 

Candidates of cluster members have been selected from the LSP3 results 
for five OCs. Details of those stars are listed in the Appendix. Note that 
for Berkeley\,17, since the stellar density is quite low, our method  
fails to yield any candidate members. To define the candidate members of this cluster,  
we have directly cross-identified our catalog with that of Krusberg \& Chaboyer (2006) directly. 
In the top five rows of Table\,3, we compare the 
average values of [Fe/H] and $V_{\rm r}$ derived from the LAMOST spectra 
with the LSP3 for member candidates of those five OCs with the literature values. 
The number of candidate stars of each cluster used in the analysis 
is listed in the last column of Table\,3.  Note that a zero point correction 
of 3.1\,km\,s$^{-1}$ has been applied to all LSP3 radial velocities derived 
from the LAMOST spectra as discussed in Section\,6.1.
The agreement is very good, except for M\,35 where the LSP3 estimates of [Fe/H] 
are on average 0.2\,dex higher than the literature value. 
The scatters of LSP3 [Fe/H] values deduced 
for individual clusters are about 0.1\,dex. 
For radial velocities, the dispersions are about 3.0--4.0\,km\,s$^{-1}$. 
 
\begin{figure*}
\centering
\includegraphics[width=160mm]{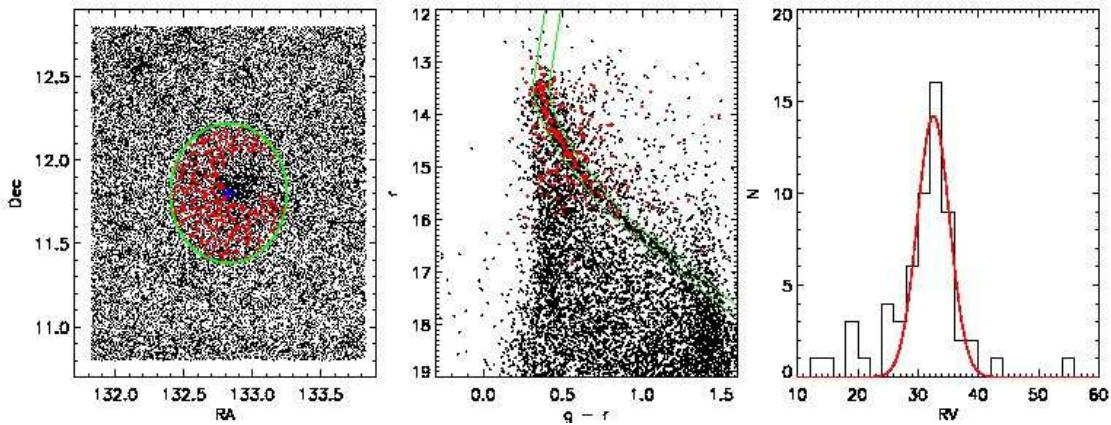}
\caption{Selecting member candidates of the open cluster M\,67 based on 
     the celestial coordinates (left panel), positions on the colour-magnitude diagram (middle panel) 
     and radial velocities derived from the LAMOST spectra (right panel). 
     The blue plus in the left panel 
     indicates the central position of M\,67, 
     while the green circle has a radius twice the angular radius of 
     M\,67. Red pluses in the left and middle panels are stars observed with the LAMOST. 
     The red curve in the right panel is a Gaussian fit to the distribution 
     of radial velocities of stars targeted with the LAMOST that locate within 
     the region delineated by the two green curves in the middle panel. 
     Stars that fall inside the green circle in the left panel and 
     within the region delineated by the two green curves in the middle panel, 
     and have radial velocities within $2\sigma$ of the mean are considered 
     candidates of the cluster members.}
\label{Fig19}
\end{figure*}

Fig.\,20 plots values of log\,$g$, [Fe/H] and $V_{\rm r}$ 
as a function of $T_{\rm eff}$ for candidate member stars of M\,67 (NGC\,2682), 
M\,35 (NGC\,2168) and NGC\,2099, derived from the LAMOST spectra with the LSP3. 
Also over-plotted in the plots of log\,$g$ versus $T_{\rm eff}$
are the Yonsei-Yale (Y$^2$) isochrones (Demarque et al. 2004) for the three clusters.
Fig.\,20 shows that in the $T_{\rm eff}$ -- log\,$g$ plane, the LSP3 
parameters generally match the isochrones. However, for the dwarf stars 
of $T_{\rm eff}$ between 5800 and 6400\,K, the LSP3 values of log\,$g$ 
may have been systematically underestimated by about 0.2\,dex, 
presumably due to a lack of templates of metal-rich dwarfs in that 
temperature range in the MILES library. For stars cooler than 7500\,K, 
the values of [Fe/H] and $V_{\rm r}$ deduced show no obvious trend with 
$T_{\rm eff}$. 
 
We have also tested the accuracy of LSP3 parameters using the SDSS spectra 
of cluster member stars. Lee et al. (2008b) present lists of member stars of 
two OCs (M\,67 and NGC\,2420) and of three globular clusters (GCs; M\,2, M\,13 and M\,15) 
that have SDSS spectra. We apply the LSP3 to the SDSS spectra of those cluster 
member stars. The results are presented in the last 
five rows of Table\,3 and compared with the literature values. 
For [Fe/H], the agreement is generally good. An exception is M\,15, a metal-poor 
([Fe/H] = $-2.26$\,dex) GC, for which the LSP3 values are on average 0.32\,dex higher. 
For M\,2, over-estimates of [Fe/H] for the hot ($T_{\rm eff} > 6000$\,K) 
stars, probably caused by the lack of hot, metal-poor templates in 
the MILES library, have led to relatively large discrepancies (0.19\,dex) with 
the literature values. The dispersions of [Fe/H] of the 2 OCs are less than 0.1\,dex, 
while those of the 3 GCs are about 0.2\,dex. 
For $V_{\rm r}$, the mean velocities are consistent with the literature 
values for all clusters, with systematic differences less than 2.5\,km\,s$^{-1}$ 
except for M\,13. The latter shows a large systematic difference of 4.0\,km\,s$^{-1}$ 
for unknown reasons.  Note that we have already applied a zero point correction 
of 3.1\,km\,s$^{-1}$ to all the LSP3 velocities.

\begin{table*}
\centering
\caption{Comparison of the LSP3 metallicities and radial velocities with 
         the literature values for open and globular clusters}
\label{}
\begin{tabular}{cccccccccc}
\hline
Cluster & [Fe/H] & Reference & $\langle$[Fe/H]$\rangle$ & $\sigma$([Fe/H]) &
   $V_{\rm r}$ (km\,s$^{-1}$) & Reference & $\langle$$V_{\rm r}$ $\rangle$ (km\,s$^{-1}$)& $\sigma$($V_{\rm r}$) (km\,s$^{-1}$)& $N$ \\
 & Literature  & & This work & This work & Literature & & This work & This work & \\
\hline
Berkeley17$^{a)}$ & $-0.1$ & F05 &$-0.06$ & 0.13 & $-73.7$ & F05 & $-73.4$ & 3.5 & 5 \\
NGC1912$^{a)}$ & $-0.11$ & L87 &$-0.12$ & 0.09 & $-1.0$ & S06 &$2.8$ & 2.4 & 14 \\
NGC2099$^{a)}$ & 0.01 & P10 & $-0.02$ & 0.09 & 8.3 & M08 & 9.8 & 4.5 & 27 \\
M35$^{a)}$ & $-0.21$ & B01 &$-0.02$ & 0.08 & $-5.0$ & S11 &$-2.8$ & 4.1 & 47 \\
M67$^{a)}$ & $-0.01$ & J11 &$-0.02$ & 0.08 & 33.5& M86,M08 & 35.5 & 2.7 & 87 \\
M67$^{b)}$ & $-0.01$ & J11 &0.05 & 0.05 & 33.5 & M86,M08 &34.7 & 2.0 & 52 \\
NGC2420$^{b)}$ & $-0.2$ & J11 &$-0.16$ & 0.07 & 73.6& J11 & 73.0 & 3.3 & 163 \\
\hline
M2$^{c)}$ & $-1.62$ & H96 &$-1.43$ & 0.26 & $-5.3$ & H96 &$-5.0$ & 12.0 & 76 \\
M13$^{c)}$ & $-1.54$ & H96 &$-1.58$ & 0.16 & $-245.6$ & H96 &$-249.6$ & 7.0 & 293 \\
M15$^{c)}$ & $-2.26$ & H96 &$-1.94$ & 0.21 & $-107.0$ & H96 &$-106.5$ & 11.0 & 98 \\
\hline
\end{tabular}
\begin{tablenotes}
\item[]$^{a)}$ An open cluster, for which the LSP3 parameters are 
        derived from the LAMOST spectra. 
\item[]$^{b)}$ An open cluster, for which the spectra analyzed with the LSP3 are from the SDSS.
\item[]$^{c)}$ A globular cluster, for which the spectra analyzed with the LSP3 are from the SDSS.
\item[]{\em References} -- B01: Barrado y Navascu\'es et al. (2001);  
            F05: Friel et al. (2005); H96: Harris (1996); J11: Jacobson et al. (2011); 
            L87: Lyngi{\aa} (1987); M86: Mathieu et al. (1986); M08: Mermilliod et al. (2008); 
            P10: Pancino et al. (2010); S06: Szab\'{o} et al. (2006); S11: Smolinski et al. (2011).
\end{tablenotes}
\end{table*}

\begin{figure*}
\centering
\includegraphics[width=160mm]{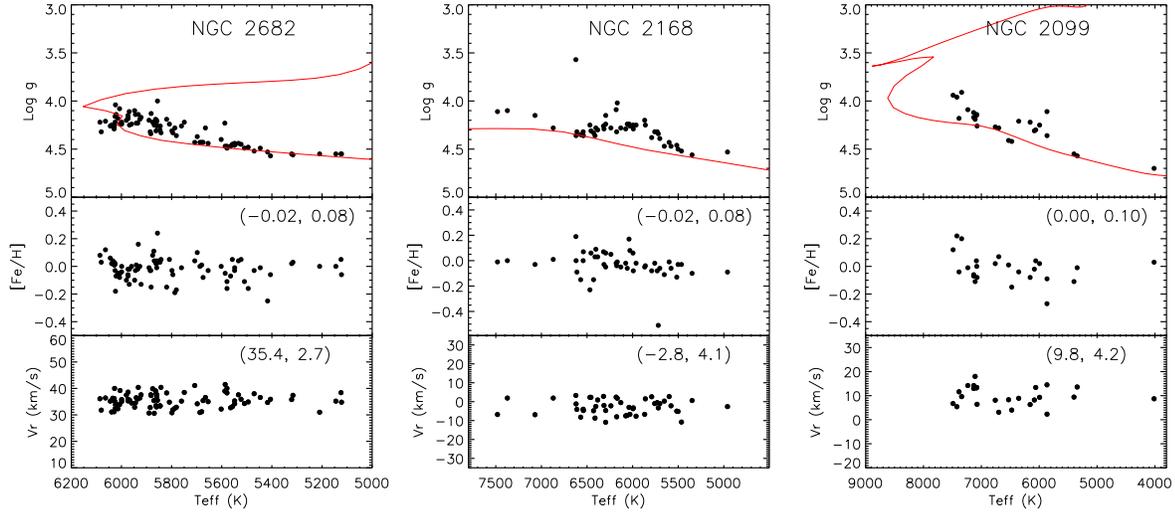}
\caption{Values of LSP3 log\,$g$, [Fe/H] and $V_{\rm r}$ plotted against $T_{\rm eff}$ 
         for member candidates of the open clusters NGC\,2682 (M\,67; left), 
         NGC\,2168 (M\,35; middle) and NGC\,2099 (right) derived from LAMOST 
         spectra. Also over-plotted are Y$^2$ isochrones (red) in the panels of log\,$g$ 
         versus $T_{\rm eff}$. The metallicity and ages of the isochrones are 
         $0.0$\,dex, 4.5\,Gyr (von Hippel 2005) for NGC\,2682, 0.0\,dex, 
         0.2\,Gyr (von Hippel 2005) for NGC\,2168 and 0.0\,dex, 0.6\,Gyr 
         (Kalirai et al. 2001) for NGC\,2099.} 
\label{Fig20}
\end{figure*}

\subsection{Comparison with the APOGEE and SDSS stellar atmospheric parameters}

\begin{figure*}
\centering
\includegraphics[width=160mm]{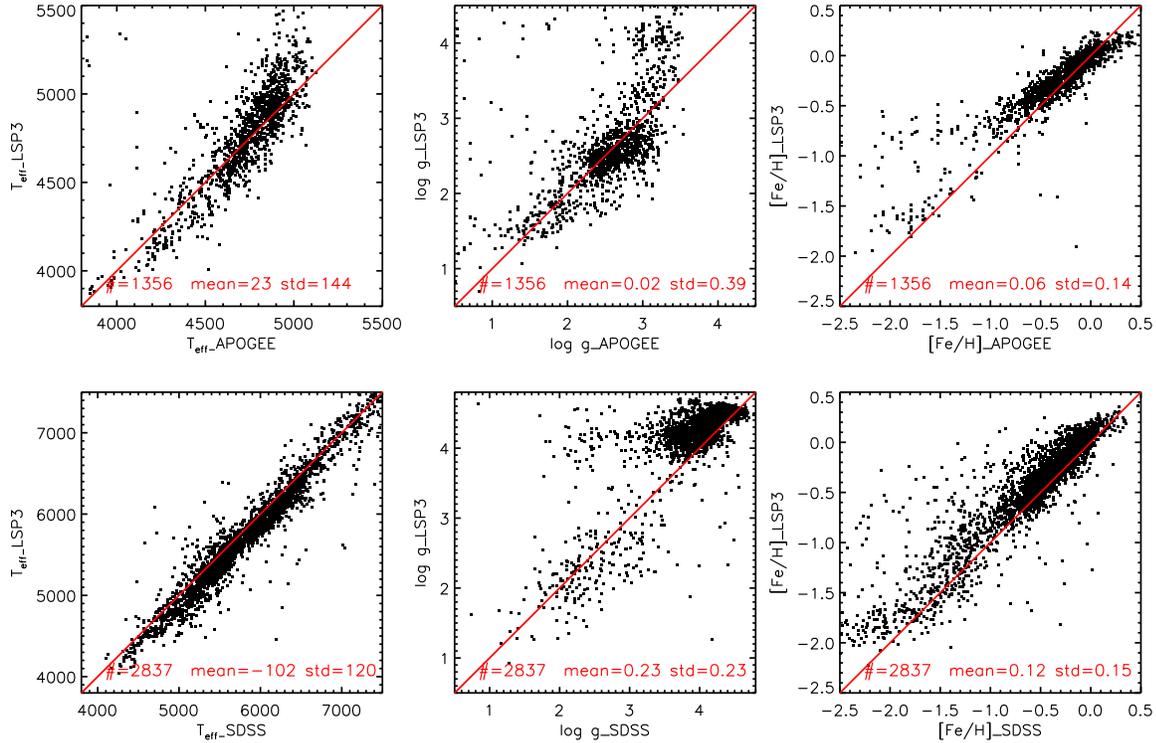}
\caption{Comparison of the LSP3 parameters with those of the APOGEE (upper panels) 
         and SDSS DR9 (lower panels). The number of stars, the mean and standard 
         deviation of the differences are marked in each plot. }
\label{Fig21}
\end{figure*}

In this Subsection, we compare the LSP3 atmospheric parameters with those 
from the APOGEE survey (Ahn et al. 2013) and the SDSS DR9 (Ahn et al. 2012).  
The APOGEE stellar atmospheric parameters are deduced by matching 
the continuum-normalized spectra with a grid of synthetic spectra and 
searching for the best fitting template via a $\chi^2$ minimization algorithm 
(Ahn et al. 2013). The accuracies of APOGEE parameters are estimated to be about 
150\,K in $T_{\rm eff}$, 0.2\,dex in log\,$g$ and 0.1\,dex in [Fe/H] (M\'esz\'aro et al. 2013). 
There are about 33,000 stars in total in the released catalog of 
APOGEE with determinations of $T_{\rm eff}$, log\,$g$ and [Fe/H], 
nearly all of them are giants. 
Cross identification with the LAMOST sources yields 1356 common stars 
for which both the LAMOST and APOGEE spectra have a SNR 
per pixel better than 15. A comparison of the APOGEE and LAMOST 
LSP3 parameters for those targets is presented in the upper panel of Fig.\,21. 
The differences of the two sets of independent determinations 
have an average value and standard deviation of $23\pm144$\,K, $0.02\pm0.39$\,dex, 
$0.06\pm0.14$\,dex for $T_{\rm eff}$, $\log\,g$ and [Fe/H], respectively. 
Among the 1356 stars, 107 sources classified as giants in the APOGEE 
catalog (log\,$g$ $< 3.5$\,dex) have LSP3 log\,$g$ values larger than 4.0\,dex. 
These are relatively hot stars, with an effective temperature around 5000\,K, 
i.e. they are either G-dwarfs or turn-off stars. 
Those stars are responsible for the relatively large discrepancies 
between the APOGEE and LSP3 results in the cases of all three parameters. 
Some small systematic discrepancies are seen in [Fe/H], 
similar to what found when applying the LSP3 to the ELODIE spectra 
[Section\,6.2, Eq. (7)].
The discrepancy is about 0.1\,dex at an APOGEE metallicity of about $-0.5$\,dex.
The scatter of differences in log\,$g$ of the two sets 
of determinations is relatively high ($\sim$\,0.4\,dex). The scatter is mainly 
contributed by hot stars for which the LSP3 yields log\,$g$ values larger than 
3.2\,dex, whereas the APOGEE finds log\,$g$ values smaller than 3.2\,dex.
 
The SDSS stellar atmospheric parameters are derived with the SSPP, 
which adopts average values yielded by a variety of methods (Lee et al. 2008a), 
including template matching with synthetic library, neural network training  
and line-index algorithm. The SSPP parameters are claimed to have a precision 
of about 130\,K, 0.21\,dex and 0.11\,dex for $T_{\rm eff}$, log\,$g$ and [Fe/H], 
respectively, with systematic uncertainties of comparable levels  
in $T_{\rm eff}$ and [Fe/H] (Allende Prieto et al. 2008). Note that the 
SDSS DR9 results differ systematically from the the earlier values 
by about 60\,K in $T_{\rm eff}$ and 0.2\,dex in log\,$g$ as a consequence of 
recalibration (Ahn et al. 2012). 
There are 2837 LSP3 sources in common with those released in the SDSS DR9, 
for which both the LAMOST and SDSS spectra have SNRs better than 15. 
Their parameters are compared in the lower panels of Fig.\,21. 
The differences have an average of 
$-102\pm120$\,K, $0.23\pm0.23$\,dex and $0.12\pm0.15$\,dex for 
$T_{\rm eff}$, log\,$g$ and [Fe/H], respectively. 
The SDSS temperatures are $\sim$\,100\,K systematically higher. 
It seems that the SSPP calibration has systematically under-estimated the 
log\,$g$ by $\sim 0.23$\,dex. The LSP3 metallicities are systematically 
0.12\,dex higher than the 
SSPP values, with no obvious trend for [Fe/H] between $-2.0$ and 0.5\,dex. 
For [Fe/H] $< -2.0$\,dex, the discrepancies become larger, reaching 0.5\,dex 
at a SSPP metallicity of $\sim$\,$-2.5$\,dex. The large discrepancies at 
very low metallicities are likely caused by uncertainties in the LSP3 
estimates due to the limited parameter coverage 
of the MILES library, in which only a few stars have [Fe/H] $ < -2.5$\,dex, 
as well as by uncertainties in the SSPP values. 

There is a group of stars with SSPP log\,$g$ values smaller than 3.5\,dex 
but the LSP3 yields estimates larger than 4.0\,dex. 
This leads to some apparent gaps in the plot comparing the log\,$g$ values 
yielded by the SSPP and by the LSP3 (the bottom middle panel of Fig.\,21). 
The majority of those stars have effective temperatures higher than 5200\,K. 
For those stars, the SSPP find that they are giants or supergiants, whereas 
the LSP3 find they are actually turn-off or dwarfs. Accurate estimates of log\,$g$ 
for these stars are difficult with the LSP3, given the sparse of templates 
at those temperatures and surface gravities. A minority of those stars are 
cooler, for which the SSPP finds they are red giants or clump stars, whereas 
the LSP3 finds they are probably subgiants or dwarfs. 
More analyses are needed to clarify the discrepancies.

\subsection{Comparison with the LAMOST DR1}
\begin{figure*}
\centering
\includegraphics[width=160mm]{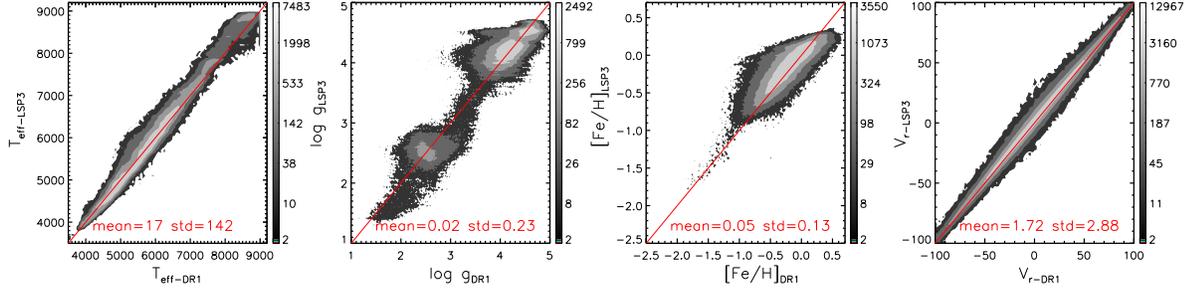}
\caption{Comparison of the LSP3 parameters with those from the LAMOST DR1. 
A total of 454,703 stars are included in the comparison (see the text). 
The colours represent the stellar number density in the parameter space, 
with a colour bar shown on the side. The black line represents  
equal values. The mean and standard deviation of differences 
are marked in each plot.}
\label{Fig22}
\end{figure*}

\begin{figure}
\centering
\includegraphics[width=80mm]{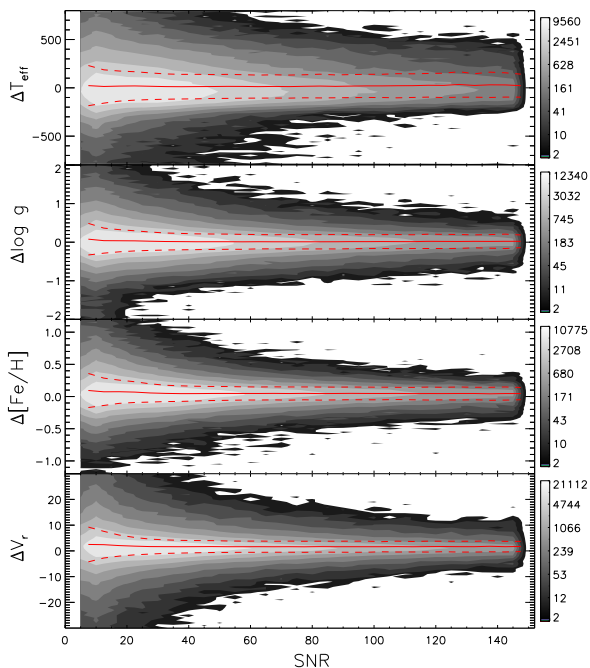}
\caption{Contour maps of LSP3 and LAMOST DR1 parameter differences 
as a function of the spectral SNR.}
\label{Fig23}
\end{figure}

\begin{figure*}
\centering
\includegraphics[width=160mm]{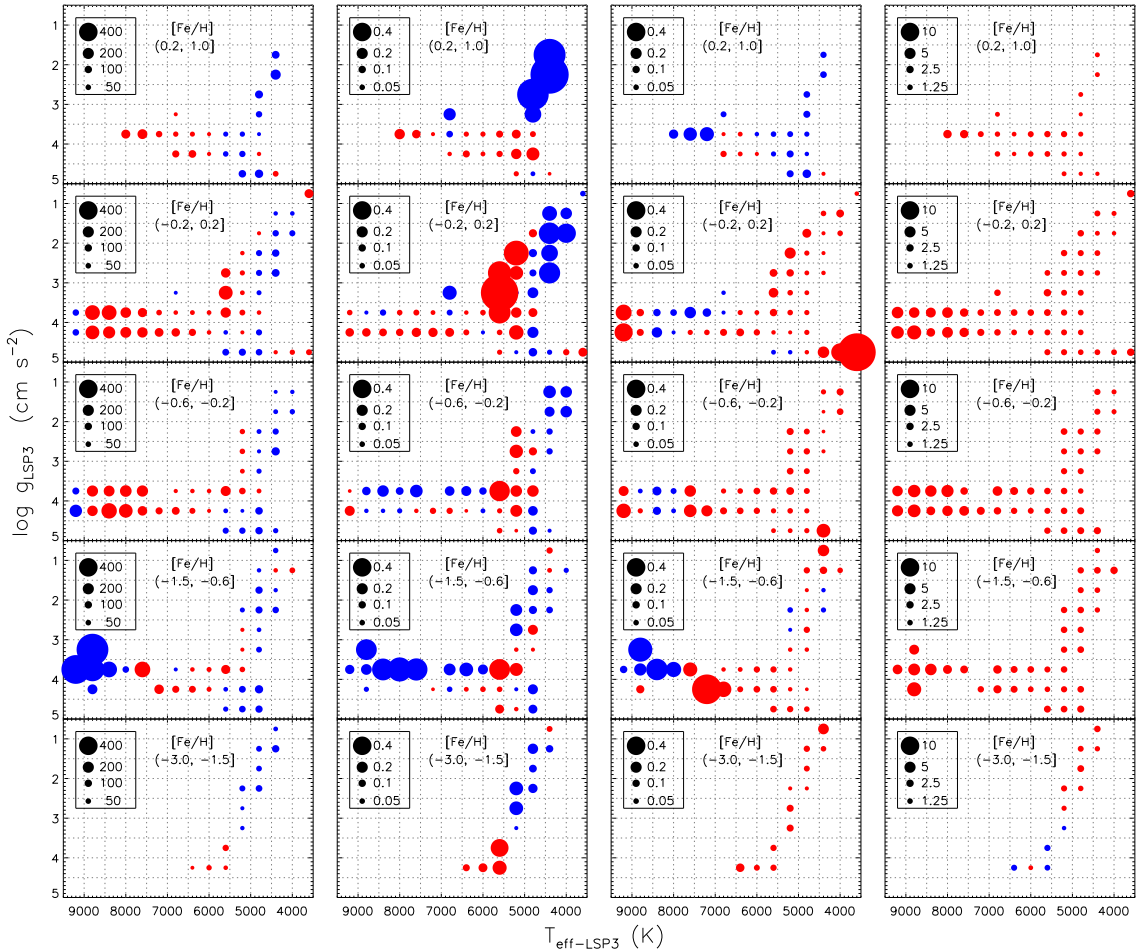}
\caption{Distributions of the averages of differences of the 
 LSP3 and LAMOST DR1 parameters in the $T_{\rm eff}$ -- log\,$g$ plane for, 
from column 1 to 4, $T_{\rm eff}$, log\,$g$, [Fe/H] and $V_{\rm r}$, respectively. 
Panels in different rows correspond to different [Fe/H] bin as marked in the plots. 
The magnitude of the difference are represented by the size of the 
symbols as marked in the plots (in units of K, dex, dex and km\,s$^{-1}$ 
for $T_{\rm eff}$, log\,$g$, [Fe/H] and $V_{\rm r}$, respectively). 
Red symbols represent that the LSP3 values are higher than those of 
LAMOST DR1 on average, while those in blue the opposite.} 
\label{Fig24}
\end{figure*}

The first data release of LAMOST (DR1; Bai et al. 2014) contains values of 
$V_{\rm r}$, $T_{\rm eff}$, log\,$g$ and [Fe/H] deduced by LASP from 1,085,405 
stellar spectra, collected by June 2013 for all components of the LAMOST 
spectroscopic surveys. The LASP derives stellar atmospheric 
parameters via template matching with the ELODIE spectral library (Wu et al. 2014). 
Among the 1,085,405 spectra of LAMOST DR1, 456,697 are from 
the LSS-GAC sources targeted by the VB (Very Bright; $r < 14$\,mag), 
B (Bright; $14 \leq r \lesssim  16.3$\,mag), 
M (Medium bright; $16.3\lesssim r \lesssim 17.8$\,mag) and 
F (Faint; $17.8\lesssim r \leq 18.5$\,mag) plates in the direction of 
the Galactic anti-center as well as the M31/M33 area (cf. Liu et al. 2014; Paper\,III). 
They have been processed with the LSP3. 
Also processed with the LSP3 are some additional spectra not included in the DR1. 
They include spectra that the LASP opts not to process because of the low SNR 
and spectra from plates that the official LAMOST 2-D pipeline fails to process 
the data due to problems related to the flux calibration but have otherwise been 
successfully processed with the flux calibration 
pipeline developed specifically for the LSS-GAC at PKU (Paper I). 
Stellar parameters for all spectra of a SNR $> 3$ are derived with the LSP3. 
The LASP adopts a more stringent SNR cut than the LSP3 does, 
in particular for VB plates collected under bright lunar conditions (Wu et al. 2014). 
In addition, plates collected as parts of a program to monitor the LAMOST 
performance (Liu et al. 2014), as well as nearly 60,000 spectra from plates 
of high Galactic latitudes ($|b| > 30^{\rm o}$), including some selected to 
study the SRCs of high Galactic latitude FoV's (and thus of low interstellar extinction), 
have been processed with the LSP3. A detailed description of the LSS-GAC sample 
can be found in Paper\,III. 

In total, approximately 570,000 spectra with stellar parameters 
released in the LAMOST DR1 have also their parameters determined 
with the LSP3. After imposing a SNR cut of 15, the number of spectra 
in common amounts to 454,703 for $T_{\rm eff}$ in the range of 3500 
-- 9000\,K. A direct comparison of the LSP3 and LAMOST DR1 parameters 
is presented in Fig.\,22. In general, the agreement is quite good, 
with average differences of $17\pm142$\,K, $0.02\pm0.23$\,dex, 
$0.05\pm0.13$\,dex and $1.7\pm2.9$\,km\,s$^{-1}$ for $T_{\rm eff}$, 
log\,$g$, [Fe/H] and $V_{\rm r}$, respectively. 
For $T_{\rm eff}$, there are some systematic discrepancies for stars 
hotter than about 6500\,K, at a level of 100 -- 200\,K. 
For dwarfs and red clump giants, the values of log\,$g$ yielded by the two 
pipelines agree well with each other. However, some subgiants 
($3 <$ log\,$g$ $< 4$ \,dex) in the DR1 are actually found to 
show log\,$g$ values typical of dwarfs by the LSP3. Some systematic 
discrepancies are also seen for the red giant branch stars of log\,$g$ $< 2$\,dex. 
The LSP3 values of [Fe/H] are 0.05\,dex systematically higher than those 
of DR1, over the whole [Fe/H] range. Note that as described earlier, 
the LSP3 metallicities may suffer from some weak boundary effects  
at high metallicities, on the level of about 0.05 -- 0.1\,dex (Fig.\,24). 
Values of LSP3 $V_{\rm r}$ are about 1.7\,km\,s$^{-1}$ higher than those 
of LAMOST DR1. Note that here the LSP3 values have not been 
corrected for the offset of 3.1\,km\,s$^{-1}$ estimated in \S{6.1}. 
The differences are small however compared to 
the estimated uncertainties of the determinations (cf. Section\,6.1). 

The differences between the LSP3 and LAMOST DR1 parameters as a function 
of the SNR are plotted in Fig.\,23. The Figure shows that the small systematic 
differences noted above between the two sets of independent determinations 
are independent of the SNR, and the scatters of the differences decrease 
with increasing SNR as expected. For $T_{\rm eff}$, the scatter decreases from 
a value of $\sim$\,200\,K at a SNR of 10 to $\sim$\,120\,K at SNRs $> 40$. 
For log\,$g$, the dispersion is about 
0.4\,dex at a SNR of 10, and drops to less than 0.2\,dex at the high end of 
the SNRs. For [Fe/H], the dispersion is about 0.3\,dex at a SNR of 10 
and about 0.1\,dex at a SNR of 40. The dispersion in $V_{\rm r}$ 
reaches 6 -- 7\,km\,s$^{-1}$ at SNRs $< 10$, and decreases to 
2 -- 3\,km\,s$^{-1}$ at SNRs $> 20$. 

To further examine any potential systematic discrepancies that might be 
present between the LSP3 and DR1 stellar parameters, we group the stars 
into different bins of $T_{\rm eff}$, log\,$g$ and [Fe/H] based on their 
LSP3 values for a detailed comparison. 
For the comparison, stars with spectral SNRs better than 15 are selected 
and divided into different bins of [Fe/H]: ($-3.0$, $-1.5$], ($-1.5$, $-0.6$], 
($-0.6$, $-0.2$], ($-0.2$, 0.2] and (0.2, 1.0]. For each [Fe/H] bin, the stars 
are further divided into 400\,K by 0.5\,dex bins in the $T_{\rm eff}$ and log\,$g$ 
space. For each bin, the average difference between the LSP3 and 
LAMOST DR1 parameters are calculated.  
The results are plotted in Fig.\,24.
Clear patterns are seen in the Figure for all the four parameters. 
For $T_{\rm eff}$, the LSP3 values of giant branch stars and G/K dwarfs 
are lower than those of LAMOST DR1, whereas for A/F stars the LSP3 gives 
higher values except for those hot stars in the ($-1.5$, $-0.6$] 
[Fe/H] bin for which the LSP3 values are lower by more than $\sim$\,400\,K. 
Typical values of the average differences range from a few tens to a hundred 
Kelvin. For hot stars of $T_{\rm eff} > 8000$\,K the values may rise to 
300 -- 400\,K. 
For log\,$g$, the LSP3 yields lower values for red giant branch stars 
than the LAMOST DR1 does. The discrepancies are more significant 
for metal-rich stars. The discrepancies are likely caused by the fact 
that there are few stars of log\,$g$ $< 2$\,dex in the ELODIE spectral library, 
the spectral template library adopted by the LAMOST DR1 for parameter determinations. 
For subgiants of $T_{\rm eff}$ $\sim$\,5500\,K, the LSP3 yields values of 
log\,$g$ that are 0.4\,dex or more higher than given by the LAMOST DR1. 
The LSP3 values of [Fe/H] for FGK stars are constantly 0.05 -- 0.1\,dex 
higher than those of the LAMOST DR1, expect for stars in the highest 
metallicity bins, where the LSP3 values for G/K stars are $\sim$\,0.05\,dex 
lower than the LAMOST DR1 values.   
For dwarfs of $T_{\rm eff} < 4000$\,K, the discrepancies in [Fe/H] are quite 
large ($>0.4$\,dex). It is difficult to be sure which set of estimates, LSP3 
or LAMOST DR1, is more accurate. It seems to us that the LSP3 values, peaking 
at $-0.1$\,dex (Fig.\,8), are more consistent with what expected for a thin-disk 
population, while as those of DR1 are probably too low.   
Finally, for $V_{\rm r}$, a small systematic difference of 1.7\,km\,s$^{-1}$ 
is found for stars in different parameter bins, except for hot stars, where 
the discrepancies reach 5\,km\,s$^{-1}$.

\subsection{Hot and cool stars}

\begin{figure}
\centering{}
\includegraphics[width=80mm]{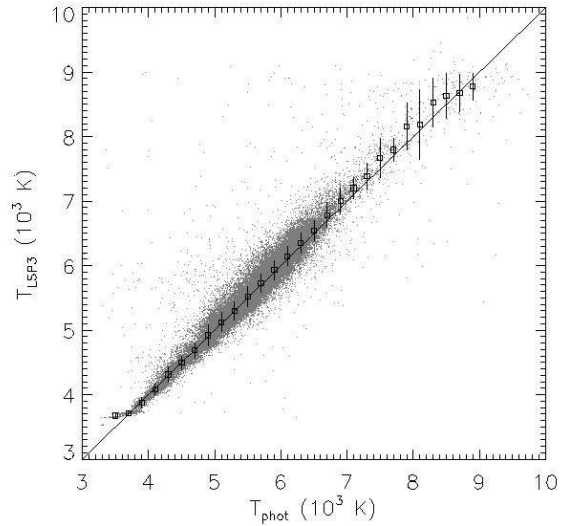}
\caption{Comparison of LSP3 estimates of $T_{\rm eff}$ with photometric values. 
Black squares and error bars are averages and standard deviations of 
the LSP3 values of $T_{\rm eff}$ for stars in the individual bins of photometric $T_{\rm eff}$, 
calculated using Eq.\,(10).} 
\label{Fig25}
\end{figure}
For cool ($T_{\rm eff} < 4000$\,K) and hot ($T_{\rm eff} > 8000$\,K) 
stars, few samples are available for comparisons. 
We thus compare the LSP3 determinations of $T_{\rm eff}$ with the predictions 
of photometric calibration to check the performance of LSP3. 

Huang et al. (2014, submitted) derive an empirical relation of $T_{\rm eff}$ as a function 
of colour $(g-K_{\rm s})$ and metallicity [Fe/H] 
based on over 100 calibration stars with $T_{\rm eff}$ inferred from the interferometric 
observations (e.g. Boyajian et al. 2013; Mozurkewich et al. 2003) and 
[Fe/H] retrieved from the PASTEL archive (Soubiran et al. 2010). 
$g$ and $K_{\rm s}$ are SDSS $g$- and 2MASS $K_{\rm s}$-band photometric magnitudes, respectively. 
The relation is, 
\begin{align}
&   \theta = a_0 + a_1 \times (g-K_{\rm s})_0 + a_2 \times {(g-K_{\rm s})_0}^2 & \nonumber\\ 
&  + a_3 \times (g-K_{\rm s})_0 \times {\rm [Fe/H]} + 
        a_4 \times {\rm [Fe/H]} + a_5 \times {\rm [Fe/H]}^2, &
\end{align}
where $\theta = 5040/T_{\rm eff}$, and $(g-K_{\rm s})_0$ is the dereddened colour. 
A best fit to the data yields coefficients of 0.56653, 0.18358, $-0.00365$, $-0.02477$, 
0.02794 and $-0.00552$ for $a_0$ -- $a_5$, respectively, with a fitting residual 
of about 1.9 per cent. The relation is deduced for dwarfs of $3000 < T_{\rm eff} < 10,000$\,K. 
In fact, an almost identical relation is also found for giants, with a difference 
of a few tens of Kelvin in the predicted $T_{\rm eff}$ for the same colour and metallicity. 
We have thus applied the above relation to all stars, both dwarfs and giants likewise. 
Note that the relation is only valid for stars of [Fe/H] $> -1.0$\,dex due to 
the lack of calibration stars of lower metallicities. For stars of 
[Fe/H] $< -1.0$\,dex, photometric $T_{\rm eff}$ values calculated assuming 
[Fe/H] = $-1.0$\,dex are adopted. 
Since there are only a small number of stars of metallicities lower than this 
in our sample, the simplification does not affect the conclusion below. 
 
For comparison, we select stars of absolute Galactic latitudes larger than 
25\,deg. and $E(B-V)$ given by the extinction map of Schlegel, Finkbeiner 
and David (1998; SFD98 hereafter) smaller than 0.05\,mag, to minimize potential 
errors caused by uncertainties of the reddening corrections. 
We further require that the stars have spectral SNRs higher than 15.
Fig.\,25 compares the LSP3 estimates of $T_{\rm eff}$ with the photometric 
values calculated using the above relation. On the whole, the agreement is good. 
The agreement is particular good for $3700 < T_{\rm eff} < 7000$\,K, 
with no systematic discrepancies and a dispersion of 150\,K only. 
For $T_{\rm eff} < 3700$\,K, the LSP3 estimates are 100 -- 300\,K 
higher than the photometric values, and the discrepancies increase 
with decreasing $T_{\rm eff}$. For $6500 < T_{\rm eff} < 9000$\,K, the LSP3 
values are again 100 -- 200\,K higher than the photometric values, and shows large 
scatters of about 300 -- 400\,K. Few data points are available beyond 9000\,K. 

\begin{figure}
\centering
\includegraphics[width=80mm]{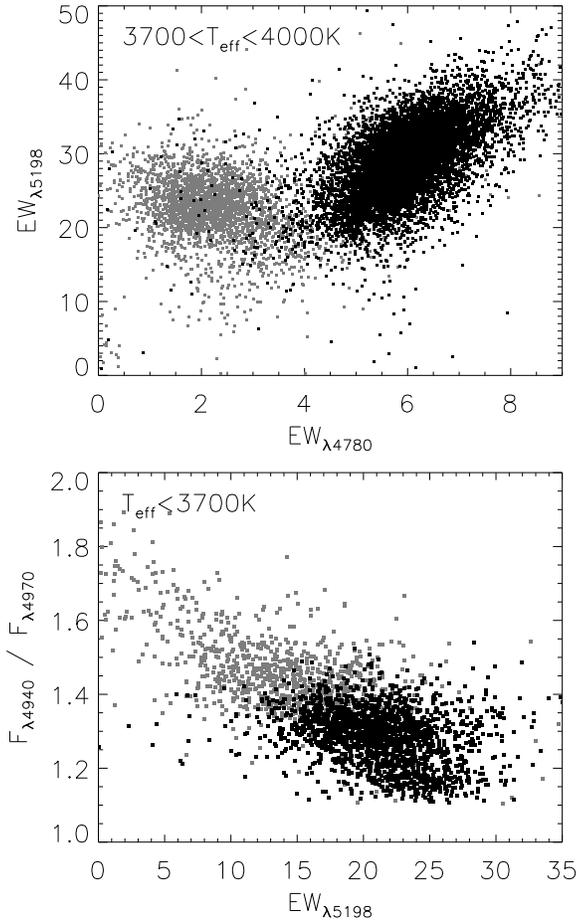}
\caption{Line indices of cool dwarfs (black) and giants (grey). 
Upper panel: EW of the MgH $\lambda$5198 band plotted against 
that of the MgH $\lambda$4780 band for stars of $3700 < T_{\rm eff} < 4000$\,K. 
Lower panel: Flux density ratio $F_{\lambda4940}$/$F_{\lambda4970}$ 
plotted against the EW of the MgH $\lambda$5198 band for stars of $T_{\rm eff} < 3700$\,K.}
\label{Fig26}
\end{figure}
As a further check of the sanity of LSP3 determinations of log\,$g$ 
for cool stars, we have calculated the equivalent widths (EWs) of the 
$\lambda\lambda$4780, 5198 MgH bands and the flux density ratio 
$F_{\lambda4940}$/$F_{\lambda4970}$. The EWs of the MgH $\lambda$4780 
band are calculated by integrating the normalized spectra between 
4760 -- 4800\,{\AA}, whereas for the MgH $\lambda$5198 band, the 
integration is from 5010 to 5260\,{\AA}. The flux density at 4940\,{\AA}, 
$F_{\lambda4940}$, is taken to be the average value between 
4930 -- 4950\,{\AA}, and $F_{\lambda4970}$ that between 4965 -- 4980\,{\AA}. 
Fig.\,26 plots the distributions of EWs of the two MgH bands, as well as 
the $F_{\lambda4940}$/$F_{\lambda4970}$ flux density ratios as a function 
of the EW of the MgH $\lambda$5198 band for cool giants and dwarfs 
that have spectral SNRs better than 10. The Figure clearly shows that 
for stars of $3700 < T_{\rm eff} < 4000$\,K, the dwarfs have larger 
EWs of the MgH $\lambda$4780 band than the giants, whereas for stars of $T_{\rm eff} < 3700$\,K, 
the giants all fall in the upper-left parts of the plot of 
$F_{\lambda4940}$/$F_{\lambda4970}$ against EW$_{\lambda5198}$. 
The results suggest that the LSP3 is capable of discriminating the dwarfs 
from the giants at $T_{\rm eff} < 4000$\,K. 

It is difficult to assess the accuracy of [Fe/H] determined from low 
resolution spectra for cool stars. The uncertainties are likely to be 
large, in particular in view that a few metal-poor cool stars of 
$T_{\rm eff} < 4000$\,K are available in the currently available spectral template 
libraries including the MILES. More calibration sources as well as  
better algorithms to estimates the metallicities of those cool stars 
are clearly needed to access the performance of LSP3.  
Similarly, more studies are needed to access the performance of LSP3 
for log\,$g$ and [Fe/H] of early type stars 
($T_{\rm eff} > 8000$\,K), and we leave them to future work.

\section{Calibration and error estimates of the final parameters} 

As discussed in Section\,6.1, after corrected for a zero-point offset 
of $-3.1$\,km\,s$^{-1}$, radial velocities determined with the LSP3 by 
cross-correlation with the ELODIE spectral templates are adopted as 
the final values. 

\begin{figure}
\centering
\includegraphics[width=90mm]{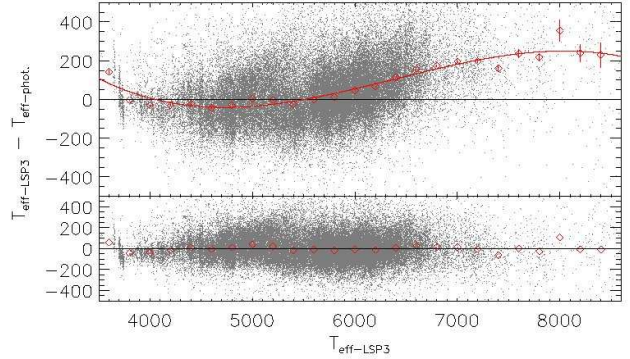}
\caption{Calibrating LSP3 $T_{\rm eff}$ against photometric values. 
The upper panel plots differences of the LSP3 and photometric estimates 
of $T_{\rm eff}$ as a function of $T_{\rm eff}$ yielded by the LSP3. 
Individual data points are represented by grey dots. Red diamonds 
and associated error bars are the average differences and standard 
errors for individual temperature bins. 
The red curve is a third-order polynomial fit to the average values. 
The residuals of the fit are shown in the lower panel, with 
the mean values of residuals for individual temperature bins represented 
by red diamonds.}
\label{Fig27}
\end{figure}
For effective temperatures, the above comparisons of LSP3 values 
with those from the PASTEL and photometric calibration, as well as 
applying the LSP3 on the ELODIE library, show that the LSP3 
over-estimates the values by 100 -- 200\,K for stars of effective 
temperatures at the hotter ($T_{\rm eff} \gtrsim$ 6500\,K) or cooler 
($T_{\rm eff}$ $\lesssim$ 3700\,K) end. 
The systematic deviations for those hot or cool stars are probably 
caused by the poor calibration of MILES parameters for those stars 
(Cenarro et al. 2007). 
We have thus decided to calibrate the LSP3 values of $T_{\rm eff}$ 
to those predicted by the photometric relation Eq.\,(10) given above.
To derive a calibration relation, we first group values of $T_{\rm eff}$ 
yielded by the LSP3 for stars of spectral SNRs better than 15 and 
with $E(B-V)$ given by the SFD98 extinction map less than 0.05\,mag  
into bins of width 200\,K, 
and then calculate the average differences between the values yielded 
by the LSP3 and those predicted by the  
photometric relation for the individual bins. 
A third-order polynomial is used to model the average 
differences as a function of $T_{\rm eff}$ yielded by the LSP3, such that 
\begin{equation}
\Delta T_{\rm eff} = a_1 + a_2 \times T_{\rm eff} + 
a_3 \times T_{\rm eff}^2 + a_4 \times T_{\rm eff}^3
\end{equation} 
\begin{equation}
T_{\rm eff}^{\rm calib} = T_{\rm eff} - \Delta T_{\rm eff}
\end{equation}
The best fit yields 3416, $-1.82$, $3.06\times10^{-4}$ 
and $-1.60\times10^{-8}$ for coefficients $a_1$, $a_2$, $a_3$ and $a_4$, respectively. 
The data and fit residuals are plotted in Fig.\,27, along with the fit. 
The model fits the data well for $T_{\rm eff}$ between 3500 and 8200\,K, with a residual of 25\,K only. 
Since there are few stars hotter than 8500\,K for the fitting, 
we simply assign values of $\Delta$$T_{\rm eff}$ for those hotter 
stars to be that at 8500\,K. 
The $T_{\rm eff}$ thus calibrated are adopted as the final results of LSP3. 
Note that as Fig.\,14 shows, at SNRs of $\sim$\,7.5, 
the LSP3 may have systematically over-estimated $T_{\rm eff}$ by $\sim$\,50\,K. 
However at such low SNRs, the uncertainties of $T_{\rm eff}$ 
($\sim$\,200 -- 400\,K) are significantly larger than the calibration uncertainties discussed 
above. We have therefore ignored possible systematic errors at extremely low SNRs.

For log\,$g$, since no obvious systematic trends are found, and also 
considering that there are no usable high quality calibration sources,  
we have opted not to apply any corrections to values yielded by the LSP3. 

Comparisons with the PASTEL and APOGEE databases, 
as well as applying the LSP3 to the ELODIE spectral library have shown that 
the LSP3 may have slightly over-estimated [Fe/H] (by about 0.04 -- 0.06\,dex on average). 
In principle, one can use the PASTEL database 
as a calibration reference considering that it provides high quality estimates of 
[Fe/H] deduced from high-resolution spectroscopy. This approach seems 
reasonably considering  that the LSP3 
have estimated values of [Fe/H] for some of the PASTEL sources using the 
LAMOST spectra of those sources directly. On the other hand, currently  
the number (81) of PASTEL stars that have been observed with the LAMOST 
is still quite limited. In view fact that a probable offset of 0.05\,dex 
of [Fe/H] yielded by the LSP3 is small compared with the typical uncertainties 
($\sim$\,0.15\,dex) estimated for the LSP3 determinations, 
we have decided not to apply any corrections to the LSP3 determinations of [Fe/H]. 
The possible causes of the small ($\sim$\,0.05\,dex) offset are unclear. 
Note that as Fig.\,14 shows, at SNRs lower than 10, 
the LSP3 may have systematically overestimated [Fe/H] by 0.05 to 0.1\,dex. 
Any such possible systematic errors are small compared to the intrinsic 
uncertainties of the method (0.2 -- 0.4\,dex) at such low SNRs, and 
they have thus been ignored. 

As the LAMOST surveys progress, more LAMOST observations of stars with 
high-resolution [Fe/H] measurements will become available. 
In addition, a project to expand the MILES spectral library, 
both in parameter space coverage by observing additional stars with quality measurements 
of stellar parameters and in wavelength coverage of the spectra and to obtain a better 
calibration of the MILES stellar parameters using interferometric measurements, 
is well under way. One can expect that values of [Fe/H] will be much 
better calibrated in the next release of LSP3 parameters. 
 
The uncertainties of the final LSP3 parameters are estimated by combining 
the systematic and random errors, 
\begin{equation}
 \sigma(X) = \sqrt{\sigma_{\rm sys}^2(X) + \sigma_{\rm ran}^2(X)},
\end{equation}
where $X$ represents $V_{\rm r}$, $T_{\rm eff}$, log\,$g$ 
and [Fe/H]. 

By systematic errors, we refer to those inherent to the LSP3 algorithms, 
including contributions from the uncertainties of stellar atmospheric parameters 
of the MILES templates or from the uncertainties of radial velocities of 
the ELODIE spectral library, as well as any potential systematics induced 
by the weighted mean and biweight mean algorithms. 
For radial velocities, the systematic errors, estimated to be at the 
level of $\sim$\,0.7\,km\,s$^{-1}$ (Section\,2.1), are much smaller than 
the potential random errors, and are therefore set to zero. 
The systematic errors of stellar atmospheric parameters are 
estimated by fitting a 2nd-order polynomial to the absolute differences 
of the MILES parameters and the LSP3 derived values as a function of the 
latter (cf. Section\,4). 
The fitting formulae are presented in Eqs.\,(14) -- (16). 
We assume that $\sigma_{\rm sys}$($T_{\rm eff}$) is a function of 
$T_{\rm eff}$ only, whereas $\sigma_{\rm sys}$(log\,$g$) depends on 
log\,$g$ as well as on $T_{\rm eff}$, and, similarly, $\sigma_{\rm sys}$([Fe/H]) 
on [Fe/H] and $T_{\rm eff}$. Coefficients of the fits are presented in Table\,4. 

The random errors are estimated from the dispersions (divided by square root of 2) 
of parameters yielded by duplicate observations (Sections\,5.2), and are functions 
of the SNR and stellar atmospheric parameters. 
To estimate the random errors, the dispersions of a given parameter 
yielded by the duplicate observations of comparable SNRs are fitted with 
a 2nd-order polynomial as a function of the SNR, $T_{\rm eff}$, log\,$g$ 
and [Fe/H]. The fitting formulae are presented in Eqs.\,(17) -- (20), 
and the coefficients are listed in Table\,4. 
Note the fitting are based on stars of SNR between 10 and 50. 
For stars of SNR better than 50, the random errors are assigned 
to be values of those at SNR = 50. 
\begin{align}
\sigma_{\rm sys}(T_{\rm eff}) = a_1 + a_2 \times T_{\rm eff} + a_3 \times T_{\rm eff}^2 
\label{eqn:tef}
\end{align}
\begin{align}
\sigma_{\rm sys}({\rm log}\,g) = \nonumber\\ 
&b_1 + b_2 \times {\rm log}\,g + b_3 \times {{\rm log}\,g}^2 + b_4 \times T_{\rm eff} + \nonumber\\ 
&b_5 \times T_{\rm eff} \times {\rm log}\,g + b_6 \times T_{\rm eff}^2 
\label{eqn:tef}
\end{align}
\begin{align}
\sigma_{\rm sys}({\rm [Fe/H]}) = \nonumber\\ 
& c_1 + c_2 \times {\rm [Fe/H]} + c_3 \times {\rm [Fe/H]}^2 + c_4 \times T_{\rm eff} + \nonumber\\ 
&c_5 \times T_{\rm eff} \times {\rm [Fe/H]} + c_6 \times T_{\rm eff}^2 
\label{eqn:tef}
\end{align}
\begin{align}
\sigma_{\rm ran}(T_{\rm eff}) = \nonumber\\ 
& d_1 + d_2 \times T_{\rm eff} + d_3 \times {\rm SNR} + d_4 \times T_{\rm eff}^2 + \nonumber\\ 
&d_5 \times {\rm SNR}^2 + d_6 \times T_{\rm eff} \times {\rm SNR}
\label{eqn:tef}
\end{align}
\begin{align}
\sigma_{\rm ran}({\rm log}\,g) = \nonumber\\ 
&e_1 + e_2 \times T_{\rm eff} + e_3 \times {\rm SNR} + e_4 \times {\rm log}\,g + \nonumber\\ 
&e_5 \times T_{\rm eff}^2 + e_6 \times {\rm SNR}^2 + e_7 \times {{\rm log}\,g}^2 + \nonumber\\ 
&e_8 \times T_{\rm eff} \times {\rm SNR} + e_9 \times T_{\rm eff} \times {\rm log}\,g + \nonumber\\ 
&e_{10} \times {\rm SNR} \times {\rm log}\,g
\end{align}
\begin{align}
 \sigma_{\rm ran} ({\rm [Fe/H]}) = \nonumber\\ 
&f_1 + f_2 \times T_{\rm eff} + f_3 \times {\rm SNR} + f_4 \times {\rm [Fe/H]} + \nonumber\\ 
&f_5 \times T_{\rm eff}^2 + f_6 \times {\rm SNR}^2 + f_7 \times {\rm [Fe/H]}^2 + \nonumber\\
&f_8 \times T_{\rm eff} \times {\rm SNR} + f_9 \times T_{\rm eff} \times {\rm [Fe/H]} + \nonumber\\ 
&f_{10} \times {\rm SNR} \times {\rm [Fe/H]}
\end{align}
\begin{align}
\sigma_{\rm ran}(V_{\rm r}) = \nonumber\\ 
&g_1 + g_2 \times T_{\rm eff} + g_3 \times {\rm SNR} + g_4 \times {\rm [Fe/H]} + \nonumber\\ 
&g_5 \times T_{\rm eff}^2 + g_6 \times {\rm SNR}^2 + g_7 \times {\rm [Fe/H]}^2 + \nonumber\\ 
&g_8 \times T_{\rm eff} \times {\rm SNR} + g_9 \times T_{\rm eff} \times {\rm [Fe/H]} + \nonumber\\ 
&g_{10} \times {\rm SNR} \times {\rm [Fe/H]}
\end{align}

In assigning the total errors to the final parameters, 
a lower limit of 30\,K, 0.05\,dex, 0.035\,dex and 3.0\,km\,s$^{-1}$ 
has been set for $T_{\rm eff}$, log\,$g$, [Fe/H] and $V_{\rm r}$, respectively.

\begin{table*}
\caption{Coefficients of fitting formulae for the estimates of systematic and random errors of 
the LSP3 parameters$^{1)}$.}
\label{}
\begin{tabular}{cccccccccc}
\hline
$a_1$ & $a_2$ & $a_3$  \\
\hline
434.41 & $-1.27$($-1$) & 1.11($-5$) \\
\hline
$b_1$ & $b_2$ & $b_3$ & $b_4$ & $b_5$ & $b_6$ \\
\hline
$-5.15$($-1$) & 2.75($-2$) & 8.88($-3$) & 2.49($-4$) & $-3.07$($-5$) & $-8.31$($-9$) \\ 
\hline
$c_1$ & $c_2$ & $c_3$ & $c_4$ & $c_5$ & $c_6$ \\ 
\hline
5.70($-1$) & $-7.05$($-3$) & $-1.81$($-4$) & $-1.65$($-4$) & $-4.86$($-6$) & 1.39($-8$) \\ 
\hline
$d_1$ & $d_2$ & $d_3$ & $d_4$ & $d_5$ & $d_6$ & & & &   \\
\hline
$-103.27$ & 6.03($-2$) & $-3.34$ & -1.53($-6$) & 6.47($-2$) & $-4.61$($-4$) & & & &  \\
\hline
$e_1$ & $e_2$ & $e_3$ & $e_4$ & $e_5$ & $e_6$ & $e_7$ & $e_8$ & $e_9$ & $e_{10}$ \\
\hline
$-6.27$($-1$) & $-$8.91($-5$) & $-$1.89($-2$) & 6.25($-1$) & $-2.45$($-9$) & 9.02($-5$) & $-$9.61($-2$) & 1.65($-7$) & $-2.15$($-5$) & 2.52($-3$)  \\ 
\hline
$f_1$ & $f_2$ & $f_3$ & $f_4$ & $f_5$ & $f_6$ & $f_7$ & $f_8$ & $f_9$ & $f_{10}$  \\
\hline
1.43($-2$) & 2.54($-5$) & $-2.82$($-3$) & 7.26($-2$) & $-1.64$($-9$) & $2.24$($-5$) & 3.40($-1$) & 6.19($-8$) & $-1.47$($-5$) & $1.37$($-3$)  \\
\hline
$g_1$ & $g_2$ & $g_3$ & $g_4$ & $g_5$ & $g_6$ & $g_7$ & $g_8$ & $g_9$ & $g_{10}$  \\
\hline
$7.24$ & $-1.66$($-3$) & $-1.06$($-1$) & $-10.29$ & 4.51($-7$) & 4.60($-3$) & 11.76 & $-5.09$($-5$) & $1.16$($-3$) & 1.85($-1$) \\
\hline
\end{tabular}
\begin{tablenotes}
\item[]$^{1)}$ The numbers in parentheses are powers of ten, thus $-1.27$($-1$) represents $-1.27$ $\times 10^{-1}$.
\end{tablenotes}
\end{table*}

\section{Error sources of the LSP3 parameters}
\subsection{Limitation of the templates}

Possible values of LSP3 parameters are obviously limited 
by the coverage of MILES spectral templates in the parameter space.
Although the MILES library has a relatively broad parameter coverage compared
with other available empirical libraries, it is nevertheless restricted 
by our current knowledge of stars in the solar neighborhood and suffers from
various observational biases. For example, there are few metal-poor late-type
(K/M) dwarfs, as well as few metal-rich AFGKM stars of [Fe/H] $ > 0.3$\,dex. 
Such stars, albeit rare, are expected to be present in the Milky Way,
but are absent in the MILES library due to either their rareness or faintness, 
and thus are difficult to find and measure. The parameter space covered by 
MILES templates is thus unlikely to encompass the whole parameter space 
occupied by stars targeted by the LSS-GAC which surveys a much larger 
volume of the Milky Way and orders of the more stars than the MILES library. 
It is therefore quite likely that the LSP3 parameters will suffer from the 
limited parameter space coverage of MILES templates in one way or other. 

In addition, within the parameter space covered by the MILES library, 
the distribution of MILES templates are inhomogeneous.
There are holes and peaks in the distribution of stars in the parameter space. 
The inhomogeneous distribution can lead to clustering artifacts 
in the resultant LSP3 parameters. To minimize such effects, we have 
interpolated the MILES spectra to fill up some of the most apparent holes in the 
$T_{\rm eff}$ -- [Fe/H] space, and introduced $w_2$, a weight reflecting 
the distribution (clustering) of templates in the $T_{\rm eff}$ -- [Fe/H] 
parameter plane, when calculating the weighted mean of parameters. 
Although the effects are much reduced, they cannot be avoided entirely. 
The effectiveness of those measures also depends strongly on the 
location of the target concerned in the parameter space. 

The MILES spectral parameters themselves are also not entirely free 
of systematics. Although significant efforts have been made 
by Cenarro et al. (2007) to homogenize the parameters 
collected from various sources and determined with a variety of methods, 
some systematic patterns remain, especially at 
$T_{\rm eff}$ $<$ 4000\,K and $T_{\rm eff}$ $>$ 6300\,K, where the parameters
are out of the range of calibration benchmarks (Cenarro et al. 2007). 
Note that Cenarro et al. homogenize the MILES 
parameters by linear regression, which may be insufficient.
For example, the giants and dwarfs from a given source may not share the 
same systematic errors. 
Any outliers, even of a small number, in the MILES can lead to 
significant systematic errors in the LSP3 parameters under certain circumstances. 
By comparing the LSP3 parameters with those predicted by the photometric relation (Section\,6.7), 
with the PASTEL archive (Section\,6.3) as well as with the ELODIE library (Section\,6.2), 
it seems that the MILES may have overestimated $T_{\rm eff}$ by 100 - 200\,K 
for stars between 6500 -- 8500\,K. 
For FGK stars, the systematic errors of MILES effective temperatures, if any, 
are likely to be small, on the level of a few tens of Kelvin. The systematic 
errors in log\,$g$ and [Fe/H] of MILES templates are also expected to be 
small and have negligible impacts on the LSP3 parameters compared with 
other potential error sources as discussed below, except for those
near the boundary of the parameters space.

\subsection{Limitation of the algorithms}
In the current version of LSP3, stellar parameters are determined 
using spectra of the same wavelength range, 4320 -- 5500\,{\AA}, 
for all types of star. The choice works well for the majority of 
FGK stars, but less so for early type or extremely metal-poor stars. 
For the latter, a blue part of the spectra (e.g. 3900 -- 4320\,\AA) 
that include the Ca~{\sc ii} $\lambda\lambda$3933,3967 K and H lines 
are more appropriate. Because for such extremely metal-poor stars, 
most metallic lines other than the prominent Ca\,{\sc ii} doublet 
become too weak to be usable as metallicity tracers. 
The Balmer decrement and jump in the blue are also important diagnostics  
of $T_{\rm eff}$ and log\,$g$ for hot stars. 
Below 4320\,{\AA}, there are also other sensitive tracers of 
$T_{\rm eff}$, log\,$g$ and [Fe/H] for stars of various types 
(e.g. the Ca\,{\sc ii} $\lambda$4226 line; cf. Gray \& Corbally 2008). 
Making use of spectra below 4320\,{\AA} in a proper way will no doubt 
improve the reliability and accuracy of the parameters deduced. 
As mentioned in Section\,3.2, the current version of LSP3 is being 
optimized in order to make a better use of 
spectra between 3900 -- 4320\,{\AA} in the blue, as well as red-arm 
spectra. 
The improvements will be included in the next release of LSP3. 

As discussed in Section\,4, some LSP3 parameters may have suffered 
from various boundary effects in the weighted mean algorithm. 
For examples, values of log\,$g$ may have been overestimated 
by $\sim$\,0.4\,dex for some subgiants, whereas for very metal-rich 
([Fe/H] $> 0.3$\,dex) stars, the LSP3 may have provided values of [Fe/H] 
that are too low. Although such stars are relatively rare, they are of interest 
for some studies. Given that a project to expand 
the MILES library, both in parameter space coverage and in spectral 
wavelength range, is well under way, we expect that those effects 
will be much reduced in the next release of LSP3. 
In the meanwhile, Flags 4 -- 6 (cf. Section\,3.8) are assigned to 
help identify stars that may have suffered from those obvious boundary effects.

\subsection{Quality of the spectra}

Another, probably the most important factor that affects the 
accuracy of parameter determinations is the quality of spectra 
used, including the quality of the raw spectra (the SNR, spectral resolution) 
as well as the quality of data reduction (sky subtraction, 
flux calibration, estimates of flux density uncertainties). Some targets are 
seriously affected by interstellar extinction, and this also 
has an impacts on the reliability and accuracy of parameters derived.  

\subsubsection{Spectral SNRs}
Limited spectral SNRs are always a major contributor to the parameter 
uncertainties. As discussed in Section\,5, the SNR 
is the dominant factor affecting the accuracy of the 
four parameters ($V_{\rm r}$, $T_{\rm eff}$, log\,$g$ and [Fe/H]) in all cases. 
The uncertainties increase rapidly as the SNR decreases (Figs.\,10 -- 13). 

A large number of spectra acquired during the LAMOST Pilot Surveys 
(September 2011 - May 2012) have low spectral SNRs. Although we 
have provided LSP3 parameters determined from spectra of SNRs lower than 
5, we strongly advice that interested users apply a minimum SNR cut 
based the scientific goals when using the LSP3 
parameters. One can refer to the extensive comparisons, presented 
in Sections\,5 and 6 to choose a suitable SNR 
threshold. Alternatively, one can exclude stars with poor parameter 
determination based on the values of flags assigned, which 
include effects of the limited SNRs (cf. Section\,7).  
\begin{figure}
\centering
\includegraphics[width=90mm]{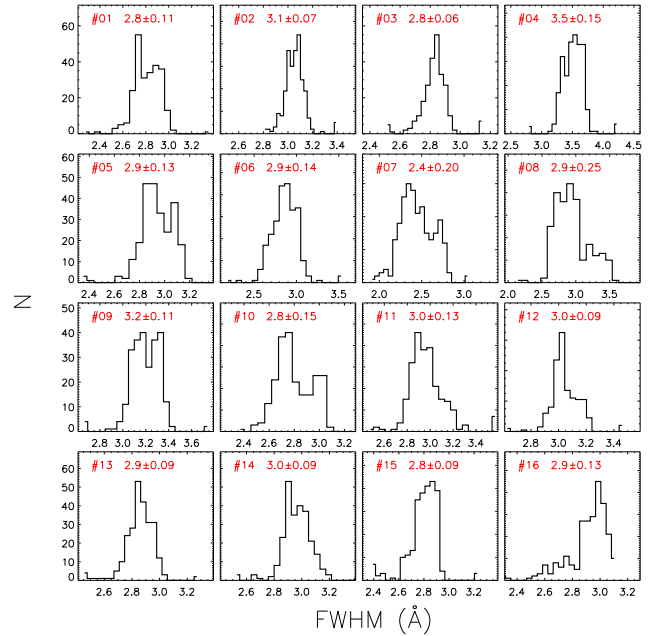}
\caption{Distributions of FWHMs of the Hg $\lambda$5460 arc line 
 of the 250 fibers of each of the sixteen spectrographs of LAMOST. 
The spectrograph ID, the mean and standard deviation of FWHMs are 
marked in each panel.}
\label{Fig28}
\end{figure}

\subsubsection{Spectral resolution}

\begin{figure}
\centering
\includegraphics[width=80mm]{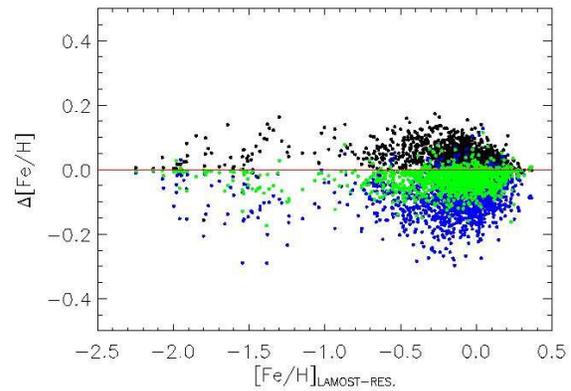}
\caption{Changes in [Fe/H] values derived with the LSP3 from the ELODIE
spectra degraded to a spectral resolution of FWHM of 2.4\,{\AA} (black), 
3.0\,{\AA} (green) and 3.5\,{\AA} (blue), compared to those derived from 
the ELODIE spectra degraded to match the average resolution of the LAMOST 
spectra, plotted as a function of the latter.}
\label{Fig29}
\end{figure}
Spectral resolution is another important factor that affects 
the determinations of stellar parameters, especially [Fe/H]. 
As described in Section\,2.1, we degrade the resolving power of 
the template spectra to match the average value of spectra from the 4000 
fibers of LAMOST. Fiber to fiber variations of the spectral resolving 
power are thus ignored, although such variations have been observed 
among the LAMOST fibers. Fig.\,28 plots the distributions of FWHMs 
of an arc line at 5460\,{\AA} for 250 fibers from each 
of the 16 spectrographs of LAMOST, as measured on a specific arc plate. 
It shows that fibers of the individual spectrographs have quite 
different distributions of spectral resolution, with mean values 
ranging from 2.4 (Spectrograph \#7) to 3.5\,{\AA} (Spectrograph \#4), 
although the majority of spectrographs have a mean FWHM between 2.8 and 3.0\,{\AA}. 
For a given spectrograph, the resolution of the individual fibers 
varies by as much as 0.2\,{\AA}.

To examine how the LSP3 results are affected by the spectral resolution, 
we degrade the ELODIE spectra to different resolutions and compare 
the resultant LSP3 parameters. The results of [Fe/H] are shown in Fig.\,29. 
It shows that as the FWHM varies from 2.4 to 3.0\,{\AA}, the resultant 
[Fe/H] can vary by as much as 0.1 -- 0.2\,dex on average. The 
variations can reach 0.2 -- 0.3\,dex as FWHM varies from 2.4 
to 3.5\,{\AA}. Therefore, the simple treatment of 
spectral resolution in the current implementation of LSP3,  
assuming a uniform wavelength-dependent spectral resolving power 
for all the 4000 fibers of LAMOST, will introduce an uncertainty 
of $\sim$0.1\,dex in the resultant values of [Fe/H] 
because of the fiber to fiber variations of the spectral resolution. 
For about 10 -- 20 per cent stars targeted by some specific spectrographs, 
the errors can even reach 0.2\,dex as a consequence. 
In the next release of LSP3, we plan to incorporate day to day, fiber 
to fiber variations of the spectral resolution to further improve the 
accuracy of our parameter determinations. 

\subsubsection{Estimates of the spectral flux density uncertainties}
Robust estimates of the spectral flux density errors for LAMOST 
spectra are difficult given the complexity of the data collection 
system and the process of data reduction. The LAMOST has an unprecedented 
large FoV of 20 sq. deg., and employs 4000 fibers to relay the light to 
16 spectrographs. Accurate flat fielding and background (sky, scatter light) 
subtraction are therefore quite difficult as the process can be easily affected 
by the potential inhomogeneity in either the background or the 
instrument sensitivity. 
As a consequence, proper and robust propagation of errors are not 
easy tasks in the implementation of data reduction pipeline. 
Fig.\,30 shows the distribution of minimum values of reduced $\chi^2$ 
for spectra collected in a specific plate as a function of the 
spectral SNR and $T_{\rm eff}$. One expects that the minimum 
values peak at unity in the ideal case where the flux density 
errors have been properly propagated and accurately modeled, 
and that the best-matching model (template) is a good approximation 
of the observation. In addition, the minimum values of reduced $\chi^2$ 
should be independent of the SNR, a natural consequence of 
the definition of reduced $\chi^2$. 
However, Fig.\,30 clearly shows that the minimum values of reduced $\chi^2$ 
peak at about 0.7 instead of unity at low SNRs. 
This is likely caused by the overestimated flux density errors by the 
current pipeline. This is one of the reasons why 
weights are assigned to the matching templates based on 
the relative values of $\chi^2$ (Section\,3.4), rather than 
the absolute ones, i.e. based on the probability of occurrence 
of that particular value of $\chi^2$ given the correct model (template). 
By such a choice, the potential effects caused by the underestimation 
of $\chi^2$ has been much reduced. 
Efforts to obtain more robust estimates of the flux density errors are in progress. 
Fig.\,30 shows that the minimum values of reduced $\chi^2$ increases 
with increasing SNR. It is likely that at high SNRs, the differences 
between the observed spectrum and the template dominate the value of $\chi^2$, 
whereas at low SNRs, the observational uncertainties dominate. 
The $\chi^2$ show little trend with $T_{\rm eff}$. 

\begin{figure}
\centering
\includegraphics[width=80mm]{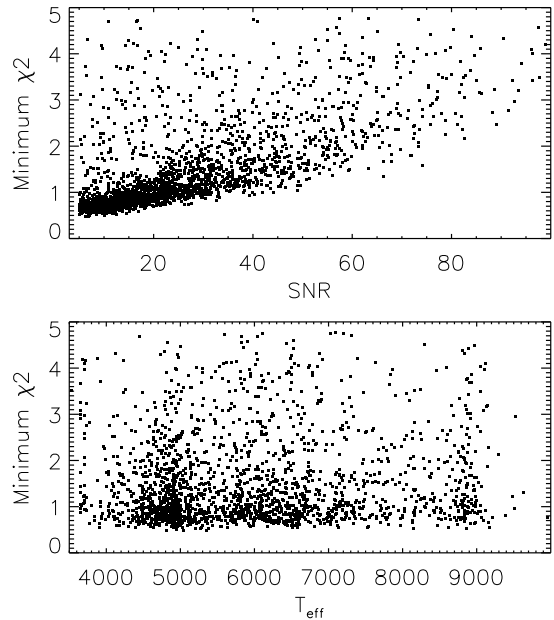}
\caption{Distribution of the minimum values of reduced $\chi^2$ as a function 
         of the spectral SNR (upper panel) and $T_{\rm eff}$ (lower panel).}
\label{Fig30}
\end{figure}

\subsubsection{Extinction and flux calibration}

As described in Section\,3.3, given that both the 
interstellar extinction and uncertainties in the spectral flux calibration could 
affect the actual shape of the observed SEDs, the LSP3 has 
used a third-order polynomial to scale the SEDs of template spectra 
to match that of the target spectrum when calculating $\chi^2$. 
The practice assumes that the effects of interstellar extinction 
and the distortions of SED caused by uncertainties in the flux calibration can 
be modeled with a third-order polynomial. Since the current implementation 
of LSP3 makes use of spectra in a limited wavelength range of 4320 -- 5500\,{\AA} 
only, a third-order polynomial is indeed found to be sufficient for the purpose. 
As a check, we use a third-order polynomial to fit the extinction curve of Fitzpatrick (1999), 
which is thought to be a good description of 
the Galactic interstellar extinction by the diffuse medium (Yuan et al. 2013), 
assuming $R_V=3.1$ and $E(B-V)=0.5$\,mag. For the wavelength range of interest, 
the residuals of the fit amount to only 0.1 per cent. 
Even for an expanded wavelength range of 3900 -- 5500\,{\AA}, 
a third-order polynomial seems to be adequate. 
However, if one uses 3900 -- 5500\,{\AA}, there are always some spectra  
for which a third-order polynomial may be insufficient to model the 
deviations of their SEDs from those of template spectra.  
  
\section{Summary}
In this work, we describe the algorithms and implementation of LSP3 -- 
the LAMOST Stellar Parameter Pipeline at Peking University, a pipeline 
developed to determine the stellar parameters (radial velocity $V_{\rm r}$, 
effective temperature $T_{\rm eff}$, surface gravity log\,$g$ and 
metallicity [Fe/H]) from LAMOST spectra based on a template matching technique. 

The LSP3 determines radial velocities by cross-correlating the target 
spectrum with templates provided by the ELODIE library. For the 
determinations of stellar atmospheric parameters, $T_{\rm eff}$, log\,$g$ 
and [Fe/H], templates from the MILES library, obtained with a spectral 
resolving power similar to that of LAMOST spectra and accurately 
flux-calibrated, are used instead. The atmospheric parameters are 
estimated via two approaches, the weighted mean and $\chi^2$ minimization. 
In the current released version, only results deduced with the former 
approach are adopted, those from the latter method are provided for 
comparison only. 
The LSP3 provides robust estimates of parameters of stars in a large 
parameter space ($-1000$ -- 1000\,km\,s$^{-1}$ for $V_{\rm r}$, 
3500 -- 9000\,K for $T_{\rm eff}$, 0.5 -- 5\,dex for log\,$g$ 
and $-2.5$ -- 0.5\,dex for [Fe/H]). In the current work, we have 
focused on the performance of LSP3 for FGK stars. 

Extensive studies have been carried out to check the performance of LSP3. 
The pipeline has been applied to spectral templates from the ELODIE 
and MILES libraries themselves, to multi-epoch LAMOST spectra of duplicate 
targets, to LAMOST and SDSS spectra of candidate member stars 
of open and globular clusters. Stellar parameters derived with the 
LSP3 are compared with independent measurements available from a 
number of external databases, including the PASTEL archive, the APOGEE, 
SDSS and RAVE surveys as well as the LAMOST DR1. The studies are used to 
characterize and quantify uncertainties of the LSP3 parameters as a function 
of the spectral SNR, and of the stellar parameters.  

The accuracy of radial velocity is found to mainly depend on the SNR 
and spectral type, and varies from $\sim$\,5, 5 -- 10 and 10 -- 20\,km\,s$^{-1}$ 
for G/K-, F- and A-type stars, respectively, for a SNR per pixel ($\sim$\,1.07\,{\AA}) 
at 4650\,{\AA} better than 10. Typical accuracies of stellar 
atmospheric parameters for FGK dwarfs and most G/K giants are better 
than 150\,K, 0.25\,dex, 0.15\,dex for $T_{\rm eff}$, log\,$g$ and [Fe/H], respectively.
For stars of effective temperatures cooler than 4000\,K or hotter than 7000\,K,
the LSP3 provides reasonable $T_{\rm eff}$ estimates, with an accuracy of
about 100 -- 200\,K for the cool ($3500 < T_{\rm eff} < 4000$\,K) stars,
and 200 -- 400\,K for the hot ($7000 < T_{\rm eff} < 8500$\,K) ones.
The LSP3 also provides robust log\,$g$ for stars of $T_{\rm eff} < 4000$\,K.
For some subgiants, the LSP3 values of log\,$g$ may have been 
overestimated by 0.4\,dex due to boundary effects of the weighted 
mean algorithm. Similarly, [Fe/H] values of some stars of super-solar 
metallicity are probably underestimated by 0.05 -- 0.1\,dex. 
Stars suffering from obvious boundary effects can be identified using 
the LSP3 flags.

More calibration sources with accurate parameters are needed 
to improve the parameter space coverage as well as the accuracy 
of parameters deduced with the LSP3. A project to expand the MILES 
spectral library, both in the parameter space coverage as well as in 
the spectral wavelength range, is well under way. Inclusion of the 
new templates, together with better calibrated parameters of the 
templates are expected to significantly improve the accuracy of LSP3 results. 
We also plan to provide estimates of [$\alpha$/Fe] and [C/Fe] 
in the next release of LSP3. 

The current version of LSP3 has been applied to over a million 
LAMOST spectra collected hitherto, mostly for sources targeted 
by the LSS-GAC. Stellar parameters deduced with the LSP3, together 
with estimates of $E(B-V)$ and distances to individual stars, 
deduced by making use of the LSP3 parameters (Paper\,III), as well 
as those deduced from multi-band photometry (Chen et al. 2014), 
are presented in the form of value-added products of LAMOST data 
release. Following the data policy of LAMOST surveys, the data 
as well as the LSP3 pipeline will be public released as value-added 
products of the first data release of LAMOST 
(LAMOST DR1; Bai et al. 2014), currently scheduled in December, 2014, 
and can be accessed via http://162.105.156.249/site/LSS-GAC-dr1/, 
along with a description file.

\vspace{7mm} \noindent {\bf Acknowledgments}{
We thank the anonymous referee for valuable suggestions.
This work is supported by National Key Basic Research 
Program of China 2014CB845700.
Guoshoujing Telescope (the Large Sky Area Multi-Object Fiber 
Spectroscopic Telescope LAMOST) is a National Major Scientific 
Project built by the Chinese Academy of Sciences. Funding for 
the project has been provided by the National Development and 
Reform Commission. LAMOST is operated and managed by the National 
Astronomical Observatories, Chinese Academy of Sciences.}

\label{lastpage}

\appendix
\section {}
\begin{table*}
\begin{minipage}[]{180mm} \centering
\caption{Candidate members of star clusters observed with the LAMOST}
\label{}
\begin{tabular}{cccccrrr}
\hline
Cluster & RA (2000.0) & Dec (2000.0) & $T_{\rm eff}$ & log\,$g$ & [Fe/H] & $V_{\rm r}$ & SNR \\
 &  (deg.) & (deg.) & (K) & (cm\,s$^{-2}$) & (dex)  & km\,s$^{-1}$ &  \\
\hline
Berkeley17 &    80.080030 &    30.523356 &  4467.3 &  2.22 & $-0.16$ &  $-71.2$ &  15.8 \\
Berkeley17 &    80.158312 &    30.578175 &  4316.3 &  2.13 &  0.07 &  $-70.3$ &  11.3 \\
Berkeley17 &    80.172651 &    30.601170 &  4994.7 &  2.43 & $-0.60$ &  $-72.5$ &  20.6 \\
Berkeley17 &    80.187027 &    30.633909 &  4786.9 &  2.51 & $-0.10$ &  $-75.3$ &  12.2 \\
Berkeley17 &    80.193692 &    30.519862 &  4416.5 &  2.01 & $-0.05$ &  $-77.8$ &  17.9 \\
   NGC1912 &    81.637042 &    35.908412 &  7116.1 &  4.26 & $-0.14$ &    2.6 &  13.5 \\
   NGC1912 &    81.922060 &    35.862551 &  7014.9 &  4.22 & $-0.05$ &    1.3 &  17.7 \\
   NGC1912 &    81.970785 &    35.587292 &  6788.1 &  4.30 & $-0.12$ &   $-3.2$ &  26.1 \\
   NGC1912 &    82.020551 &    35.880601 &  6899.6 &  4.31 & $-0.14$ &    2.9 &  29.2 \\
   NGC1912 &    82.048509 &    35.761982 &  6925.4 &  4.29 & $-0.11$ &    1.2 &  20.2 \\
   NGC1912 &    82.060197 &    35.528253 &  7403.7 &  4.09 &  0.02 &    4.2 &  33.5 \\
   NGC1912 &    82.100735 &    35.863465 &  7061.8 &  4.30 & $-0.04$ &    4.4 &  13.5 \\
   NGC1912 &    82.118035 &    35.684713 &  6572.6 &  4.34 & $-0.11$ &    3.0 &  15.9 \\
   NGC1912 &    82.150155 &    35.819408 &  7178.1 &  4.29 & $-0.14$ &    6.4 &  35.9 \\
   NGC1912 &    82.198105 &    35.888803 &  7493.5 &  4.00 &  0.08 &   $-2.6$ &  34.3 \\
   NGC1912 &    82.219693 &    35.639864 &  7369.5 &  4.01 &  0.03 &    3.2 &  53.1 \\
   NGC1912 &    82.278458 &    36.271576 &  6814.5 &  4.32 & $-0.14$ &    1.6 &  39.7 \\
   NGC1912 &    82.326016 &    36.097890 &  6805.3 &  4.31 & $-0.12$ &    0.8 &  23.1 \\
   NGC1912 &    82.517774 &    35.812384 &  6974.2 &  4.23 & $-0.20$ &    1.8 &  36.7 \\
   NGC2099 &    87.593707 &    32.352967 &  5865.0 &  4.36 & $-0.09$ &    2.3 &  19.3 \\
   NGC2099 &    87.727658 &    32.584512 &  7085.5 &  4.15 &  0.04 &   13.2 &  58.4 \\
   NGC2099 &    87.741285 &    32.368901 &  6060.1 &  4.30 &  0.04 &   13.4 &  36.4 \\
   NGC2099 &    87.805791 &    32.824768 &  7488.5 &  3.94 &  0.12 &    6.7 &  34.3 \\
   NGC2099 &    87.806249 &    32.465110 &  7078.9 &  4.14 &  0.00 &   13.4 &  61.0 \\
   NGC2099 &    87.939731 &    32.498580 &  6759.4 &  4.27 &  0.02 &    8.1 &  75.0 \\
   NGC2099 &    87.945725 &    32.536423 &  7131.8 &  4.17 & $-0.06$ &   14.2 &  73.1 \\
   NGC2099 &    87.959815 &    32.429468 &  5993.5 &  4.25 &  0.02 &    9.3 &  24.8 \\
   NGC2099 &    87.963059 &    32.639557 &  5399.1 &  4.55 & $-0.11$ &    9.4 &  11.6 \\
   NGC2099 &    87.979722 &    32.693774 &  7385.9 &  4.18 & $-0.04$ &   11.6 &  32.8 \\
   NGC2099 &    87.999774 &    32.329985 &  6529.4 &  4.41 &  0.01 &    8.3 &  31.2 \\
   NGC2099 &    88.019759 &    32.621234 &  5344.0 &  4.57 & $-0.01$ &   13.6 &  16.1 \\
   NGC2099 &    88.032838 &    32.444156 &  5865.9 &  4.11 & $-0.27$ &   14.5 &  34.4 \\
   NGC2099 &    88.058791 &    32.549114 &  4014.8 &  4.70 &  0.03 &    8.7 &  11.2 \\
   NGC2099 &    88.060874 &    32.630768 &  6699.3 &  4.28 &  0.07 &    3.1 &  62.1 \\
   NGC2099 &    88.077153 &    32.666130 &  7133.0 &  4.12 & $-0.07$ &   13.0 &  69.9 \\
   NGC2099 &    88.096656 &    32.586605 &  6083.6 &  4.31 & $-0.02$ &    8.2 &  30.8 \\
   NGC2099 &    88.164077 &    32.340684 &  6475.0 &  4.42 & $-0.15$ &    4.0 &  26.9 \\
   NGC2099 &    88.166683 &    32.707965 &  6156.6 &  4.22 & $-0.08$ &    6.3 &  27.1 \\
   NGC2099 &    88.187577 &    32.798827 &  7340.6 &  3.91 &  0.20 &    9.6 &  84.9 \\
   NGC2099 &    88.190944 &    32.591233 &  6354.3 &  4.21 & $-0.04$ &    8.9 &  23.9 \\
   NGC2099 &    88.198069 &    32.406829 &  7422.6 &  3.96 &  0.22 &    5.4 &  48.8 \\
   NGC2099 &    88.238835 &    32.623684 &  7073.9 &  4.26 & $-0.08$ &    6.4 &  70.8 \\
   NGC2099 &    88.331256 &    32.462140 &  7233.4 &  4.09 & $-0.01$ &   14.2 &  30.9 \\
   NGC2099 &    88.402562 &    32.446199 &  7107.5 &  4.19 & $-0.11$ &   18.0 &  66.9 \\
       M35 &    90.812715 &    24.047386 &  5652.9 &  4.47 & $-0.11$ &    0.3 &  13.3 \\
       M35 &    91.056838 &    24.446490 &  6870.5 &  4.28 &  0.01 &    1.9 &  42.9 \\
       M35 &    91.057711 &    23.531181 &  7481.9 &  4.11 & $-0.01$ &   $-6.8$ &  49.7 \\
       M35 &    91.079898 &    24.035402 &  5718.7 &  4.34 & $-0.51$ &   $-3.3$ &  13.8 \\
       M35 &    91.156781 &    24.785329 &  6039.3 &  4.24 &  0.17 &   $-3.0$ &  26.5 \\
       M35 &    91.170843 &    23.500702 &  5516.3 &  4.46 & $-0.13$ &   $-5.0$ &  11.5 \\
       M35 &    91.226171 &    25.216736 &  7373.5 &  4.10 &  0.00 &    1.9 &  42.9 \\
       M35 &    91.242804 &    24.765560 &  5791.3 &  4.38 & $-0.08$ &    2.6 &  28.2 \\
       M35 &    91.370669 &    23.923172 &  6456.6 &  4.31 &  0.06 &    2.3 &  32.8 \\
       M35 &    91.429181 &    24.870560 &  7070.4 &  4.15 & $-0.03$ &   $-6.9$ & 108.2 \\
       M35 &    91.502833 &    23.891186 &  6295.0 &  4.15 &  0.06 &   $-4.7$ &  30.8 \\
       M35 &    91.570607 &    24.002285 &  6295.6 &  4.24 & $-0.05$ &  $-10.9$ &  27.3 \\
       M35 &    91.576787 &    24.316003 &  6620.2 &  4.36 &  0.00 &   $-1.2$ &  16.6 \\
       M35 &    91.631412 &    24.938741 &  6383.1 &  4.30 &  0.03 &    0.9 &  24.1 \\
       M35 &    91.637225 &    23.993000 &  6569.5 &  4.35 & $-0.15$ &   $-8.2$ &  55.4 \\
       M35 &    91.655450 &    24.214521 &  6540.1 &  4.36 &  0.07 &   $-4.7$ &  66.8 \\
 \hline
\end{tabular}
\end{minipage}
\end{table*}

\setcounter{table}{0}
\begin{table*} \begin{minipage}[]{180mm} \centering
\caption{\it -- continued}
\label{}
\begin{tabular}{cccccrrr}
\hline
Cluster & RA (2000.0) & Dec (2000.0) & $T_{\rm eff}$ & log\,$g$ & [Fe/H] & $V_{\rm r}$ & SNR \\
 &  (deg.) & (deg.) & (K) & (cm\,s$^{-2}$) & (dex)  & km\,s$^{-1}$ &  \\
\hline
       M35 &    91.672723 &    24.807495 &  5579.1 &  4.47 & $-0.06$ &   $-2.2$ &  13.0 \\
       M35 &    91.684405 &    24.280308 &  5992.4 &  4.29 & $-0.08$ &   $-3.3$ &  28.9 \\
       M35 &    91.744527 &    25.133462 &  6428.3 &  4.33 & $-0.15$ &   $-4.8$ &  21.0 \\
       M35 &    91.792861 &    23.956444 &  6622.3 &  3.57 &  0.19 &    3.3 &  32.5 \\
       M35 &    91.951777 &    24.766640 &  5599.6 &  4.43 & $-0.01$ &    2.7 &  16.1 \\
       M35 &    92.109718 &    24.459892 &  6076.0 &  4.29 & $-0.02$ &   $-7.5$ &  45.3 \\
       M35 &    92.122290 &    24.483560 &  6129.6 &  4.28 & $-0.05$ &   $-0.5$ &  42.9 \\
       M35 &    92.123130 &    24.235860 &  6610.8 &  4.32 & $-0.09$ &   $-4.1$ &  37.0 \\
       M35 &    92.158984 &    24.370499 &  6412.6 &  4.36 &  0.03 &   $-8.7$ &  61.5 \\
       M35 &    92.175891 &    24.539354 &  6314.0 &  4.29 & $-0.03$ &   $-2.0$ &  47.1 \\
       M35 &    92.182670 &    24.286720 &  6175.1 &  4.32 & $-0.04$ &   $-8.0$ &  55.1 \\
       M35 &    92.199540 &    24.336500 &  6542.5 &  4.32 &  0.00 &   $-4.0$ &  87.5 \\
       M35 &    92.234890 &    24.451880 &  5348.3 &  4.56 & $-0.10$ &    0.6 &  21.1 \\
       M35 &    92.315396 &    24.513493 &  5702.4 &  4.39 & $-0.06$ &   $-1.0$ &  25.7 \\
       M35 &    92.330580 &    24.319080 &  6058.7 &  4.24 & $-0.06$ &   $-7.3$ &  34.2 \\
       M35 &    92.349589 &    24.388638 &  6030.1 &  4.26 &  0.08 &   $-6.7$ &  33.3 \\
       M35 &    92.380897 &    24.463773 &  5730.3 &  4.32 & $-0.08$ &   $-0.5$ &  13.1 \\
       M35 &    92.438109 &    23.577000 &  5867.9 &  4.20 & $-0.05$ &   $-6.7$ &  58.5 \\
       M35 &    92.440868 &    24.756354 &  5996.1 &  4.25 &  0.06 &   $-3.7$ &  33.4 \\
       M35 &    92.482379 &    24.504451 &  5502.2 &  4.50 & $-0.03$ &   $-5.2$ &  16.9 \\
       M35 &    92.529745 &    25.631488 &  6319.2 &  4.26 &  0.07 &    2.8 &  24.4 \\
       M35 &    92.569606 &    24.340346 &  4959.9 &  4.53 & $-0.09$ &   $-2.6$ &  16.6 \\
       M35 &    92.576234 &    25.305559 &  5962.3 &  4.25 & $-0.02$ &   $-7.7$ &  12.2 \\
       M35 &    92.698655 &    24.300170 &  5757.8 &  4.32 & $-0.02$ &   $-1.0$ &  30.5 \\
       M35 &    92.748451 &    25.319479 &  6168.3 &  4.02 & $-0.01$ &    1.8 &  18.2 \\
       M35 &    92.913694 &    24.970778 &  5860.5 &  4.25 & $-0.04$ &    1.5 &  19.7 \\
       M35 &    92.984542 &    25.291350 &  5464.7 &  4.52 & $-0.03$ &  $-10.8$ &  27.7 \\
       M35 &    93.036766 &    25.323962 &  6180.7 &  4.09 & $-0.02$ &    2.4 &  33.1 \\
       M35 &    93.070419 &    25.104811 &  6238.2 &  4.28 &  0.05 &   $-2.2$ &  47.6 \\
       M35 &    93.113441 &    23.558530 &  6403.2 &  4.28 &  0.09 &   $-2.5$ &  17.9 \\
       M35 &    93.406855 &    24.992339 &  6468.2 &  4.25 & $-0.23$ &    2.3 &  16.2 \\
       M67 &   132.255818 &    11.982972 &  6080.5 &  4.32 &  0.03 &   31.8 &  56.0 \\
       M67 &   132.270552 &    11.696430 &  5144.9 &  4.55 & $-0.00$ &   35.2 &  14.4 \\
       M67 &   132.284122 &    12.334856 &  6029.3 &  4.29 &  0.03 &   31.2 &  22.9 \\
       M67 &   132.313167 &    11.737440 &  5885.3 &  4.32 & $-0.03$ &   32.8 &  35.8 \\
       M67 &   132.366239 &    11.706203 &  5586.3 &  4.47 & $-0.05$ &   41.5 &  23.9 \\
       M67 &   132.400701 &    11.363308 &  5417.1 &  4.53 & $-0.25$ &   34.7 &  10.3 \\
       M67 &   132.433580 &    12.367976 &  5209.6 &  4.55 &  0.00 &   31.0 &  12.2 \\
       M67 &   132.438195 &    12.040859 &  5969.0 &  4.11 & $-0.13$ &   35.5 &  72.4 \\
       M67 &   132.440581 &    11.979247 &  5446.3 &  4.49 & $-0.01$ &   36.6 &  19.7 \\
       M67 &   132.467429 &    11.546726 &  5944.2 &  4.14 & $-0.03$ &   34.5 &  58.4 \\
       M67 &   132.485962 &    11.948136 &  5973.7 &  4.18 & $-0.07$ &   35.9 &  51.1 \\
       M67 &   132.503509 &    11.702729 &  5861.6 &  4.24 &  0.02 &   38.0 &  18.1 \\
       M67 &   132.541439 &    11.998321 &  5870.8 &  4.23 &  0.11 &   35.4 &  46.3 \\
       M67 &   132.550622 &    11.714708 &  5316.1 &  4.56 &  0.03 &   37.4 &  20.5 \\
       M67 &   132.561035 &    11.741567 &  5321.2 &  4.55 &  0.02 &   35.8 &  10.9 \\
       M67 &   132.567596 &    12.326211 &  5781.0 &  4.36 & $-0.17$ &   32.9 &  14.9 \\
       M67 &   132.570682 &    12.450383 &  6044.5 &  4.26 &  0.06 &   35.2 &  55.6 \\
       M67 &   132.576618 &    11.906679 &  5548.5 &  4.44 & $-0.01$ &   34.2 &  25.7 \\
       M67 &   132.577855 &    12.266950 &  5860.3 &  4.20 &  0.02 &   33.0 &  13.6 \\
       M67 &   132.583854 &    11.819741 &  6034.7 &  4.26 &  0.04 &   36.3 &  68.4 \\
       M67 &   132.583890 &    11.819694 &  6039.1 &  4.25 &  0.02 &   31.1 &  43.8 \\
       M67 &   132.589041 &    11.985801 &  5494.4 &  4.48 & $-0.16$ &   34.9 &  13.3 \\
       M67 &   132.589891 &    11.839766 &  5680.0 &  4.43 &  0.01 &   31.2 &  20.2 \\
       M67 &   132.590068 &    11.682061 &  5940.9 &  4.23 &  0.01 &   36.5 &  53.2 \\
       M67 &   132.620890 &    11.635912 &  5999.3 &  4.22 & $-0.04$ &   35.8 &  42.8 \\
       M67 &   132.658436 &    11.911304 &  5557.7 &  4.45 &  0.05 &   33.1 &  28.4 \\
       M67 &   132.658636 &    11.331912 &  5707.6 &  4.43 &  0.04 &   41.1 &  37.0 \\
       M67 &   132.661141 &    11.203579 &  5875.5 &  4.22 &  0.08 &   39.9 &  41.3 \\
       M67 &   132.675421 &    11.439848 &  6021.5 &  4.14 & $-0.07$ &   34.1 &  67.4 \\
       M67 &   132.677078 &    11.663713 &  6063.8 &  4.21 &  0.12 &   36.4 &  42.1 \\
       M67 &   132.685551 &    11.990308 &  5749.3 &  4.22 & $-0.53$ &   38.5 &  21.9 \\
       M67 &   132.696363 &    11.715231 &  5842.2 &  4.33 &  0.05 &   40.4 &  31.8 \\
       M67 &   132.702588 &    12.061477 &  6025.5 &  4.16 & $-0.03$ &   32.3 &  41.3 \\
\hline
\end{tabular}
\end{minipage} \end{table*}

\setcounter{table}{0}
\begin{table*} \begin{minipage}[]{180mm} \centering
\caption{\it -- continued}
\label{}
\begin{tabular}{cccccrrr}
\hline
Cluster & RA (2000.0) & Dec (2000.0) & $T_{\rm eff}$ & log\,$g$ & [Fe/H] & $V_{\rm r}$ & SNR \\
 &  (deg.) & (deg.) & (K) & (cm\,s$^{-2}$) & (dex)  & km\,s$^{-1}$ &  \\

\hline
       M67 &   132.717930 &    11.750991 &  5579.2 &  4.47 & $-0.11$ &   40.0 &  15.0 \\
       M67 &   132.721149 &    11.667290 &  5926.7 &  4.19 & $-0.00$ &   36.4 &  59.4 \\
       M67 &   132.721518 &    11.792825 &  5973.6 &  4.18 & $-0.02$ &   38.6 &  51.8 \\
       M67 &   132.726414 &    12.259466 &  6005.5 &  4.24 & $-0.05$ &   39.2 &  39.2 \\
       M67 &   132.733352 &    11.897774 &  5947.4 &  4.10 & $-0.10$ &   32.9 &  70.2 \\
       M67 &   132.735524 &    11.635607 &  5795.0 &  4.33 & $-0.06$ &   31.9 &  12.3 \\
       M67 &   132.738335 &    12.093445 &  5869.2 &  4.25 &  0.04 &   30.6 &  16.3 \\
       M67 &   132.741071 &    11.587439 &  5862.0 &  4.31 &  0.03 &   33.5 &  42.5 \\
       M67 &   132.749192 &    11.853498 &  5509.5 &  4.49 & $-0.11$ &   34.3 &  24.2 \\
       M67 &   132.783816 &    12.018475 &  5819.7 &  4.19 & $-0.15$ &   38.3 &  55.5 \\
       M67 &   132.787759 &    12.050611 &  5980.3 &  4.20 & $-0.10$ &   37.5 &  36.2 \\
       M67 &   132.791306 &    11.771373 &  5674.4 &  4.43 & $-0.08$ &   35.4 &  40.7 \\
       M67 &   132.796550 &    11.443022 &  5787.0 &  4.28 & $-0.19$ &   32.5 &  46.6 \\
       M67 &   132.801343 &    11.310762 &  5859.4 &  4.26 & $-0.01$ &   36.8 &  32.4 \\
       M67 &   132.823736 &    12.461495 &  5697.6 &  4.37 &  0.10 &   34.2 &  13.1 \\
       M67 &   132.836630 &    11.661145 &  5858.8 &  4.21 &  0.04 &   37.6 &  40.6 \\
       M67 &   132.840996 &    11.721588 &  5687.0 &  4.43 &  0.00 &   30.9 &  15.6 \\
       M67 &   132.846624 &    11.654122 &  5533.5 &  4.44 &  0.04 &   37.6 &  17.3 \\
       M67 &   132.860338 &    11.643576 &  5881.6 &  4.13 & $-0.15$ &   37.3 &  62.2 \\
       M67 &   132.875371 &    11.458759 &  5564.7 &  4.47 & $-0.07$ &   32.7 &  13.9 \\
       M67 &   132.911410 &    11.710382 &  5883.0 &  4.34 & $-0.03$ &   37.6 &  36.8 \\
       M67 &   132.915392 &    12.203025 &  5124.3 &  4.55 &  0.05 &   38.4 &  12.2 \\
       M67 &   132.921686 &    12.409107 &  5959.7 &  4.24 & $-0.00$ &   34.5 &  19.0 \\
       M67 &   132.937204 &    11.649700 &  6025.4 &  4.22 &  0.02 &   35.0 &  51.5 \\
       M67 &   132.940772 &    11.645889 &  5665.1 &  4.28 & $-0.79$ &   36.6 &  26.5 \\
       M67 &   132.969325 &    11.513024 &  5921.7 &  4.17 & $-0.13$ &   37.6 &  38.7 \\
       M67 &   132.995892 &    11.696994 &  5523.0 &  4.45 &  0.05 &   35.9 &  19.0 \\
       M67 &   133.005656 &    12.186139 &  6014.3 &  4.18 & $-0.05$ &   32.3 &  44.7 \\
       M67 &   133.006516 &    12.065050 &  5846.5 &  4.30 &  0.03 &   33.4 &  25.9 \\
       M67 &   133.034143 &    12.262911 &  5798.6 &  4.31 & $-0.03$ &   30.8 &  13.1 \\
       M67 &   133.041442 &    12.175305 &  5932.8 &  4.14 &  0.16 &   40.4 &  11.1 \\
       M67 &   133.058637 &    12.212279 &  5999.1 &  4.24 &  0.00 &   35.1 &  47.0 \\
       M67 &   133.072200 &    12.334044 &  5548.7 &  4.46 & $-0.03$ &   35.3 &  22.0 \\
       M67 &   133.091003 &    11.779630 &  5974.0 &  4.15 & $-0.06$ &   35.3 &  22.3 \\
       M67 &   133.104021 &    11.174182 &  5808.7 &  4.24 &  0.05 &   34.8 &  32.5 \\
       M67 &   133.104041 &    11.174156 &  5863.8 &  4.19 &  0.03 &   39.1 &  17.7 \\
       M67 &   133.107283 &    12.384580 &  5890.0 &  4.20 & $-0.01$ &   30.7 &  16.1 \\
       M67 &   133.126162 &    11.714005 &  5845.2 &  4.26 &  0.04 &   33.4 &  25.0 \\
       M67 &   133.137844 &    11.756676 &  6027.4 &  4.25 &  0.01 &   40.0 &  46.4 \\
       M67 &   133.149849 &    11.623530 &  5470.8 &  4.52 & $-0.03$ &   38.0 &  14.7 \\
       M67 &   133.186029 &    11.430369 &  5935.1 &  4.23 & $-0.02$ &   36.0 &  44.6 \\
       M67 &   133.244947 &    11.757958 &  5122.3 &  4.55 & $-0.06$ &   34.8 &  13.7 \\
       M67 &   133.282203 &    11.803569 &  6007.5 &  4.08 & $-0.08$ &   33.9 &  48.9 \\
       M67 &   133.291121 &    12.004910 &  5760.5 &  4.26 & $-0.01$ &   35.2 &  11.3 \\
       M67 &   133.301250 &    11.674549 &  5602.2 &  4.40 & $-0.00$ &   32.2 &  26.9 \\
       M67 &   133.327231 &    12.366576 &  5587.6 &  4.23 & $-0.55$ &   39.0 &  11.9 \\
       M67 &   133.340579 &    11.287396 &  5967.9 &  4.25 & $-0.02$ &   35.6 &  39.2 \\
       M67 &   133.354071 &    12.417892 &  5654.2 &  4.44 & $-0.03$ &   35.0 &  22.1 \\
       M67 &   133.384039 &    12.186201 &  6023.8 &  4.04 & $-0.18$ &   33.8 &  51.3 \\
       M67 &   133.464535 &    11.292121 &  5856.3 &  4.00 &  0.24 &   35.5 &  63.2 \\
       M67 &   133.490143 &    12.262763 &  6028.8 &  4.28 &  0.01 &   36.2 &  51.2 \\
       M67 &   133.510279 &    11.875851 &  5405.7 &  4.57 & $-0.06$ &   35.9 &  16.3 \\
       M67 &   133.511237 &    12.134510 &  6085.8 &  4.22 &  0.08 &   36.1 &  10.9 \\
       M67 &   133.525382 &    11.753270 &  5579.1 &  4.49 & $-0.16$ &   38.3 &  19.9 \\

\hline
\end{tabular}
\end{minipage} \end{table*}

\end{document}